%% file: 3DPixelSensors_CMSTracker.tex
\documentclass[final,3p,times,twocolumn]{elsarticle}

\usepackage{amssymb}
\usepackage{supertabular,booktabs}
\usepackage{subfig}
\usepackage{graphicx}
\newcommand{\figsubref}[2]{\ref{#1}\subref{#2}}
\usepackage{appendix}
\usepackage{hyperref}
\hypersetup{
    colorlinks=true,
}
\usepackage[table,xcdraw,dvipsnames,x11names]{xcolor}
\usepackage[load-configurations = abbreviations]{siunitx}
\usepackage{lineno}
\usepackage{afterpage}
\newcommand{\dhyphen}{\mbox{--}}

\newcommand{\chndof}{\ensuremath{\chi^2/\text{ndof}}}

\newlength{\subcolumnwidth}
\newenvironment{subcolumns}[1][0.45\columnwidth]
 {\valign\bgroup\hsize=#1\setlength{\subcolumnwidth}{\hsize}\vfil##\vfil\cr}
 {\crcr\egroup}
\newcommand{\nextsubcolumn}[1][]{%
  \cr\noalign{\hfill}
  \if\relax\detokenize{#1}\relax\else\hsize=#1\setlength{\subcolumnwidth}{\hsize}\fi
}
\newcommand{\nextsubfigure}{\vfill}

\newcommand{\cmsAuthorMark}[1]
{\hbox{\textsuperscript{\normalfont#1}}}

\journal{Nuclear Physics A}

\begin{document}

\begin{frontmatter}

\title{Evaluation of 3D pixel silicon sensors for the CMS Phase-2 Inner Tracker}

\author{The Tracker Group of the CMS Collaboration\corref{cor1}\fnref{label2}}
\cortext[cor1]{\begin{minipage}[t]{\linewidth}
Clara Lasaosa Garc\'ia, clara.lasaosa.garcia@cern.ch\\
Davide Zuolo, davide.zuolo@cern.ch\\
\end{minipage}}
\fntext[label2]{Complete author list at the end of the document}

\begin{abstract}
The high-luminosity upgrade of the CERN LHC requires the replacement of the CMS tracking detector to cope with the increased radiation fluence while maintaining its excellent performance. An extensive R\&D program, aiming at using 3D pixel silicon sensors in the innermost barrel layer of the detector, has been carried out by CMS in collaboration with the FBK (Trento, Italy) and CNM (Barcelona, Spain) foundries. The sensors will feature a pixel cell size of $25\times\SI{100}{\micro m^2}$, with a centrally located electrode connected to the readout chip. The sensors are read out by the RD53A and CROCv1 chips, developed in 65~nm CMOS technology by the RD53 Collaboration, a joint effort between the ATLAS and CMS groups. This paper reports the results achieved in beam test experiments before and after irradiation, up to a fluence of approximately $2.6\times10^{16}$~n$_{\text{eq}}$/cm$^2$. Measurements of assemblies irradiated to a fluence of \mbox{\SI{1e16}{n_{eq}/\cm^{2}}} show a hit detection efficiency higher than 96\% at normal incidence, with fewer than 2\% of channels masked, across a bias voltage range greater than $\SI{50}{V}$. Even after irradiation to a higher fluence of \mbox{\SI{1.6e16}{n_{eq}/\cm^{2}}}, similar performance is maintained over a bias voltage range of $\SI{30}{V}$, remaining well within CMS requirements.
 
\end{abstract}


\end{frontmatter}

%\begin{linenumbers}
\section{Introduction}
\label{sec:intro}
The High-Luminosity LHC (HL-LHC)~\cite{highlumi} is an upgrade of the CERN LHC expected to operate for a period longer than 10 years starting in 2030. It will run at a nominal center-of-mass energy of $\SI{14}{TeV}$ with a $\SI{25}{ns}$ bunch spacing. The peak luminosity will reach \mbox{\SI{5e34}{cm^{-2}s^{-1}}} with an average of up to 140 overlapping proton-proton collisions per bunch crossing. This will enable the CMS experiment~\cite{CMS_Phase0,CMS_Phase1} to collect an integrated luminosity of up to $\SI{3000}{\femto b^{-1}}$ over its lifetime. The increased luminosity will greatly expand the physics potential of the LHC, allowing precision studies of the standard model and searches for statistically rare processes beyond the standard model. The unprecedented conditions require upgrades to many components of the CMS detector. In particular, the innermost sub-detector, which is made of silicon pixel modules, will be completely replaced with a new system called the Inner Tracker~(IT), to withstand the foreseen radiation levels with improved performance~\cite{Phase2TrackerTDR}. Successful operation under HL-LHC conditions requires pixel sensors with high radiation tolerance, a low material budget, increased granularity to improve the spatial resolution, high single-hit detection efficiency, and minimal power dissipation.

\begin{figure}[htb!]
    \centerline{\includegraphics[width=3.1in]{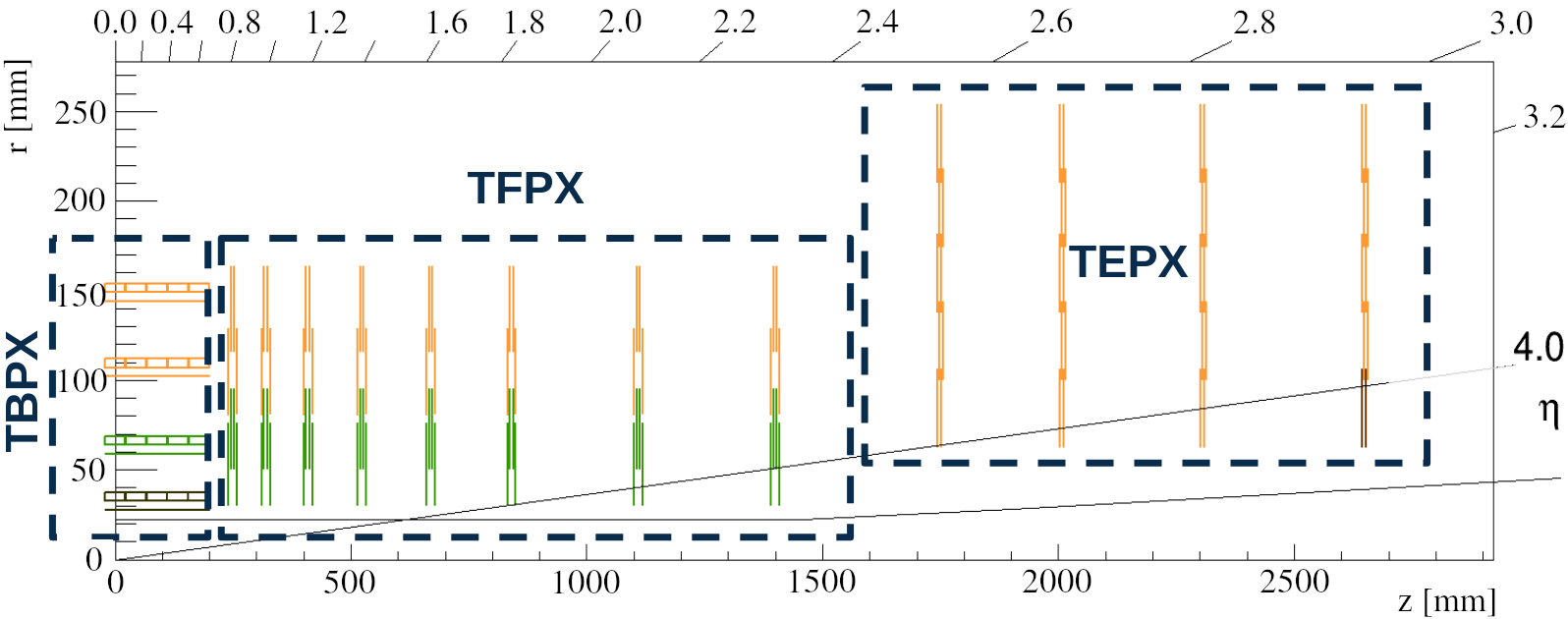}}
    \caption{A slice of a quarter of the upgraded CMS IT layout in the longitudinal view. It consists of three substructures: TBPX, TFPX, and TEPX. Planar pixel modules with two and four readout chips are depicted in green and orange, respectively. The 3D pixel modules with two readout chips for the innermost TBPX layer are shown in black. The innermost ring of the last TEPX disc (shown in brown) features modules with four readout chips and sends data to the beam luminosity system. The thin lower black line represents the outer radius of the beam pipe.}
    \label{fig:ITlayout}
\end{figure}

The current CMS pixel detector~\cite{PixelPhase1} is made of modules with $n$-in-$n$ planar silicon sensors with a thickness of $\SI{285}{\micro m}$ and a cell size of $100\times150$~$\SI{}{\micro m^{2}}$, providing acceptance up to pseudorapidities of $|\eta| = 3.0$. The IT extends the tracking coverage up to pseudorapidities of $|\eta| = 4.0$ with an optimized layer arrangement and improves the pixel granularity by a factor of six. One quarter of the IT layout is shown in Fig.~\ref{fig:ITlayout}. It consists of three substructures: the Tracker Barrel Pixel (TBPX) with four layers, the Tracker Forward Pixel (TFPX) with eight small double-discs at each end, and the Tracker Endcap Pixel (TEPX) with four large double-discs per end. The IT structure has been designed to allow for maintenance and possible repairs during sufficiently long accelerator shutdown periods.
The IT will be constructed from hybrid pixel modules with two or four readout chips, referred to as double and quad modules, respectively. The pixel cell size and nominal active thickness have been reduced to $25\times100$~$\SI{}{\micro m^{2}}$ and $\SI{150}{\micro m}$, respectively, in order to meet the above-mentioned requirements. Square pixel cells with a $\SI{50}{\micro m}$ pitch were initially considered as an option for the endcap and forward regions of the detector. However, simulations of track and jet reconstruction, as well as heavy flavor tagging efficiency, showed only marginal gains in performance. This marginal improvement did not justify the additional cost of producing two different sensor geometries, so this option was not pursued. 

Semiconductor tracking devices suffer radiation damage when exposed to high particle fluences. The main effects are an increase in leakage current (and thus noise), changes in resistivity, and reduced charge collection due to carrier trapping. Because of the high fluences expected at the HL-LHC, it is essential to irradiate samples and study the effects as part of the sensor qualification program. 

To mitigate these effects, the baseline design uses n-in-p planar silicon pixel sensors bump-bonded to readout chips. This configuration was chosen primarily because, in irradiated n-in-p silicon, charge is induced mainly by electrons, which are less susceptible to trapping due to their higher mobility. This sensor type also does not undergo effective type inversion of the bulk material after irradiation. Moreover, its fabrication process is less expensive compared to the double-sided n-in-n planar sensors currently used in CMS~\cite{CMS_Phase1}.

In the layer closest to the interaction point — located at a radius of only $\SI{3}{cm}$ and expected to receive a $\SI{1}{MeV}$ neutron equivalent fluence ($\mathrm{n_{eq}}/\mathrm{cm}^2$) of \mbox{\SI{2.6e16}{n_{eq}/\cm^{2}}} and a total ionizing dose (TID) of $\SI{13.4}{MGy}$ — 3D pixel sensors have been shown to be the best option owing to their radiation tolerance~\cite{3datlas} and lower power consumption. The qualification of these sensors is the subject of this paper.

These 3D pixel sensors have columnar electrodes penetrating the bulk perpendicular to the sensor surface (Fig.~\ref{fig:3D}). Unlike in planar pixel sensors, where the electrodes are parallel to the surface, the inter-electrode distance is independent from the active thickness of the device. This results in a lower depletion voltage and, consequently, lower power dissipation, as well as a shorter collection path for the charge carriers and, therefore, reduced trapping probability after irradiation. However, there are some drawbacks with respect to planar pixel sensors, such as the loss of hit detection efficiency within the electrodes at normal particle incidence and a lower production yield. Additionally, 3D pixel sensors exhibit lower homogeneity in the electric field~\cite{Loi_2021}, with regions of zero field within the pixel cell, which affects charge collection.

\begin{figure}[htb!]
    \centerline{\includegraphics[width=3.1in]{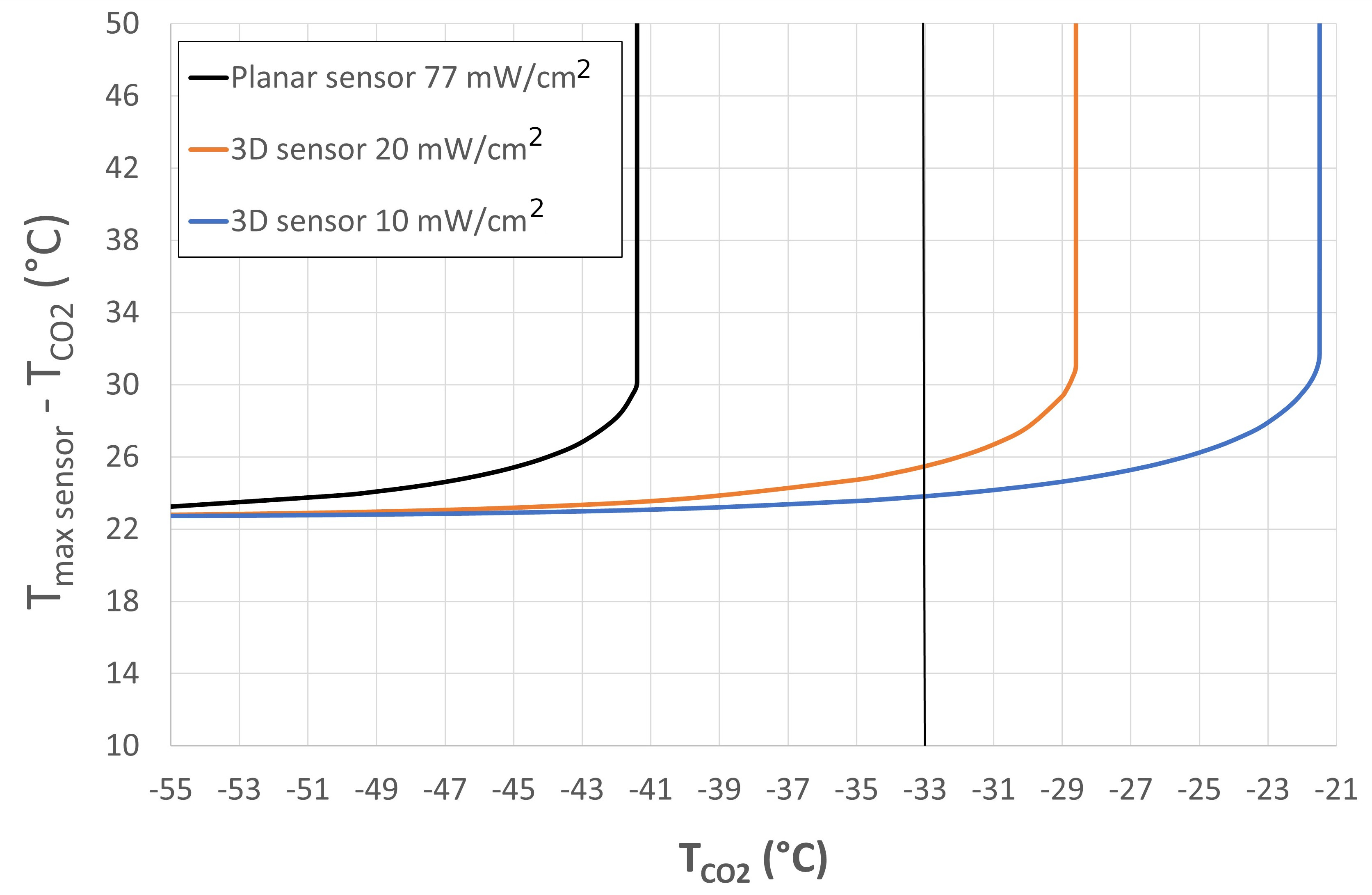}}
    \caption{Thermal simulation for planar (black) and 3D (orange/blue) pixel sensor modules in the innermost layer of the barrel. The comparison is done for a simulated fluence of \mbox{\SI{2e16}{n_{eq}/\cm^{2}}}, assuming operation voltages of $\SI{600}{V}$ and $\SI{140}{V}$ for planar and 3D pixel sensors, respectively. The minimum temperature the coolant can reach at the modules is depicted with a vertical black line (T${_\mathrm{CO2}} = \SI{-33}{\degree C}$).}
    \label{fig:thermal_simulation_L1}
\end{figure}

The thermal properties of the first layer of the barrel section have been extensively simulated using Ansys Fluent~\cite{ansys} for planar and 3D pixel sensors under different fluence and power consumption scenarios. Figure~\ref{fig:thermal_simulation_L1} shows the outcome of the thermal simulation for a fluence of \mbox{\SI{2e16}{n_{eq}/\cm^{2}}}, which is expected to be reached after approximately seven years of HL-LHC operation. The power consumption for planar pixel modules was determined from extensive laboratory measurements, while for 3D pixel sensors, two different values were selected based on leakage current measurements during test beam campaigns. The curves in the figure turn into vertical lines when the $\mathrm{CO_2}$ coolant temperature is no longer low enough to effectively dissipate the heat generated by the sensor, leading to increased and uncontrolled heat generation, a regime known as thermal runaway. The temperature needed to prevent thermal runaway in the planar pixel modules is significantly below the minimum $\mathrm{CO_2}$ temperature ($\SI{-33}{\degree C}$), while for the 3D pixel sensors, there is a margin of more than $\SI{10}{\degree C}$ in the low-power scenario and $\SI{4}{\degree C}$ in the high-power scenario.

The requirements specified for the production of 3D pixel sensors are summarized in Table~\ref{tab:sensorRequirements}. The qualification program foresees measurements after irradiation of the sensors up to \mbox{\SI{1.5e16}{n_{eq}/\cm^{2}}}, following the performance established in the test beam campaigns presented in this paper. It has been estimated that this fluence will be reached after the first six years of operation. At the end of this period the innermost layer will be replaced, as was done for the current CMS pixel detector.

This article presents the performance of prototype 3D pixel silicon sensors, arranged in single-chip modules, which were tested to validate their use in the IT. A detailed description of the sensors and readout ASICs is given in Sections~\ref{sec:sensor_description} and~\ref{sec:readout_chip}, respectively. Section~\ref{sec:facilities_setups} describes the irradiation campaigns and beam test infrastructure. The configuration and calibration of the ASIC is explained in Section~\ref{sec:tuning} and the methodology for the data analysis is described in Section~\ref{sec:analysis}. The main results for non-irradiated and irradiated modules are reported in Sections~\ref{sec:fresh_results} and~\ref{sec:irradiated_results}.
\begin{table}[htb!]
    \vspace{-0.2cm}
    \centering
    \caption{Requirements for 3D pixel sensors for the IT before and after irradiation. The depletion voltage is denoted as $V_{\text{depl}}$ and the minimum voltage at which the irradiated sensors achieve the hit efficiency specification is referred to as $V_{\text{op}}$. The fluence is denoted by $\Phi$ and measured in units of $\mathrm{n_{eq}}/\/\mathrm{cm^{2}}$.}
    \label{tab:sensorRequirements}
    \begin{tabular}{lcc}
    \hline\hline
     Parameter &\hspace{-0.25cm} Value &\hspace{-0.2cm} Notes\\ 
    \hline\hline
     Depletion voltage &\hspace{-0.25cm} $<\SI{10}{V}$ &\hspace{-0.2cm} non-irradiated\\
     Breakdown voltage &\hspace{-0.25cm} $>V_{\text{depl}}+\SI{25}{V}$ &\hspace{-0.2cm} non-irradiated\\
     Leakage current at\\ room temperature\\ and $V_{\text{depl}}+\SI{25}{V}$ &\hspace{-0.25cm} $<\SI{2.5}{\micro Acm^{-2}}$ &\hspace{-0.2cm} non-irradiated\\
     Maximum $V_{\text{op}}$ &\hspace{-0.25cm} $<\SI{200}{V}$ &\hspace{-0.2cm} $\Phi=\SI{1.5e16}{}$\\
     Maximum current\\
     at $V_{\text{op}}$ and $\SI{-25}{\degree C}$ &\hspace{-0.43cm}  $<\SI{150}{\micro Acm^{-2}}$ &\hspace{-0.2cm} $\Phi=\SI{1.5e16}{}$\\
     Hit efficiency at\\
     normal incidence &\hspace{-0.25cm} $\geq96\%$ &\hspace{-0.2cm} $\Phi=\SI{1.5e16}{}$\\
     Hit efficiency at\\
     10º incidence &\hspace{-0.25cm} $\geq97\%$ &\hspace{-0.2cm} $\Phi=\SI{1.5e16}{}$\\
     Masked channels &\hspace{-0.25cm} $<2\%$ &\hspace{-0.2cm} $\Phi=\SI{1e16}{}$\\
     \hline\hline
    \end{tabular}
\end{table}

%%%%%%%%%%%%%%%%%%%%%%%%%%%%%%%%%
%%%%%%%%%%%%%%%%%%%%%%%%%%%%%%%%%

\vspace{-0.35cm}
\section{Sensor description}
\label{sec:sensor_description}

The 3D pixel silicon sensors described in this article were manufactured by two companies: Fondazione Bruno Kessler (FBK)~\cite{FBK} and Centro Nacional de Microelectrónica (CNM)~\cite{CNM}. Both companies use silicon wafers composed of two \mbox{$p$-type} layers, one with low and one with high resistivity, bonded together by means of the Direct Wafer Bonding technique developed by ICEMOS~\cite{ICEMOS}. The low resistivity layer provides mechanical support and ohmic contact to the high resistivity layer, which is the active region of the sensor. The high resistivity layer has a nominal thickness of $\SI{150}{\micro m}$ and a resistivity greater than $\SI{3}{k\Omega cm}$, while the low resistivity layer has an initial thickness of $\SI{500}{\micro m}$, which is then thinned down to $\SI{100}{\micro m}$, and a resistivity of $0.1-1$~$\SI{}{\Omega cm}$. The doping concentration in the high resistivity layer is lower than $\SI{4.5e12}{cm^{-3}}$.  The columns are produced by a single-sided Deep Reactive Ion Etching (DRIE) process. The $n^{+}$ readout columns on $p$-type bulk material have a spatial density of about $\SI{4e4}{cm^{-2}}$.

\begin{figure}[htb!]
  \centering
  \subfloat[]{\includegraphics[width=0.36\textwidth]{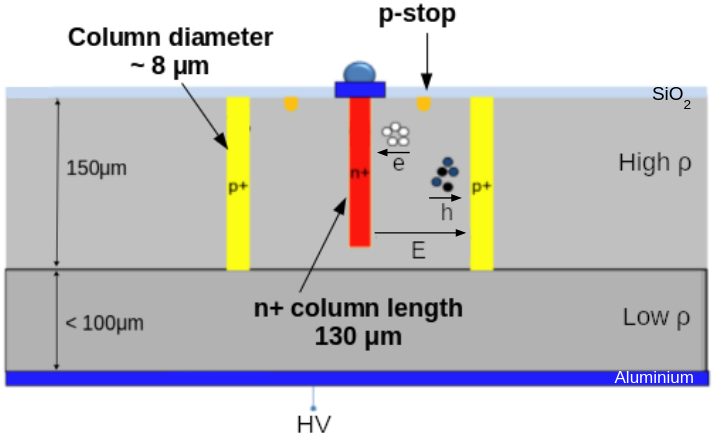}}
  \nextsubfigure
  \subfloat[]{\includegraphics[width=0.475\textwidth]{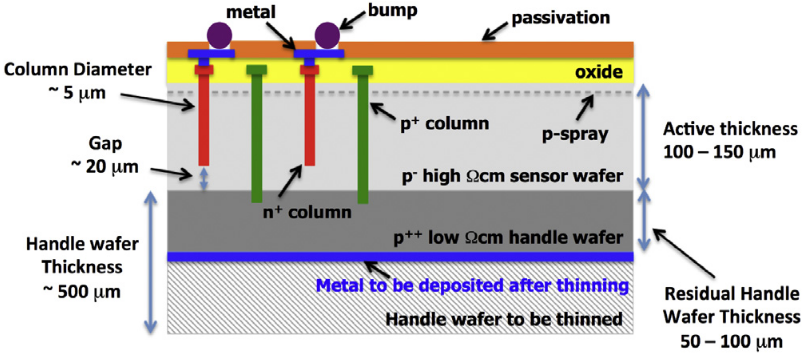}}
  \caption{Cross-section of a typical (a) CNM and (b) FBK 3D pixel silicon sensor, showing the high resistivity and low resistivity layers, together with the $p^+$ and $n^+$ columns~\cite{3D,3D_2}.}
  \label{fig:3D}
\end{figure}

Figure~\ref{fig:3D} shows sketches of the 3D pixel sensors produced by the two foundries. The main differences between the FBK and CNM processes are the \mbox{$n^+$ column} geometry and the inter-column isolation. The \mbox{$n^+$ columns} electrically connect the ASIC to the sensor through the bumps. FBK sensors have \mbox{$n^+$ columns} with a diameter of $\SI{5}{\micro m}$ and a length of $\SI{130}{\micro m}$ or $\SI{115}{\micro m}$, corresponding to two production versions referred to as Stepper-1 and Stepper-2, respectively. The shorter columns in Stepper-2 were introduced in more recent production campaigns aiming to reduce noise observed in irradiated sensors, as discussed later. CNM sensors have $n^+$ columns with $\SI{8}{\micro m}$ diameter and $\SI{130}{\micro m}$ length. The isolation structures for both producers consist of implantations on the surface of the high resistivity wafer. CNM employs the $p$-stop technique for inter-column isolation (an additional $p^+$ implantation between the $n^+$ ones), while FBK uses the $p$-spray method (a continuous deposition of low dose $p$-dopants on the sensor surface).

\begin{figure}[htb!]
\centering
  \subfloat[]{\includegraphics[width=4.5cm]{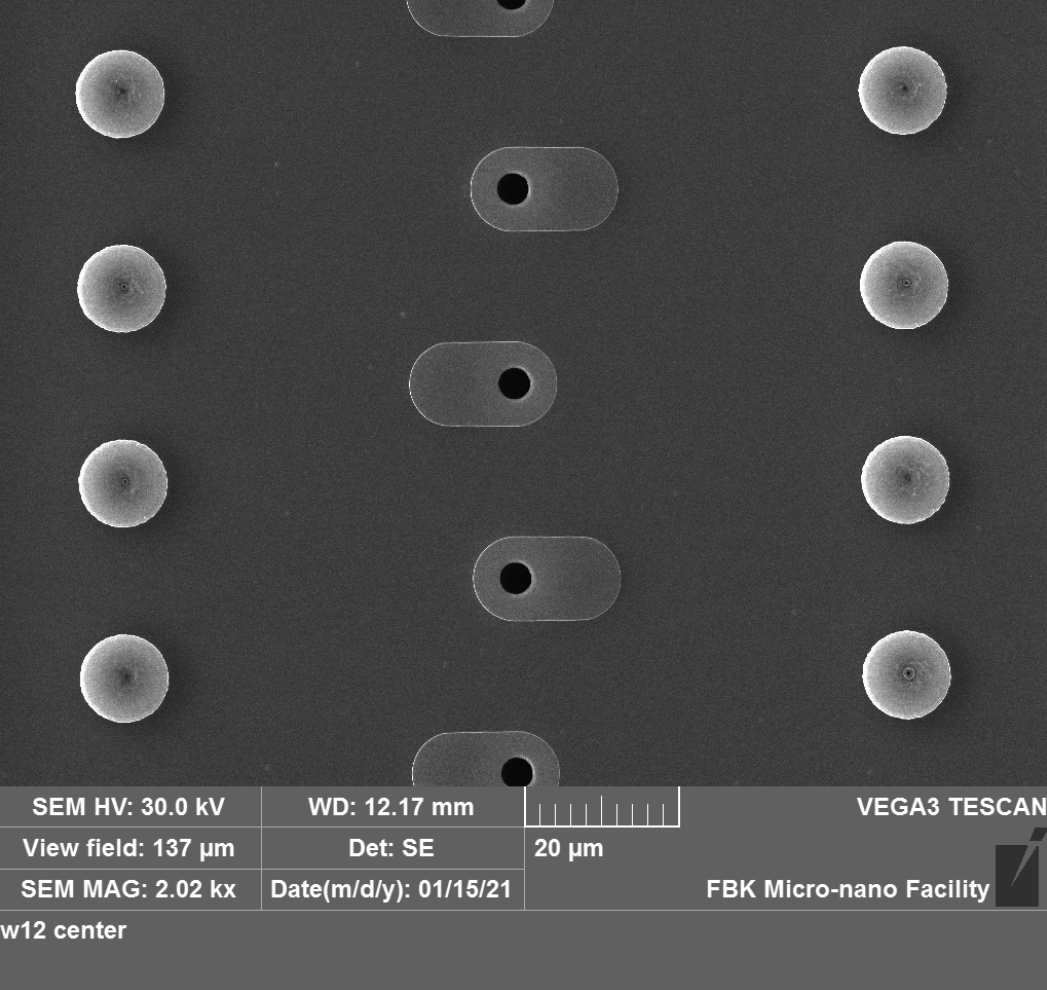}\label{fig:top_view}}
  \nextsubfigure
  \subfloat[]{\includegraphics[width=4.5cm]{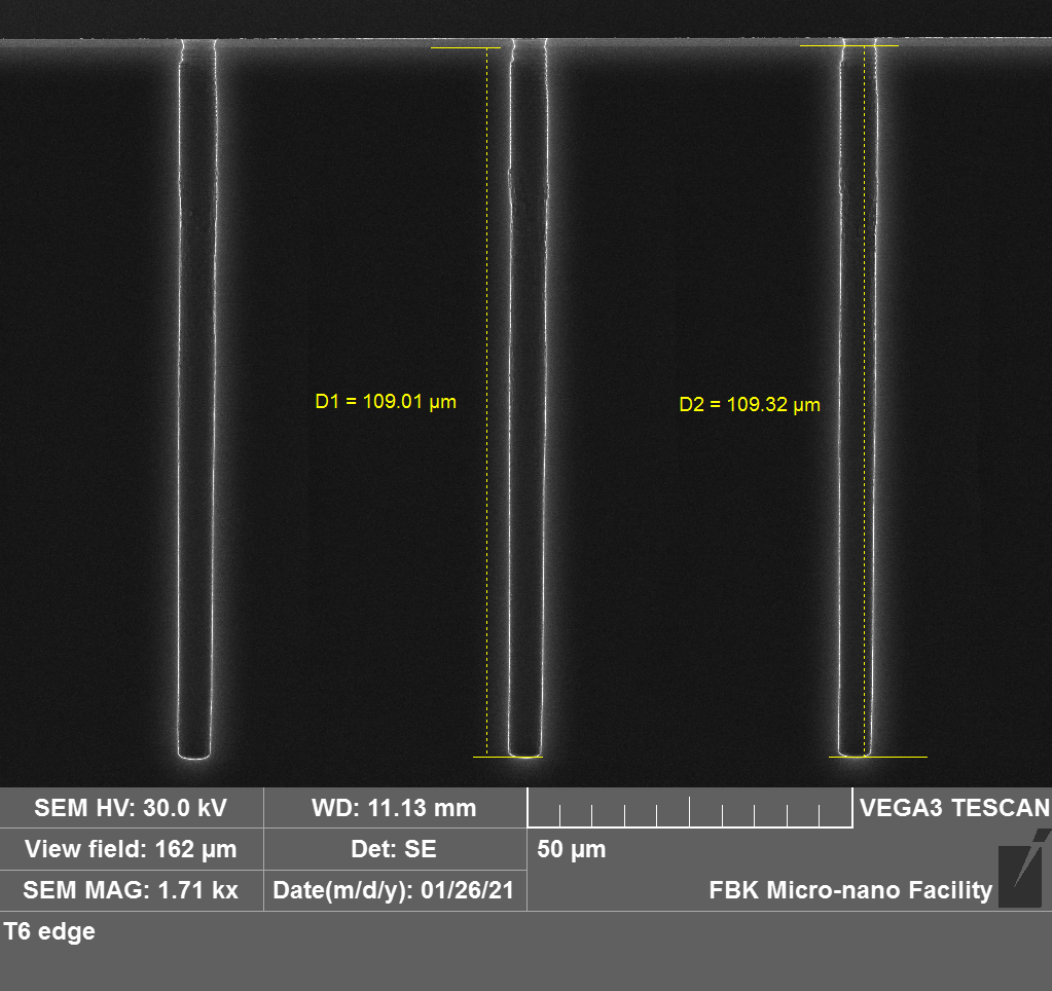}\label{fig:side_view}}
  \caption{Electron microscope images of an FBK sensor: (a) top view and (b) cross-section. Only the $n^+$ columns are visible in the latter.}
  \label{fig:sem}
\end{figure}

Two electron microscope images of an FBK sensor are presented in Fig.~\ref{fig:sem}, showing the top view and cross-section of a silicon wafer with rectangular pixel cells. Each pixel cell features a single $n^+$ columnar electrode at the center and four $p^+$ columns at the corners.
The sensor size is approximately $22\times\SI{17}{mm^{2}}$. The nominal thickness of the active layer is chosen to be $\SI{150}{\micro m}$, but its effective thickness is reduced by about $\SI{10}{\micro m}$ due to boron diffusion from the carrier wafer, i.e., the low resistivity layer~\cite{DOPING}. The expected Most Probable Value (MPV) of the collected charge for a Minimum Ionizing Particle (MIP) traversing a nominal $\SI{150}{\micro m}$ thick active layer sensor is about $\SI{10200}{electrons}$~\cite{Bichsel}.

%%%%%%%%%%%%%%%%%%%%%%%%%%%%%%%%%
%%%%%%%%%%%%%%%%%%%%%%%%%%%%%%%%%

\section{Readout chip description}
\label{sec:readout_chip}
The RD53 Collaboration has been working since 2014 on developing pixel readout chips for the ATLAS and CMS Phase-2 upgrades with $50\times\SI{50}{\micro m^2}$ readout cells that can sustain very high hit rates (up to $\SI{3.5}{GHz/\cm^2}$) in an extremely hostile radiation environment. The choice of the readout cell size was based on the fact that both experiments planned to use sensors with either $50\times\SI{50}{\micro m^2}$ or $25\times\SI{100}{\micro m^2}$ pixel cells.

The first prototype, named RD53A, became available in 2018.  Its readout matrix has 400 columns and 192 rows, resulting in a size of $20\times\SI{11.6}{mm^{2}}$. Three different analog front-end (AFE) designs \dhyphen ``Synchronous", ``Linear", and ``Differential"\dhyphen were implemented, with each connected to one third of the pixel matrix. Following an extensive characterization campaign of bench and beam tests, the CMS Collaboration chose the linear AFE~\cite{AFEChoice}. 

The first version of the CMS-specific chip, named CROC (CMS ReadOut Chip), was submitted in 2021.  This ``CROCv1" chip features a larger pixel matrix with 432 columns and 336 rows, as well as a revised linear AFE with improved comparator and threshold-trimming Digital-to-Analog Converter (DAC)~\cite{linafe}. The final version of the chip, referred to as CROCv2, includes minor bug fixes and additional features, and was first released in 2024. The size of the final chip is $22\times\SI{19}{mm^{2}}$.

All chips are equipped with an internal charge injection circuit that is used to measure the threshold and noise of each pixel. The circuit includes a capacitor connected to the outputs of two 12-bit voltage DACs. The amplitude of the voltage step supplied to the capacitor, $\Delta$VCal, is the difference between the voltage outputs of the two DACs.

\begin{figure}[htb!]
  \centering
  \includegraphics[width=0.475\textwidth]{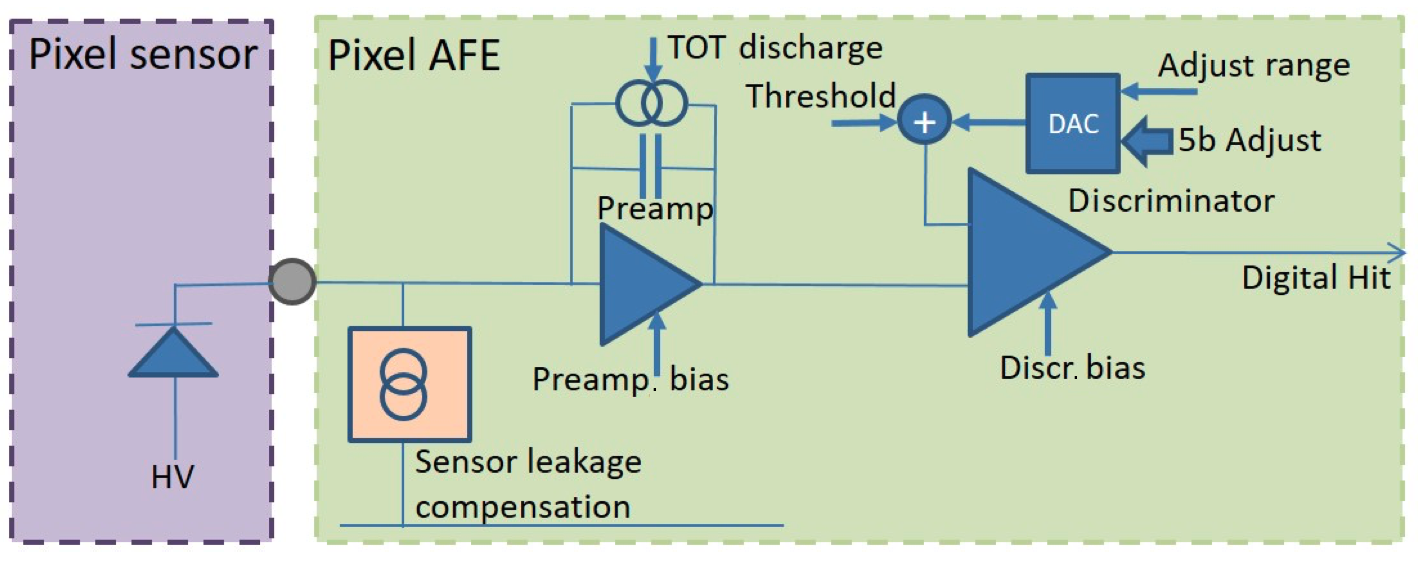}
  \caption{Schematic view of the linear AFE.}
  \label{fig:linafe}
\end{figure}

The readout chain of these chips~\cite{RD53Amanual, CROC}, shown in Fig.~\ref{fig:linafe}, includes a charge-sensitive amplifier (CSA) with Krummenacher feedback, designed to compensate for the expected large radiation-induced increase in detector leakage current and ensure a linear discharge of the feedback capacitor. The signal from the CSA is fed to a low-power comparator. This comparator provides a time-to-digital conversion together with a Time-over-Threshold (ToT) counter that measures time in integer $\SI{40}{MHz}$ clock cycles. Further conversion of the measured charge from ToT units to electrons is possible once a calibration of the gain has been performed. Channel-to-channel threshold dispersion is addressed by means of a local circuit for adjustment, based on a 4-bit (RD53A) or 5-bit (CROCv1 and CROCv2) current-mode, binary-weighted DAC.

\newpage
To achieve a faster discharge, the current feeding the Krummenacher feedback circuit can be increased at the cost of higher power dissipation in the chip. ROCs bonded to 3D pixel sensors in the innermost layer of the IT will be operated in this fast discharge mode. The expected hit rate in this layer is so high that two particles coming from different bunch crossings may hit the same pixel within the Krummenacher circuit discharge time, leading to a loss of hit detection efficiency. Dedicated runs have been included in the test beam campaigns to verify that the performance of the sensors remains unaltered when the ROCs are operated in fast discharge mode.

\begin{figure}[htb!]
\vspace{-0.7cm}
  \centering
  \includegraphics[width=0.38\textwidth]{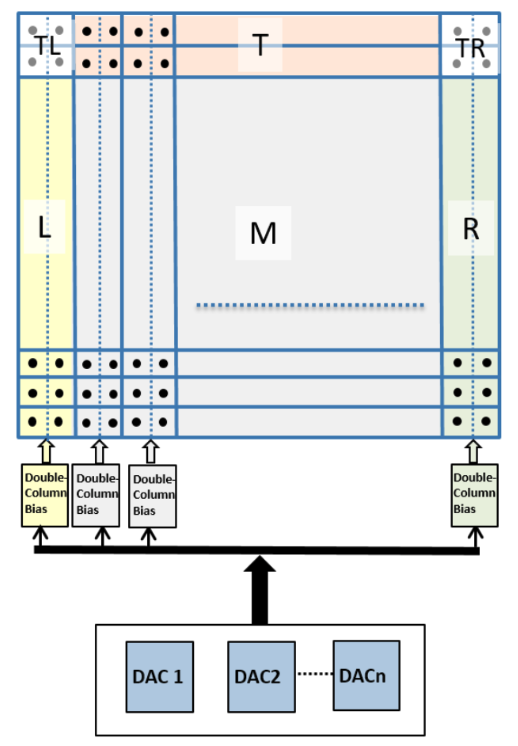}
  \caption{Schematic view of the CROCv1 and CROCv2 power distribution, highlighting the regions available for different tunings.}
  \label{fig:CROC_Special_Columns}
\end{figure}

The improved design of the CROCv1 and CROCv2 chips provides the possibility to set independent thresholds and bias voltages for the AFE across six different regions, as shown in Fig.~\ref{fig:CROC_Special_Columns}: the interior of the chip (M), left edge (L), right edge (R), top (T), top-left corner (TL) and top-right corner (TR).

This design choice arises from the fact that double and quad planar pixel modules will feature a single sensor bump-bonded to two or four readout chips, with a $1\times2$ and $2\times2$ configuration, respectively. The assembly procedure requires a separation of few hundred microns between the ROCs, leading to large interchip regions. In order to avoid the presence of inactive areas, pixel cells between ROCs are larger, resulting in higher capacitance. A dedicated tuning of the threshold in the peripheral pixels can be achieved by exploiting the regions described above. This feature will not be used in 3D pixel modules, which will consist of two ROCs bump-bonded to two individual sensors. In this configuration, an uninstrumented gap of approximately $\SI{600}{\micro m}$ exists between the sensors. This decision is based on the lack of a reliable process for producing large-area 3D pixel sensors at the moment.

\begin{figure}[htb!]
\vspace{-0.8cm}
  \centering
  \includegraphics[width=0.4\textwidth]{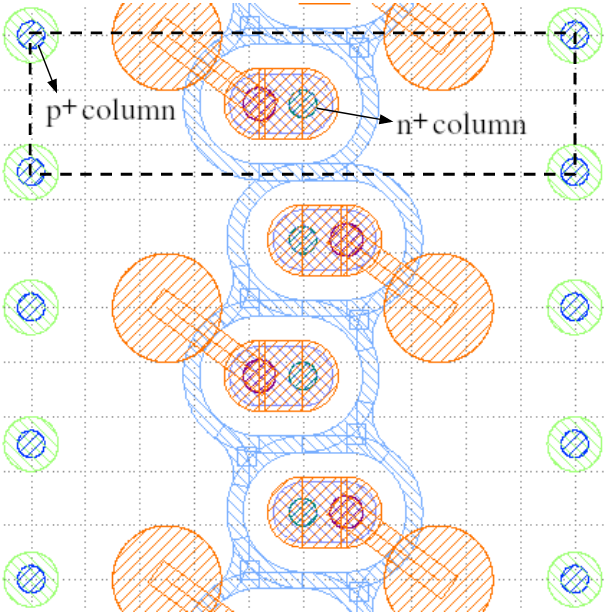}
  \caption{Schematic top view of four adjacent $25 \times \SI{100}{\micro m^2}$ pixel cells. One of the cells is delimited by a rectangle with dashed black lines for reference. The blue circles at the corners represent the $p^+$ columns, while the light blue circle in the center represents the $n^+$ column. The orange elliptical pad on top of the $n^+$ column represents the metal contact for bump-bonding. Its counterpart on the readout chip is shown as a large orange circle. The two metal contacts are connected by a diagonal orange rail.}
  \label{fig:pads}
\end{figure}

To match the readout chip pitch, a diagonal routing scheme is used for bump-bonding in 3D pixel sensors with a $25\times\SI{100}{\micro m^2}$ cell size, as illustrated in Fig.~\ref{fig:pads}.

%%%%%%%%%%%%%%%%%%%%%%%%%%%%%%%%%
%%%%%%%%%%%%%%%%%%%%%%%%%%%%%%%%%

\clearpage
\section{Irradiation and beam test setup}
\label{sec:facilities_setups}

\subsection{Irradiation: facilities and setup}
\label{sec:irradiation}

Irradiation campaigns of 3D pixel sensors for use in the IT have been carried out at the Institut Pluridisciplinaire Hubert Curien (IPHC)~\cite{iphc}, Zyklotron AG at the Karlsruhe Institute of Technology (KIT)~\cite{kit}, the Fermilab Irradiation Test Area (ITA), and the CERN PS~\cite{cernps}. The energy of the proton beam used at each irradiation facility and its corresponding hardness factor are provided in Table~\ref{tab:irradiation_details}. The TID received by the modules depends on the proton energy.  For the IPHC and KIT facilities, a fluence of \mbox{\SI{1e16}{n_{eq}/\cm^{2}}} corresponds to roughly $\SI{15}{MGy}$, while for the ITA and CERN PS, it corresponds to about $\SI{5}{MGy}$.  For comparison, about $\SI{13.4}{MGy}$ is expected over the duration of the HL-LHC running period at the innermost layer of the TBPX. Therefore, $\SI{6.7}{MGy}$ would be sufficient for testing, considering the replacement of this layer after 6 years. Radiation damage studies using RD53 chips~\cite{Papadopoulos_2023} confirm their functionality up to a total ionizing dose of $\SI{10}{MGy}$. The requested target fluences for the devices in this paper were in the range \mbox{1\dhyphen\SI{2e16}{n_{eq}/\cm^{2}}}.
\begin{table}[htb!]
    \centering
    \caption{Proton energy of the beams used at the irradiation facilities and their corresponding hardness factor or k-factor~\cite{Allport_2019}. This scaling parameter is used to express the radiation damage in terms of the $\SI{1}{MeV}$ neutron equivalent fluence.}
    \label{tab:irradiation_details}
    \begin{tabular}{lcc}
    \hline\hline
     Facility &\hspace{-0.25cm} Proton energy &\hspace{-0.2cm} k-factor\\
    \hline\hline
     IPHC &\hspace{-0.25cm} $\SI{23}{MeV}$ &\hspace{-0.2cm} $\sim2.2$\\
     KIT &\hspace{-0.25cm} $\SI{23}{MeV}$ &\hspace{-0.2cm} $\sim2.2$\\
     ITA &\hspace{-0.25cm} $\SI{400}{MeV}$ &\hspace{-0.2cm} $\sim0.83$\\
     CERN PS &\hspace{-0.25cm} $\SI{23}{GeV}$ &\hspace{-0.2cm} $\sim0.62$\\
     \hline\hline
    \end{tabular}
\end{table}

\begin{figure}[htb!]
\centering
\subfloat[]{\includegraphics[width=7.8cm]{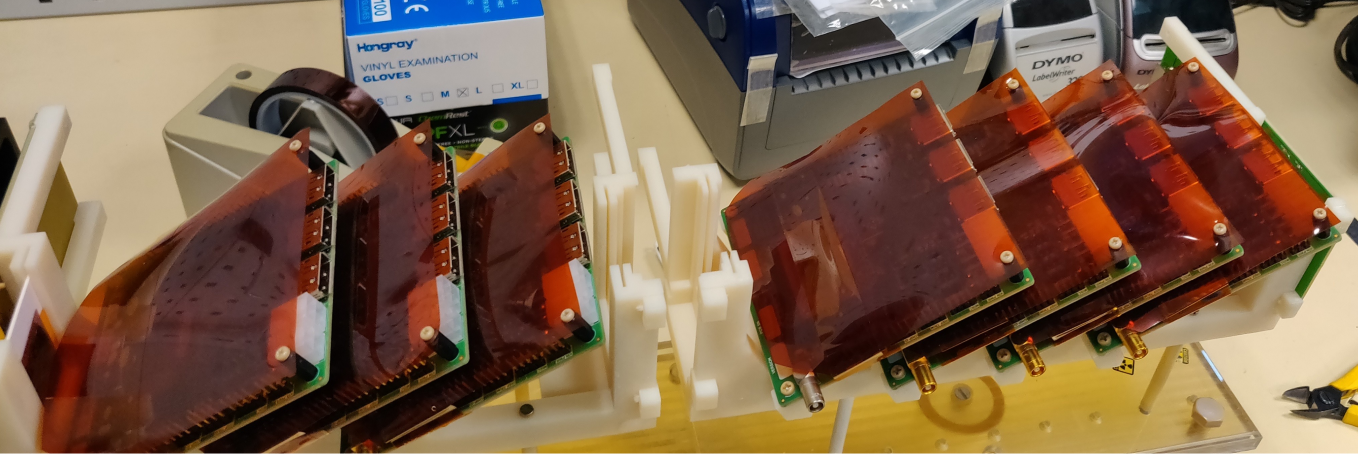}\label{fig:longitudinal_view_setup}}
\hspace{0.5cm}
\begin{subcolumns}[0.26\textwidth]
  \subfloat[]{\includegraphics[angle=-90,width=\subcolumnwidth]{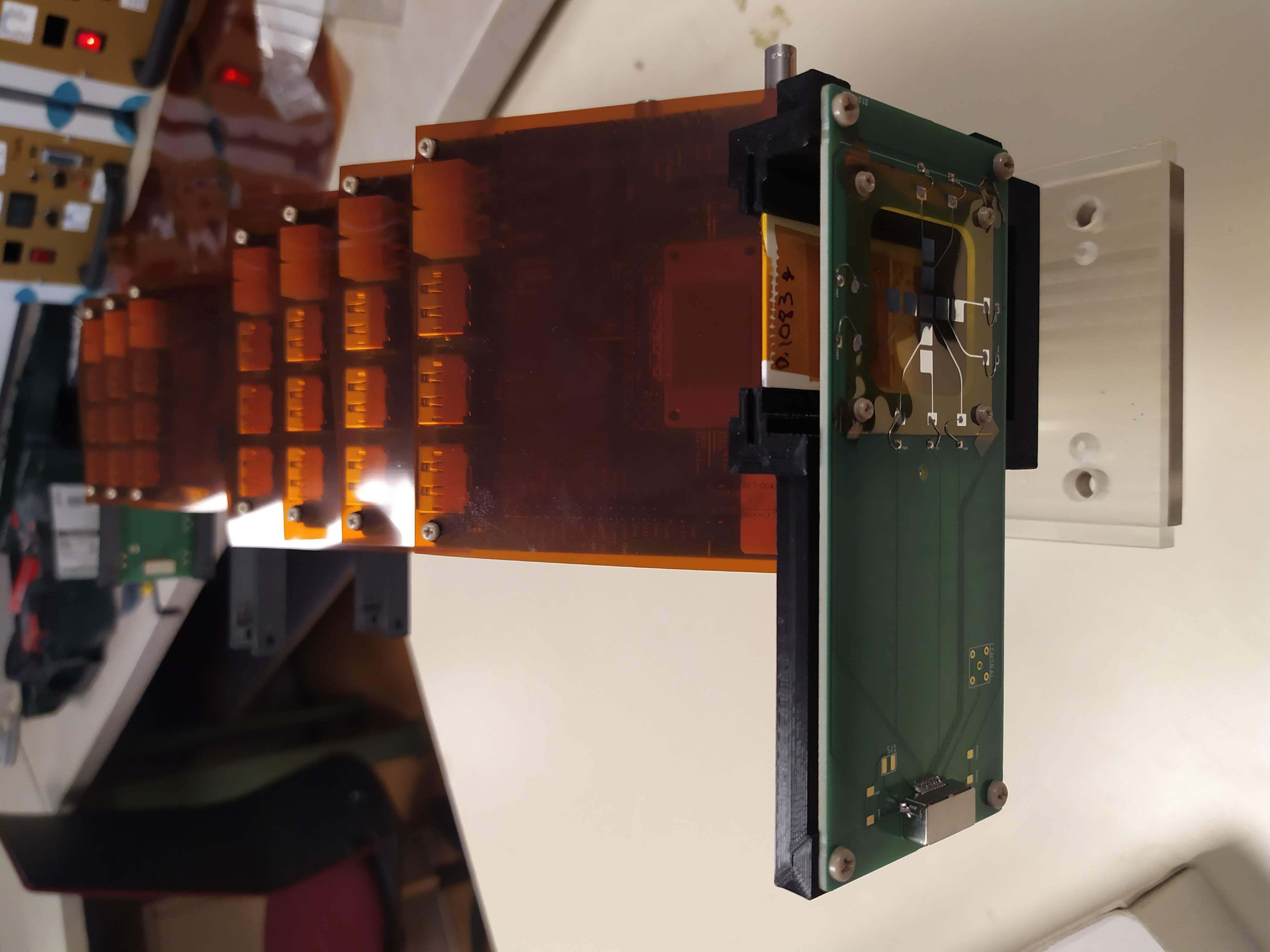}\label{fig:front_view_setup}}
\nextsubcolumn[0.19\textwidth]
  \subfloat[]{\includegraphics[width=\subcolumnwidth]{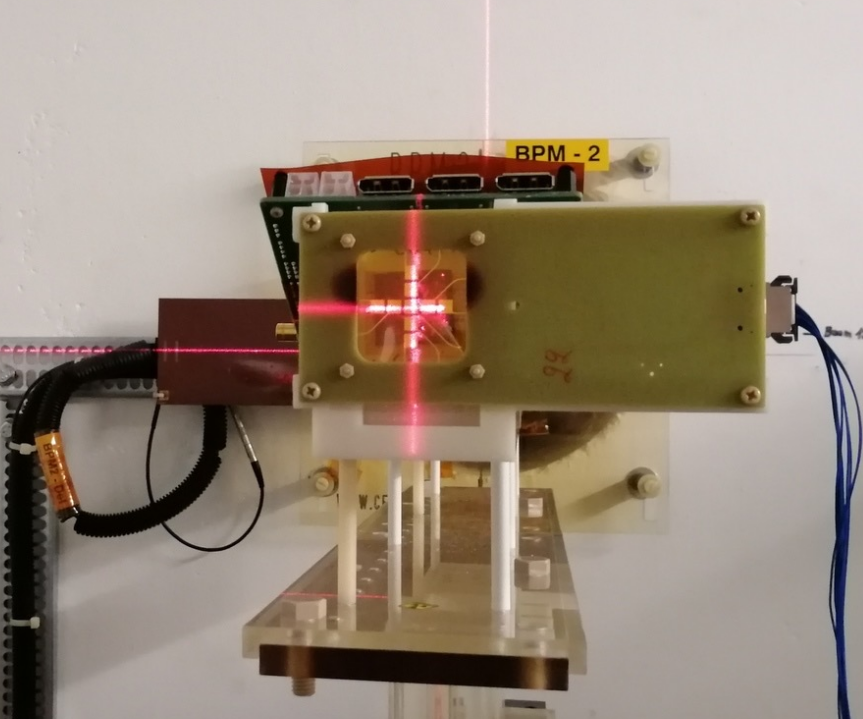}\label{fig:beam_alignment}}
\nextsubfigure
  \subfloat[][]{\includegraphics[width=\subcolumnwidth]{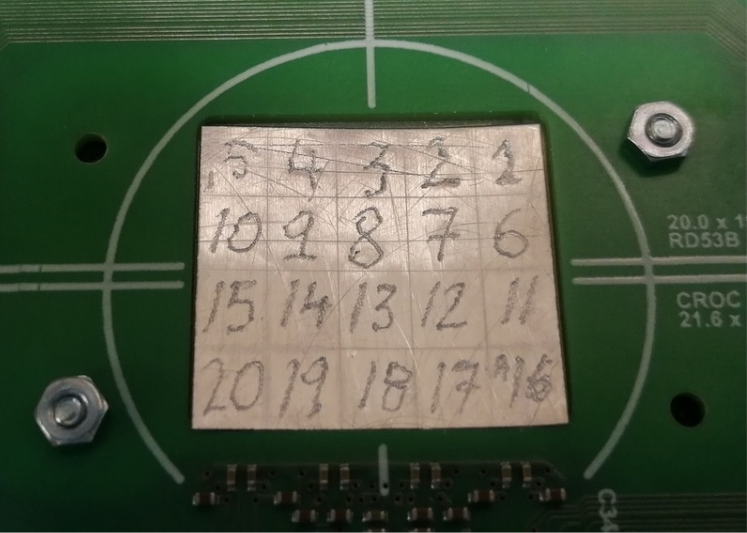}\label{fig:al_foil}}
\end{subcolumns}
\caption{The support used for the irradiation campaigns at the CERN PS is shown in (a) and (b). The samples are tilted by $\SI{30}{\degree}$ as seen from the longitudinal view. The beam alignment with respect to the modules using a laser and the aluminum foil placed on the backside of a CROCv1 3D pixel sensor are shown in (c) and (d), respectively.}
\label{fig:irradiation_setup}
\end{figure}

The fluence profile on the surface of sensors irradiated at KIT and IPHC is uniform, while this is not the case at ITA and the CERN PS. The beam profile at the CERN PS is approximately Gaussian with a Full Width at Half Maximum (FWHM) of about $\SI{14.5}{mm}$ in the horizontal direction and $\SI{6.5}{mm}$ in the vertical direction (to be compared with a sensor size of $22\times\SI{17}{mm^{2}}$). A specific support, shown in Figs.~\figsubref{fig:irradiation_setup}{fig:longitudinal_view_setup} and~\figsubref{fig:irradiation_setup}{fig:front_view_setup}, has been designed to hold the modules tilted by $\SI{30}{\degree}$ with respect to the horizontal plane, ensuring a more uniform irradiation along the vertical direction. To cover the whole horizontal dimension of the modules, the support is moved along a $\SI{26}{mm}$-long path. The beam alignment with respect to the modules is checked using a laser, as shown in Fig.~\figsubref{fig:irradiation_setup}{fig:beam_alignment}. Modules are coated with parylene before irradiation to protect the wire bonds. Additionally, aluminum foils are placed behind some of them to obtain a precise dose profile. After irradiation, the foils are cut into small pieces, as shown in Fig.~\figsubref{fig:irradiation_setup}{fig:al_foil}, and the activity of each piece is measured using HPGe spectroscopy. The estimated uncertainty in the measured fluence is 7\%.

The beam profile at the ITA is also approximately Gaussian, with a standard deviation of about \SI{1}{cm} in both the horizontal and vertical directions. To improve the uniformity of the irradiation, the devices are tilted by $\SI{30}{\degree}$ in the horizontal direction, similar to the setup at the CERN PS, with the modules supported on kapton held in an aluminum frame. Four circular foils of aluminum are mounted in line with the beam to enable the measurement of the fluence and beam profile. The estimated fluence uncertainty is 17\%, due to the activity measurement and the relative position of the foils and modules. This is higher than for the CERN PS (7\%), reflecting the fact that the ITA facility is newly commissioned and its procedures are not yet fully stabilized.

All irradiations were performed at room temperature, except at KIT, where the devices were cooled during the process. Afterwards, the samples were stored at low temperature to prevent annealing. In all facilities, the devices were unpowered during irradiation.

\subsection{Beam test: facilities and setup}
\label{sec:beamtest}
The beam test measurements have been performed in the period 2019--2023 at the following facilities: at Deutsches Elektronen-Synchrotron (DESY)~\cite{DESY_TB} with an electron beam at around $5\dhyphen\SI{6}{GeV/c}$, at the Fermilab Test Beam Facility (FTBF)~\cite{FTBF} with a proton beam at $\SI{120}{GeV/c}$, and at the CERN Super Proton Synchrotron (SPS)~\cite{cerntb} with a pion beam at $\SI{120}{GeV/c}$.

\begin{figure}[htb!]
    \centering
    \subfloat[]{\includegraphics[width=0.47\textwidth]{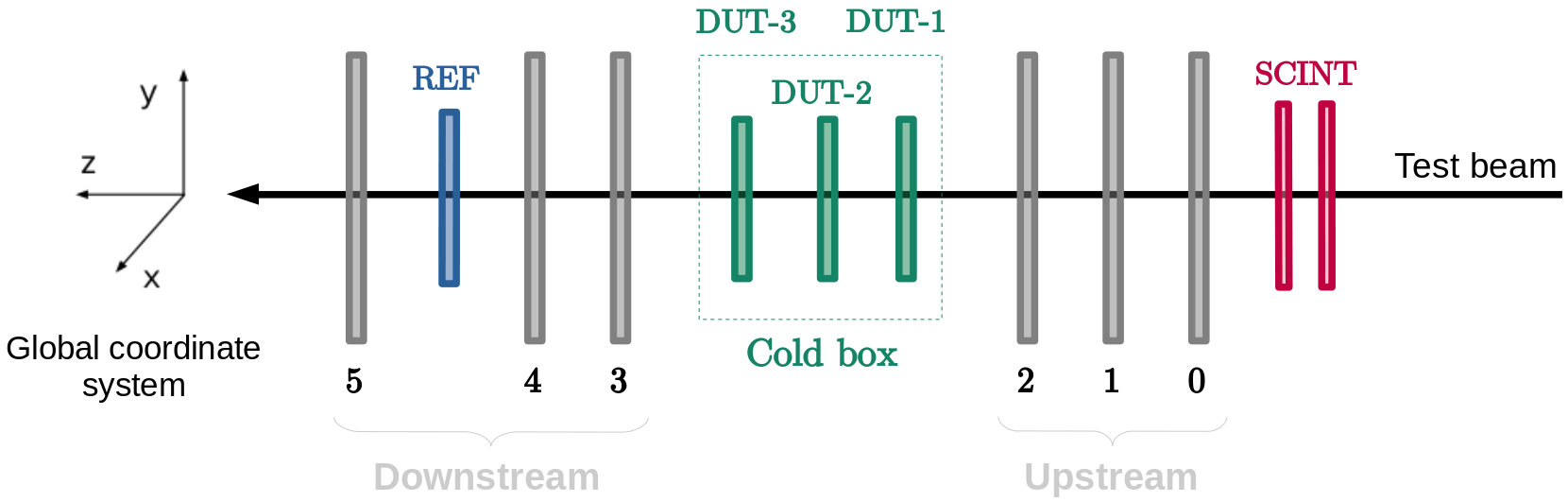}\label{fig:telescopeSPS}}
    \hspace{0.5cm}
    \centering
    \subfloat[]{\includegraphics[width=0.47\textwidth]{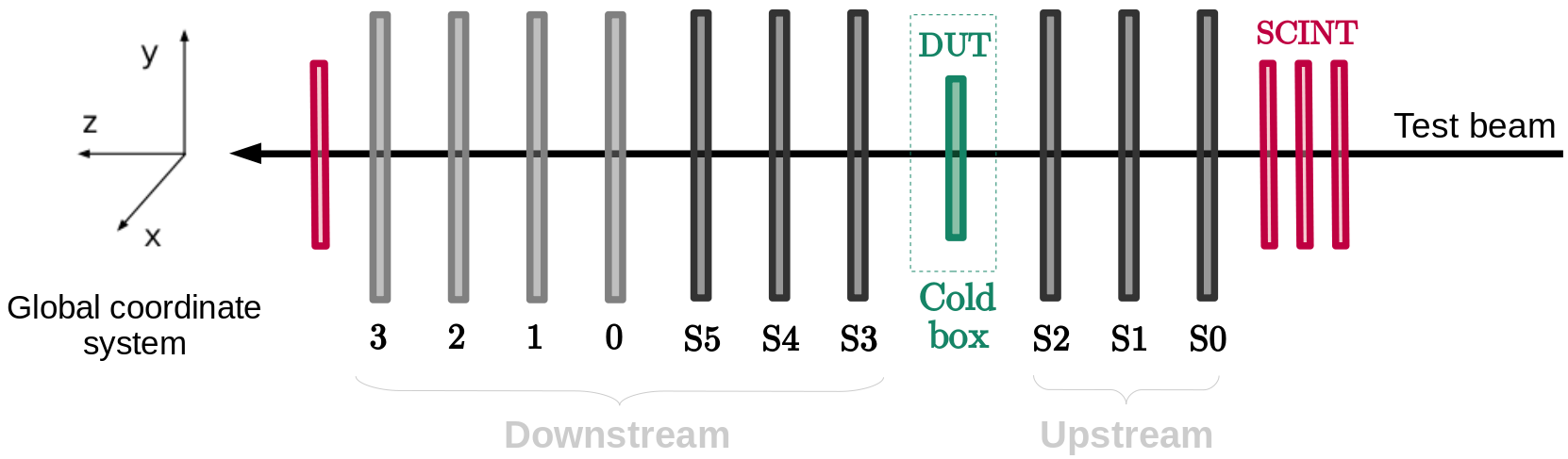}\label{fig:telescopeFTBF}}
    \caption{Sketch of the test beam setup at (a) SPS/DESY and (b) FTBF. The timing-reference pixel module is labeled as REF, the scintillators are indicated as SCINT, and the devices under test are labeled as DUT. The telescope pixel planes (0--5) are shown in pale gray, while the double layers of telescope strip planes (S0--S5) are depicted in dark gray. Further details are described in the text. Note: the position of REF and SCINT in the EUDET-type telescopes may differ between facilities. }
    \label{fig:telescopeSchema}
\end{figure}

The test beam setups at CERN and DESY are equipped with an EUDET-type telescope~\cite{AIDA_TELESCOPE} as shown in Fig.~\figsubref{fig:telescopeSchema}{fig:telescopeSPS}. The telescope consists of six sensor planes made of MIMOSA26 monolithic active pixel devices, each with an integration time of $\SI{115}{\micro s}$. The telescope planes, which feature a pixel size of \mbox{$18.4\times\SI{18.4}{\micro m^{2}}$} and a thickness of $\SI{50}{\micro m}$, are arranged in upstream and downstream triplets with respect to the position of the Device Under Test (DUT). The telescope is complemented by a Trigger Logic Unit (TLU) providing the trigger signal and a timing-reference (REF) pixel module that is used to select a subsample of the reconstructed tracks as explained in Section~\ref{sec:offline_analysis}. The average resolution depends on the particle type and energy used in the two test beam facilities and on the telescope configuration; on average it ranges between 3 and $\SI{5}{\micro m}$.

The FTBF provides a telescope composed of twelve silicon strip planes, six upstream and six downstream of the DUT station, and four downstream pixel planes, along with a trigger based on upstream and downstream scintillator planes. The strip planes are arranged in pairs, with strips orthogonal to each other, to provide six \mbox{two-dimensional} measurement points. The active thickness of the sensors is $\SI{320}{\micro m}$ and the strip pitch is $\SI{30}{\micro m}$. The readout pitch is $\SI{60}{\micro m}$ because only every other channel is read out. The intermediate strips are capacitively coupled to the readout strips. The four pixel planes feature planar sensors with a pixel size of \mbox{$100\times\SI{150}{\micro m^{2}}$} and a sensor thickness of $\SI{285}{\micro m}$, bump-bonded to PSI46V2 readout chips. This configuration results in a track pointing resolution at the DUT position ranging between 4 and $\SI{5}{\micro m}$. The trigger signal is generated by a scintillator located in front of the first telescope plane. The Off The Shelf DAQ (OTSDAQ)~\cite{otsdaq}, developed at Fermilab, is employed to read out both the telescope and DUT planes. 

At all three facilities, the DUTs are mounted on PCB carrier boards, which are placed inside a thermally insulated cold box flushed with dry air to prevent condensation. The cold box is connected to a chiller that circulates a coolant to maintain a stable temperature. Additionally, the cold box is attached to a set of stages that allow remote control of its position and orientation. Translation in the two directions orthogonal to the beam axis and rotation around the vertical axis are possible.

%%%%%%%%%%%%%%%%%%%%%%%%%%%%%%%%%
%%%%%%%%%%%%%%%%%%%%%%%%%%%%%%%%%

\section{Tuning: RD53A and CROCv1}
\label{sec:tuning}
The Ph2\_ACF data acquisition system~\cite{Ph2_ACF} was used to configure, tune, and operate the RD53A and CROCv1 modules. A complete description of the DAC registers used to configure the linear AFE can be found in~\cite{CROC}. A short description of the most relevant and routinely calibrated parameters is given below:
\begin{itemize}
\item The global threshold of the linear AFE, corresponding to the DC threshold voltage applied to the discriminator input. Increasing the value of this register results in an increased global threshold.
\item The output dynamic range of the in-pixel threshold trimming DAC (TDAC). Increasing the value of this register results in a coarser resolution of the TDAC and a wider output range. Operation of irradiated modules typically requires an increase of this register value.
\item The current in the Krummenacher feedback, which discharges the preamplifier feedback capacitance at a constant rate. Increasing this current leads to a faster return to baseline and a reduced Time-over-Threshold (ToT). Most test beam campaigns were performed using a low current (slow discharge), while performance at higher current (fast discharge) was verified in one campaign and found to be compatible.
\item The current in the preamplifier input stage, which represents the main contribution to the linear AFE current consumption. Irradiated modules can be operated with a higher current to reduce noise at the cost of an increased power consumption.
\end{itemize}

\begin{figure}[htb!]
  \centering
  \subfloat[]{\includegraphics[height=0.34\textwidth,width=0.45\textwidth]{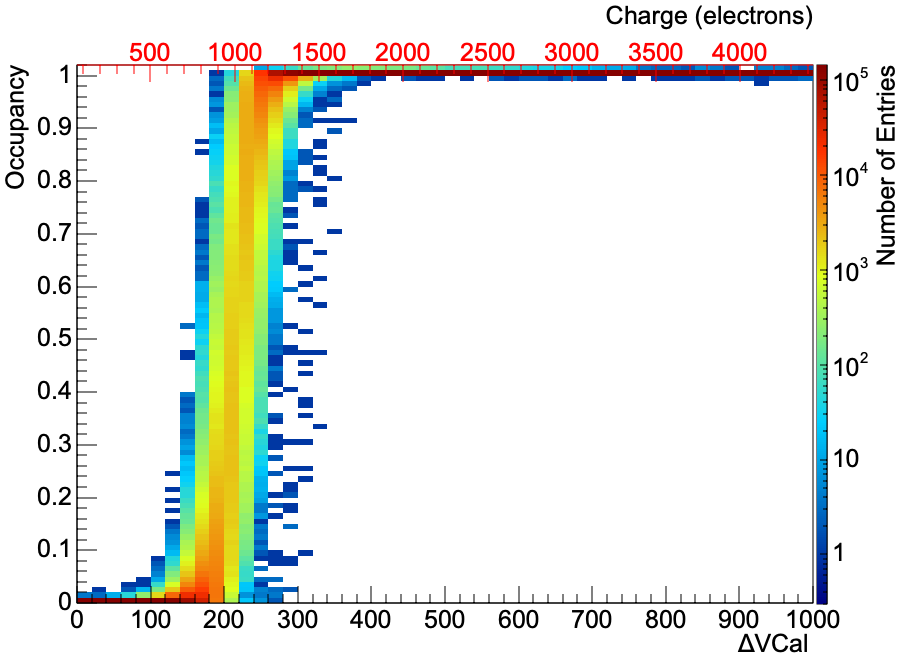}\label{fig:scurve}}\\
  \centering
  \hspace{-0.25cm}\subfloat[]{\includegraphics[height=0.34\textwidth,width=0.44\textwidth]{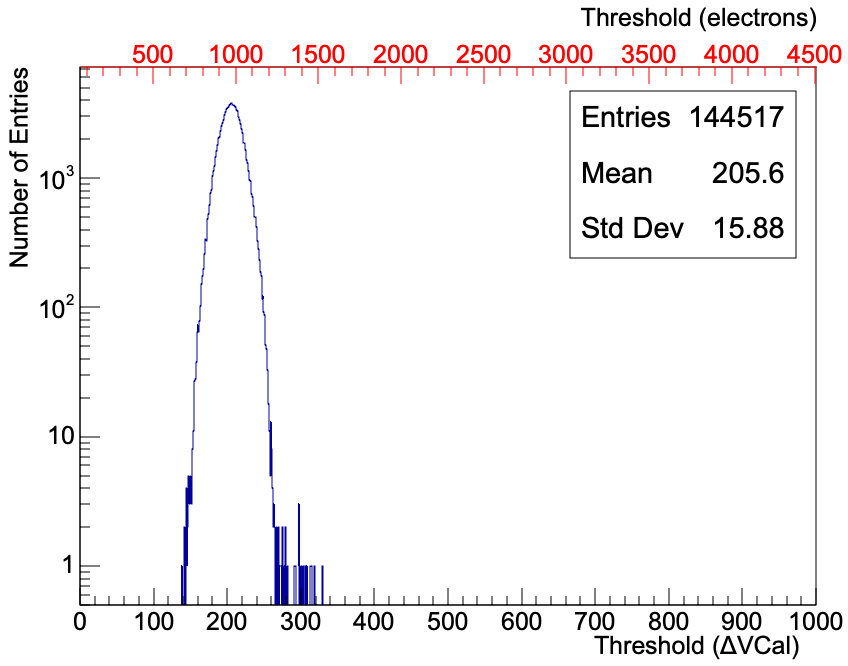}\label{fig:thr}}\\
  \centering
  \hspace{-0.3cm}
  \subfloat[]{\includegraphics[height=0.34\textwidth,width=0.435\textwidth]{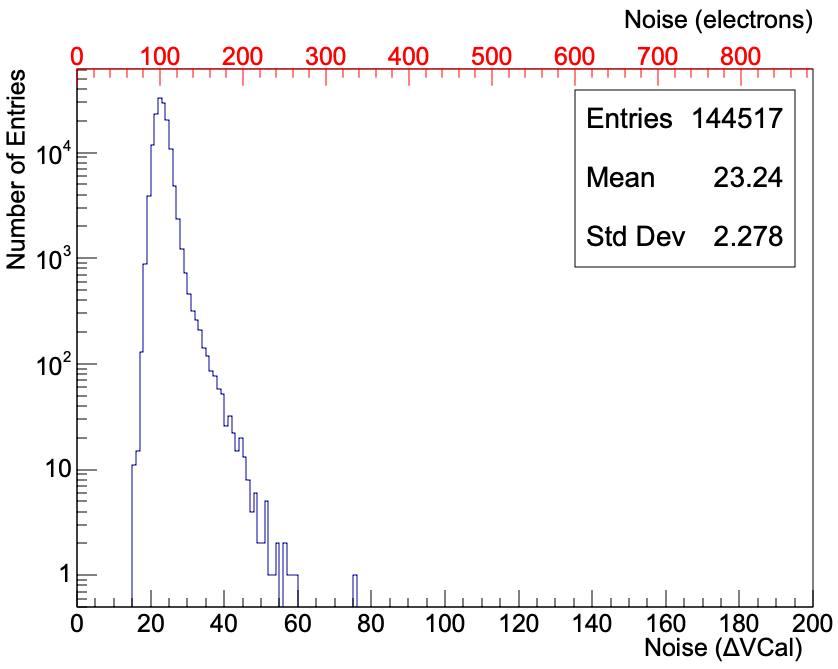}\label{fig:noise}}\\
  \caption{Examples of distributions obtained during the tuning procedure of a non-irradiated CROCv1 chip bonded to a 3D pixel sensor. Subfigure (a) represents the superposition of the S-curve of each pixel, while subfigures (b) and (c) represent the threshold and noise distributions, respectively.}
  \label{fig:croc_tuning_irrad}
\end{figure}

\begin{figure}[htb!]
  \centering
{\includegraphics[height=0.32\textwidth,width=0.44\textwidth]{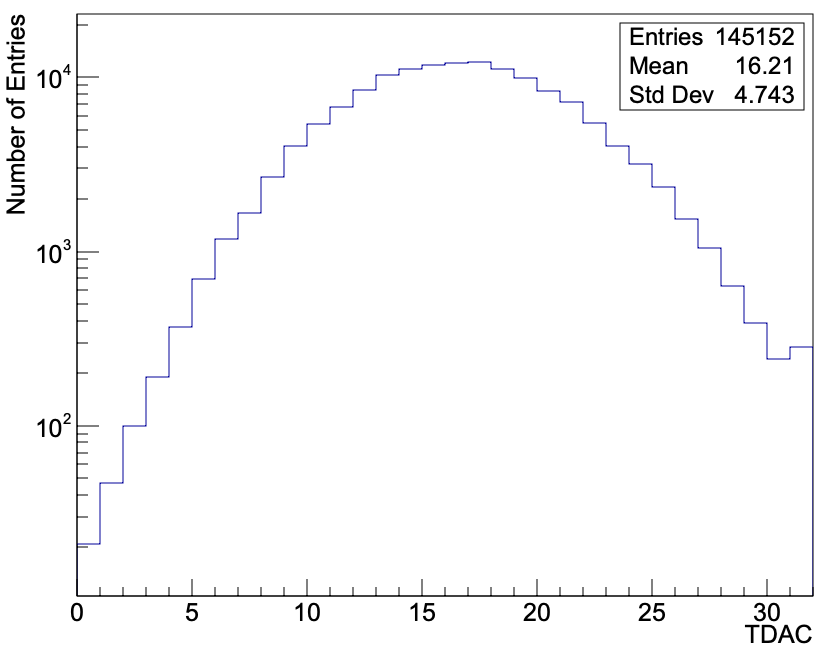}}
  \caption{Example of distribution obtained from the per-pixel threshold adjustment of a non-irradiated CROCv1 chip bonded to a 3D pixel sensor.}
  \label{fig:tdac}
\end{figure}

\begin{figure}[htb!]
  \centering
{\includegraphics[height=0.33\textwidth,width=0.44\textwidth]{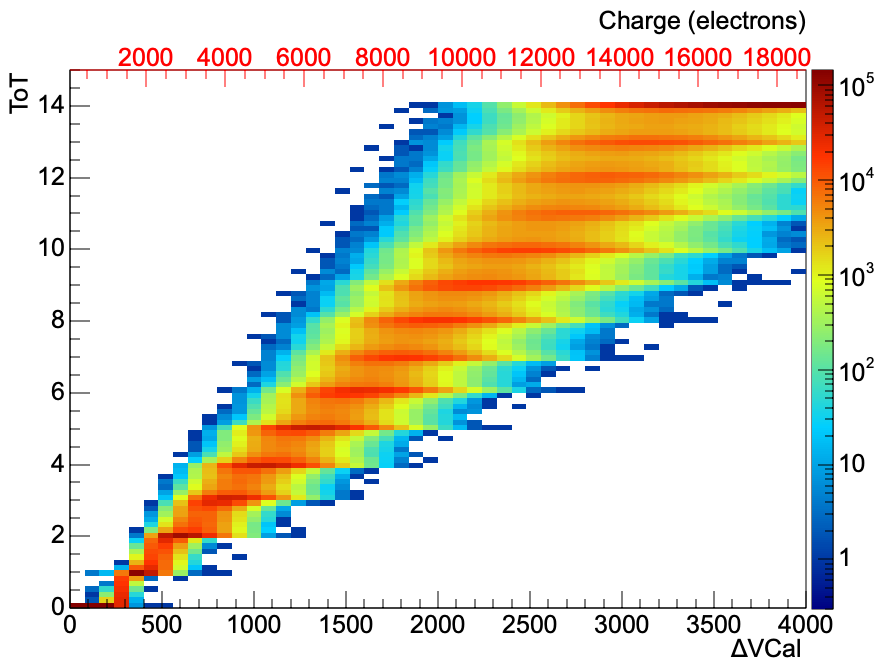}}
  \caption{Example of distribution obtained from the gain calibration of a non-irradiated CROCv1 chip bonded to a 3D pixel sensor. A linear dependence between ToT and $\Delta$VCal is clearly visible. The results for all the pixels are superimposed on the same histogram.}
  \label{fig:gain}
\end{figure}

The tuning procedure is described below, and Figs.~\ref{fig:croc_tuning_irrad}--\ref{fig:gain} show representative distributions from the tuning of a non-irradiated CROCv1 chip bonded to a 3D pixel sensor, with a target threshold of 1000 electrons.

To measure the threshold of each individual pixel, voltage steps with increasing $\Delta$VCal are supplied, and for each voltage the corresponding charge is injected into the pixel 100 times. The occupancy is defined as the ratio between the number of registered hits ($N_{\rm{hits}}$) and the number of injections ($N_{\rm{inj}}$). Plotting the occupancy as a function of $\Delta$VCal results in a characteristic ``S-curve'', which can be fitted with an error function. The value of $\Delta$VCal at which 50\% occupancy is reached is taken as the pixel threshold, while the sigma of the error function represents its noise. The S-curves for all pixels of a CROCv1 chip are shown in Fig.~\figsubref{fig:croc_tuning_irrad}{fig:scurve}. The distributions in Figs.~\figsubref{fig:croc_tuning_irrad}{fig:thr} and~\figsubref{fig:croc_tuning_irrad}{fig:noise} are derived from the threshold and noise evaluations of all pixels, based on their corresponding S-curves.

Low threshold, low noise, and good charge collection are fundamental requirements for good detector performance, and there can be trade-offs among them, particularly for irradiated devices. The global threshold is individually tuned for each module to achieve a target value after per-pixel adjustment between 1000 and 1200 electrons, or the lowest value that ensures the number of masked pixels remains below 1\% for non-irradiated modules and below 2\% for irradiated modules. For irradiated modules, the resulting global threshold is typically around \mbox{1500--1800} electrons for RD53A assemblies and \mbox{1000--1200} electrons for CROCv1 assemblies. 

The per-pixel threshold adjustment is performed in two steps. First, the threshold of each pixel is trimmed targeting a global threshold of approximately 2000 (3000) electrons for non-irradiated (irradiated) modules. A binary search algorithm is applied to find the TDAC value that minimizes the distance of the pixel threshold from the average threshold across the matrix. For irradiated modules the procedure is repeated targeting a threshold of 2000 electrons. Finally, an additional threshold trimming is performed targeting a global threshold as specified above.  In this step, the TDAC is incremented or decremented by a single unit to minimize changes to the local threshold. For non-irradiated modules, if the desired threshold is lower than 1200 electrons, no further trimming has to be performed. An average TDAC distribution obtained at the end of the tuning procedure is shown in Fig.~\ref{fig:tdac}: successful tuning should lead to a peak around 16 without accumulations at the extreme ends of the distribution.

At the end of the tuning procedure ``stuck" and ``noisy" pixels are masked. Stuck pixels do not respond to charge injections while noisy pixels have an occupancy higher than $2\times10^{-5}$ without injecting charge. This threshold corresponds to 1\% of the occupancy expected in the first layer of TBPX ($2 \times 10^{-3}$).

Finally, a gain scan is performed to determine the pixel-by-pixel relation between the ToT of the output signal and the amplitude of the voltage step supplied to the capacitor, in $\Delta$VCal units. The results of this scan are shown in Fig.~\ref{fig:gain}. The conversion of the charge, $Q$, from $\Delta$VCal units to electrons is performed using the following relation:
\begin{equation}
Q\,[\mathrm{electrons}] = \frac{V_{\mathrm{ref}}}{2}\frac{Q\,[\mathrm{\Delta VCal}]\, C_{\mathrm{inj}}}{4096\,e}
\label{eq:charge-conversion}
\end{equation}
where $V_{\mathrm{ref}}$ is the ROC internal reference voltage (typically around $\SI{0.8}{V}$) , $e$ is the elementary charge, $C_{\mathrm{inj}}$ is the injection capacitor capacitance ($\SI{8}{fF}$), and 4096 represents the number of possible charge values for the 12-bit ADC of the charge injection circuit. The gain of the CROCv1 charge injection circuit can be doubled to inject larger charges into the pixel front-end. In this regime, the charge values computed from the formula above must be \mbox{multiplied by 2}.

The formula from Eq.~\ref{eq:charge-conversion} has been measured to still work well also for irradiated devices and is thus assumed to be valid throughout the rest of the paper.

%%%%%%%%%%%%%%%%%%%%%%%%%%%%%%%%%
%%%%%%%%%%%%%%%%%%%%%%%%%%%%%%%%%

\section{Data analysis}
\label{sec:analysis}
Data collected at the CERN SPS test beam facility have been analyzed using the Corryvreckan framework~\cite{Dannheim_2021}. Corryvreckan provides a flexible offline event building facility to combine data from detectors with different readout schemes, with or without trigger information, and includes the possibility to correlate data from multiple devices based on timestamps. While the most recent data collected at the DESY test beam facility were analyzed using Corryvreckan, previous DESY test beam data and data collected at FTBF were processed using dedicated software frameworks. 

All frameworks follow the same key steps: data decoding, noisy and stuck pixel masking, hit clustering, track reconstruction, telescope alignment, DUT alignment, and DUT analysis. A more detailed description of these steps is provided below, using the Corryvreckan example.

\subsection{Offline reconstruction and alignment}
\label{sec:offline_analysis}
The first step of the offline reconstruction is the decoding of the data coming from the MIMOSA26 telescope planes and the DUTs. The six MIMOSA26 sensors are read out in binary mode, meaning that a global threshold is applied on the chip and only the positions of the pixels exceeding the threshold are recorded. A threshold of 5 or 6 times the individual pixel noise is used, as explained in Ref.~\cite{AIDA_TELESCOPE}. For the DUTs the response of pixels above threshold is digitized with a 4-bit precision ToT counter and stored together with the pixel positions.

The ``local density" noise estimation method, taken from the Proteus framework~\cite{kiehn_moritz_2019_2586736}, is exploited to identify noisy pixels on the telescope planes. It uses a local estimate of the expected hit rate to find pixels that are a certain number of standard deviations away from this estimate. The local density method is not applied to the DUTs, and only the pixels identified during the tuning procedure described in Section~\ref{sec:tuning} are masked.

The decoded hits in the DUT and telescope are then clustered. For the telescope planes, the cluster position is calculated using a simple arithmetic mean of the pixel coordinates. For the DUTs, a ``center-of-gravity'' algorithm is applied, computing the cluster position as the weighted mean of the pixel positions, with the reconstructed charges serving as weights. 

The next step is the alignment, which determines the position and orientation of the telescope planes and the DUTs relative to an alignment-reference plane, which can be either one of the telescope planes or the timing-reference plane. This alignment is necessary to ensure that the positions and orientations of all planes are accurately known, allowing for a reliable reconstruction of the observables defined in Section~\ref{sec:observables}.

A preliminary telescope alignment is achieved by translating each plane along the $x$ and $y$ axes to minimize the average distance between the hit positions on the telescope planes and those on the alignment-reference plane in both directions. After this coarse alignment, tracks are reconstructed using a wide matching distance for track-hit association on each plane. This matching distance corresponds to the spatial region defined by an ellipse, whose semi-major and semi-minor axes are each set to ten times the hit position uncertainty in the respective directions. Two tracking algorithms are applied to the data presented in this paper: a ``straight-line" fit and the General Broken Line (GBL) fit~\cite{gbl}. The straight-line algorithm starts from a seed formed by clusters on the first and last telescope planes and adds intermediate clusters that satisfy the previously defined track-hit association criteria. Assuming a linear trajectory, it fits a straight line through the associated clusters by minimizing the $\chndof$, where $\mathrm{ndof}$ is the number of degrees of freedom. In contrast, the GBL algorithm more accurately accounts for multiple scattering effects along the track, which is especially important in the case of electron beams. This algorithm is used for the analysis of data collected at DESY.

Two additional alignment steps are performed using the reconstructed tracks to minimize the distance between the track impact point and the position of the associated cluster on each plane. The first step is performed using the ``Millepede"~\cite{blobel2002new} algorithm, which simultaneously adjusts the positions and rotations of the telescope planes and the alignment-reference plane. A final refinement is then performed using a simple ``straight-line" algorithm to refit all the tracks minimizing their $\chndof$.

After the telescope alignment, tracks are reconstructed with tighter selection criteria for use in the DUT alignment. The main requirements include a track-hit matching distance within an ellipse whose semi-axes are set to five times the hit position uncertainty in the respective directions, and a $\chndof$ below 10. The DUTs have a much faster readout time of $\SI{25}{ns}$ compared to the $\SI{115}{\micro s}$ integration time of the telescope planes. To reject out-of-time tracks, all reconstructed tracks are required to have an associated hit on the timing-reference plane.

The reconstructed tracks are extrapolated to the position of the DUTs for track-hit association and DUT alignment, which is performed by minimizing the distance between the track and the hit. For modules with a $25 \times \SI{100}{\micro m^2}$ pixel size, the association is based on a matching region defined by an ellipse with semi-major and semi-minor axes of $\SI{200}{\micro m}$ and $\SI{100}{\micro m}$ along the long and short pixel directions, respectively. This asymmetric window is chosen to account for the finite telescope resolution, which has a greater impact in the short pixel direction. In the case of modules with a $50 \times \SI{50}{\micro m^2}$ pixel size, a symmetric window is applied, with semi-axes of $\SI{100}{\micro m}$ in both directions.

\subsection{Final track selection}
\label{sec:selections}
After completing the telescope and DUT alignment, a final selection of tracks is made to determine DUT properties, including hit detection efficiency, spatial resolution, and charge collection. The selected tracks must fulfill the following requirements:
\begin{itemize}
\item at least five associated hits on the telescope planes and one associated hit on the timing-reference plane. The track-hit matching distance must lie within an ellipse whose semi-major and semi-minor axes are set to three times the hit position uncertainty in the respective directions;
\item $\chndof < 5$;
\item one associated hit on the DUT within a matching region defined as an ellipse with semi-major and semi-minor axes set to five times the hit position uncertainty in the respective directions;
\item for the determination of hit detection efficiency, tracks are additionally required to point to a pixel on the DUT that is at least one pixel away from a masked one, to avoid introducing spurious inefficiencies.
\end{itemize}

\subsection{Definition of observables}
\label{sec:observables}

\begin{figure}[htb!]
  \centering
  \begin{subcolumns}[0.22\textwidth]
  \subfloat[]{\includegraphics[width=\subcolumnwidth]{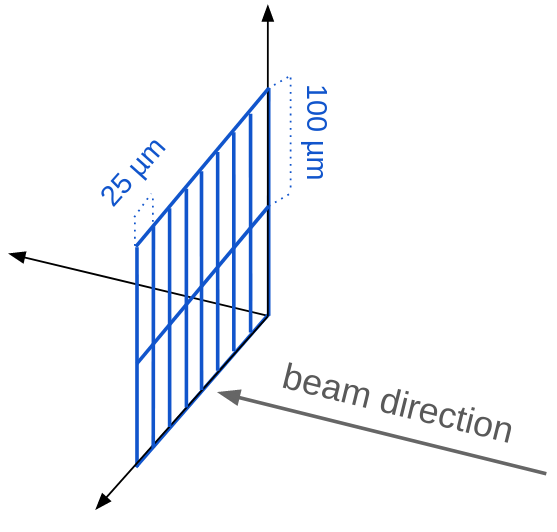}}
  \nextsubcolumn[0.22\textwidth]
  \hspace{-0.5cm}
  \subfloat[]{\includegraphics[width=\subcolumnwidth]{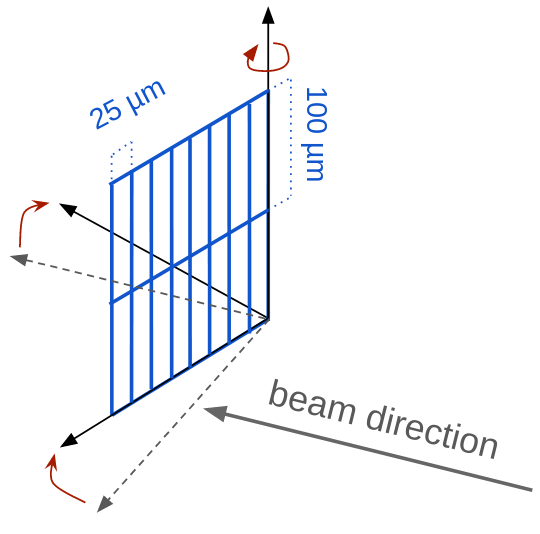}}
  \end{subcolumns}
  \caption{Schematic of a sensor with $25\times \SI{100}{\micro m^{2}}$ pixel size, tested (a) at normal beam incidence and (b) under a rotation angle around an axis parallel to the long edge of the pixel.}
  \label{fig:rotation_schema}
\end{figure}

The observables described in this section have been studied both at normal beam incidence and as a function of the rotation angle, as shown in Fig.~\ref{fig:rotation_schema}. Throughout this paper, rotations of the DUT refer to a rotation around an axis parallel to the long edge of the pixel and perpendicular to the beam, with zero degrees corresponding to normal beam incidence.

The pixel cell directions are defined as follows: the \mbox{$x$-direction} corresponds to the $\SI{100}{\micro m}$ pitch, and the \mbox{$y$-direction} corresponds to the $\SI{25}{\micro m}$ pitch. When observables are plotted within the pixel cell, the plot represents the average value across all pixel cells in the sensor matrix that registered a hit, rather than that of a single pixel.

\subsubsection{Hit detection efficiency}

The hit detection efficiency, $\epsilon$, and its uncertainty, $\sigma_{\epsilon}$, are defined as:
\begin{equation}
\epsilon = \frac{N^{\text{track}}_{\text{hit}}}{N^{\text{track}}}
\label{eq:1}
\end{equation}
and
\begin{equation}
\sigma_{\epsilon} = \sqrt{\frac{\epsilon\,(1-\epsilon)}{N^\mathrm{track}}}
\label{eq:2}
\end{equation}
where $N^\mathrm{track}$ denotes the number of telescope tracks that pass the selection described in Section~\ref{sec:selections}, and $N^{\mathrm{track}}_{\mathrm{hit}}$ is the subset of those tracks that are matched with a hit on the DUT. In order to account for the fraction of masked pixels, the hit detection efficiency can be multiplied by the acceptance, $a$, defined as:
\begin{equation}
a = 1 - \frac{N_{\mathrm{masked}}}{N_{\mathrm{total}}}
\label{eq:3}
\end{equation}
where $N_{\mathrm{masked}}$ is the number of masked pixels and $N_{\mathrm{total}}$ is the total number of pixels on the DUT considered in the analysis.

\subsubsection{Collected charge}
For each of the $N^{\text{track}}_{\text{hit}}$ tracks, the charge of the DUT cluster associated to the track is taken. The charge spectrum is then fitted using a Landau convoluted with a Gaussian to account for pixel-to-pixel variations in gain and electronic noise. The MPV of the Landau distribution is taken as the measurement of the collected charge.

\subsubsection{Spatial resolution}
The spatial resolution of the DUT is extracted from the track residuals distribution. The track residual is defined as the difference between the coordinate of the impact point of the track on the DUT and the coordinate given by the DUT cluster position. The width of this distribution, typically obtained from a Gaussian fit, represents the sum in quadrature of the DUT resolution and the track position resolution. The track position resolution is estimated differently at each facility:
\begin{itemize}
\item The track reconstruction of the FTBF telescope is performed using a Kalman fit based algorithm, which computes the track position uncertainty at each telescope plane and DUT.
\item The telescope resolution at SPS can be determined by fitting the edge of the DUT residual distribution along the long pitch to the cumulative distribution function (CDF) of a Gaussian distribution. This is possible because the effects of telescope resolution and charge sharing on the DUT residuals along the \mbox{$\SI{100}{\micro m}$ pitch} are decoupled due to low charge sharing across the $\SI{25}{\micro m}$ pixel boundary.
\item The telescope resolution at DESY is estimated using the two tracklets from the upstream and downstream arms. The difference between the impact point of the two tracklets is computed for each event and the telescope resolution is estimated dividing the width of the resulting distribution by~$\sqrt{2}$. When irradiated sensors are tested, only the upstream arm is used because the material of the cold box causes multiple scattering, which degrades the resolution of the downstream planes. In this case, the telescope resolution is estimated from a simulation~\cite{Spannagel} based on the GBL algorithm, which takes into account the material budget of the telescope and the beam energy.
\end{itemize}

%%%%%%%%%%%%%%%%%%%%%%%%%%%%%%%%%
%%%%%%%%%%%%%%%%%%%%%%%%%%%%%%%%%

\section{Results for non-irradiated single-chip modules}
The performances of non-irradiated sensors bump-bonded to RD53A and CROCv1 chips are presented separately below. A summary and description of each device is given in~\ref{sec:DUTsummary}. The nomenclature used to identify the modules indicates the manufacturer that produced the sensor as well as the ROC flavor.

The FBK sensors connected to RD53A chips are from two different productions, referred to as Stepper-1 and Stepper-2. Both have the same pixel design; Stepper-2 features shorter n+ columns ($\SI{115}{\micro m}$) compared to Stepper-1 ($\SI{130}{\micro m}$), as introduced to prevent discharges with the underlying low-resistivity support wafer, which is kept at bias voltage. FBK sensors bump-bonded to CROCv1 chips come from a subsequent production that followed the same design parameters as Stepper-2.

\label{sec:fresh_results}
\subsection{RD53A modules}

\begin{figure}[htb!]
\centering
\hspace*{0.1cm}\includegraphics[width=0.47\textwidth]{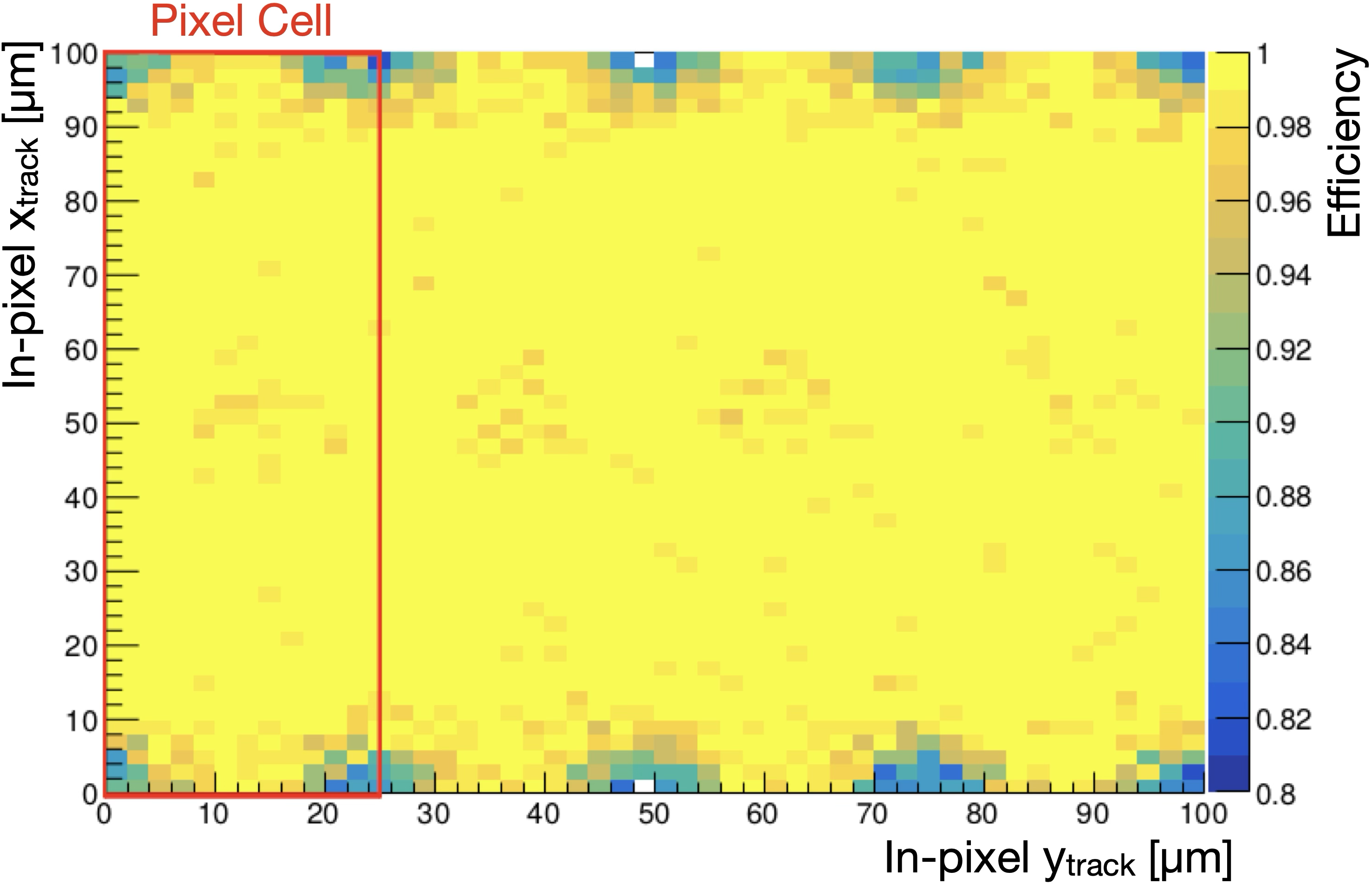}
\caption{\label{3D25eff} Hit detection efficiency map at normal beam incidence for a \mbox{$4\times1$} pixel grid of a non-irradiated FBK sensor biased at $\SI{30}{V}$. The red rectangle indicates a pixel cell for reference.}
\end{figure}

Several FBK 3D pixel sensors, with $25\times \SI{100}{\micro m^{2}}$ cell size and connected to the RD53A readout chip, were investigated at the DESY test beam. These devices, tuned to an average pixel threshold of 1000 electrons at room temperature, have a hit detection efficiency greater than 97\% with a bias voltage as low as $\SI{5}{V}$. Figure~\ref{3D25eff} shows the hit detection efficiency map in a \mbox{$4\times 1$} pixel grid at a bias voltage of $\SI{30}{V}$ and normal beam incidence. Lower efficiency is observed in the corners of the pixel cells, corresponding to the locations of the \mbox{$p^+$ columns}. This occurs for two reasons: the intrinsic inefficiency of the columns, which are made of passive material, and the effect of charge sharing, driven by diffusion, which is most relevant at the corners of the pixels. The efficiency near the \mbox{$n^+$ columns} at the pixel cell center is close to 100\% since they are the collecting electrodes and the charge sharing is strongly suppressed in this region. The \mbox{$n^+$ columns} are also shorter than the active thickness (Fig.~\ref{fig:3D}), and therefore charge can additionally be generated underneath the column, which is not the case for the \mbox{$p^+$ electrodes}.

\begin{figure}[htb!]
\centering
\subfloat[][]{
\includegraphics[width=0.47\textwidth]{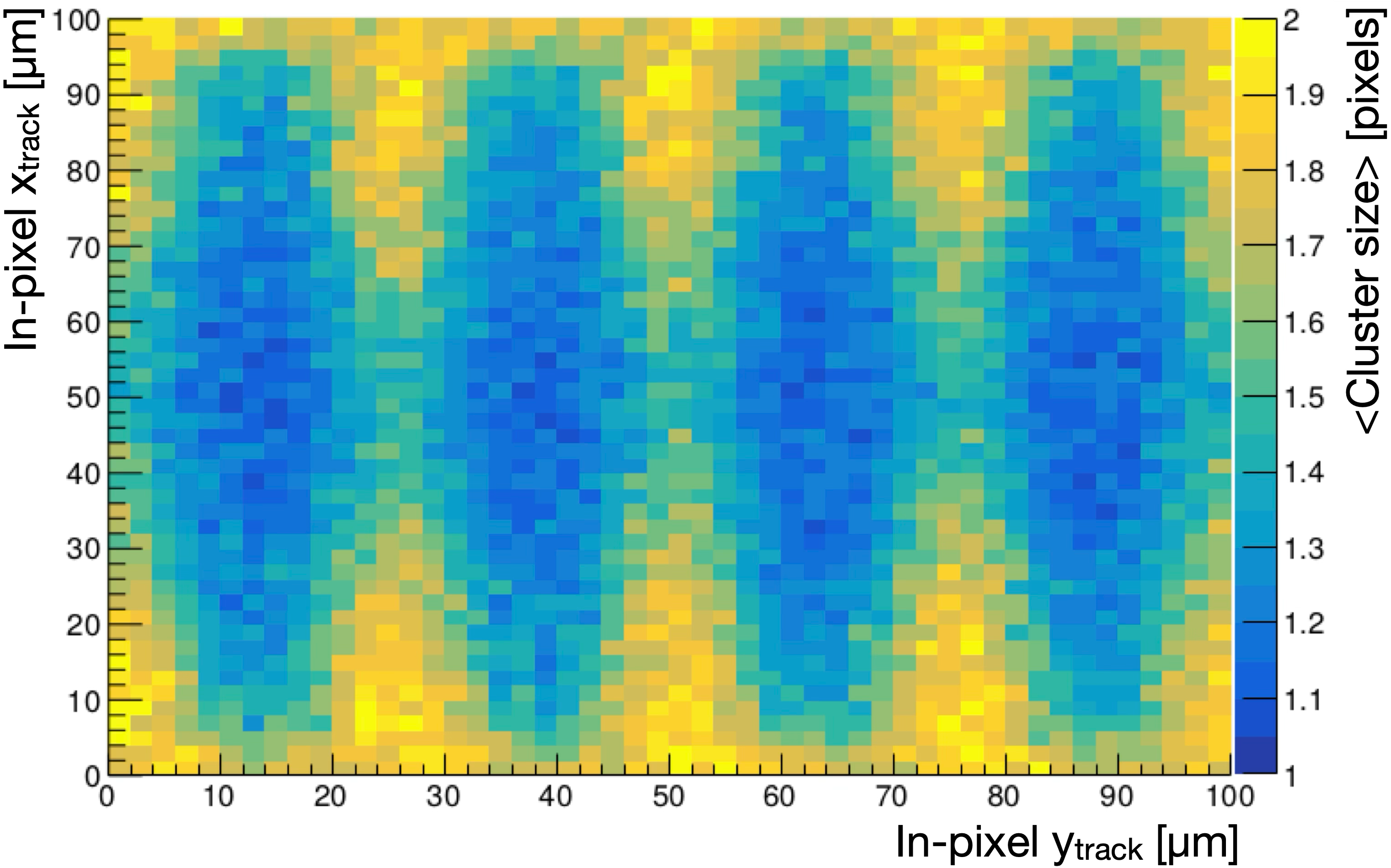}}\\
\subfloat[][]{
\includegraphics[width=0.47\textwidth]{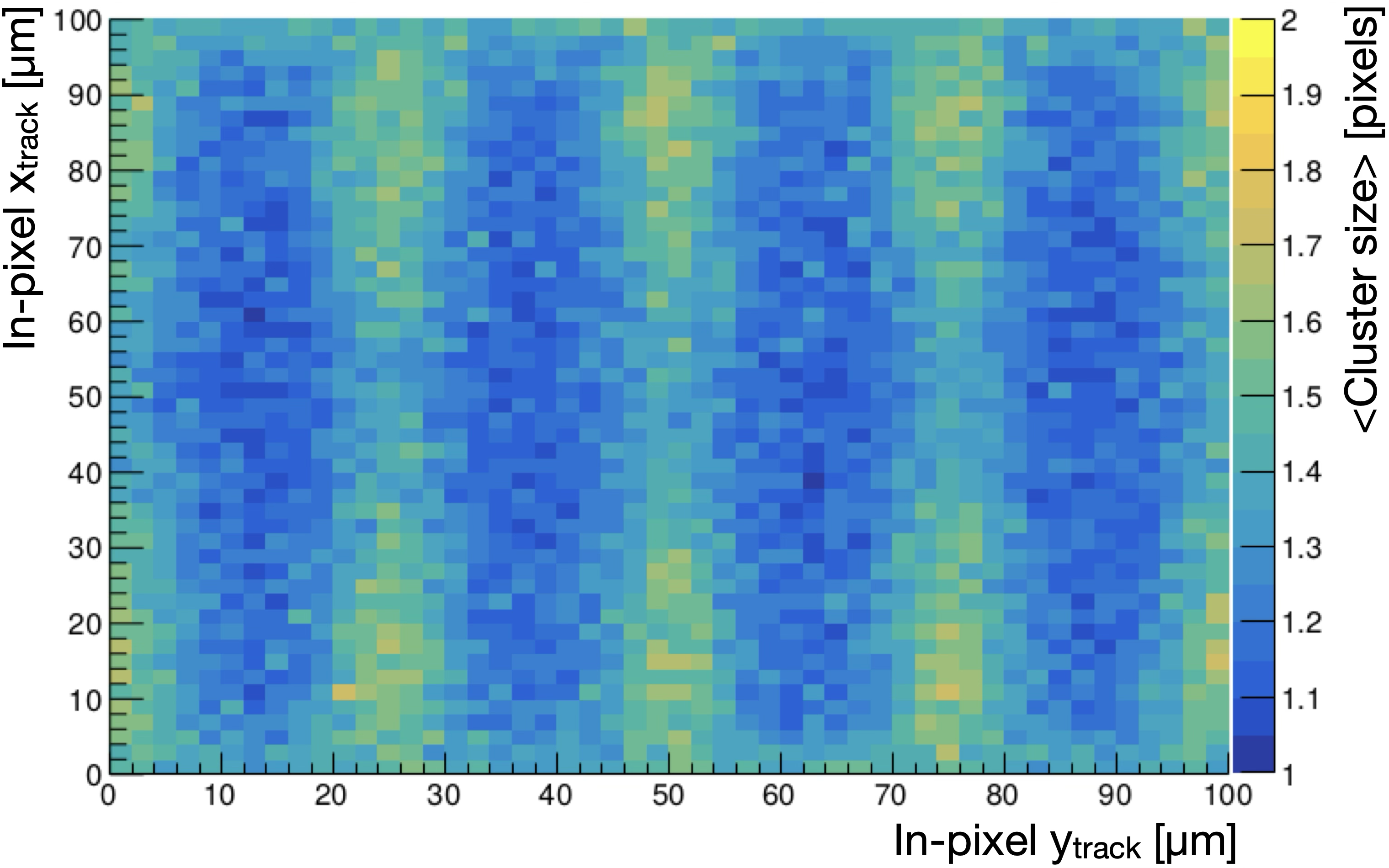}}
\caption{\label{3D25cls} Average cluster size maps at normal beam incidence for a \mbox{$4\times1 $} pixel grid of a non-irradiated FBK sensor biased at (a) $\SI{5}{V}$ and (b) $\SI{30}{V}$.}
\end{figure}

Figure~\ref{3D25cls} again shows a \mbox{$4\times1$} pixel grid with normal beam incidence, now presenting the average cluster size at two different bias voltages. As expected, the cluster size, mainly driven by diffusion, increases with distance from the central electrode, reaching its maximum for particles passing near the pixel edges. As the bias voltage increases, the strength of the electric field is enhanced, which reduces the charge sharing due to diffusion.
\vspace{0.5cm}

\begin{figure}[htb!]
\centering
\includegraphics[width=0.47\textwidth]{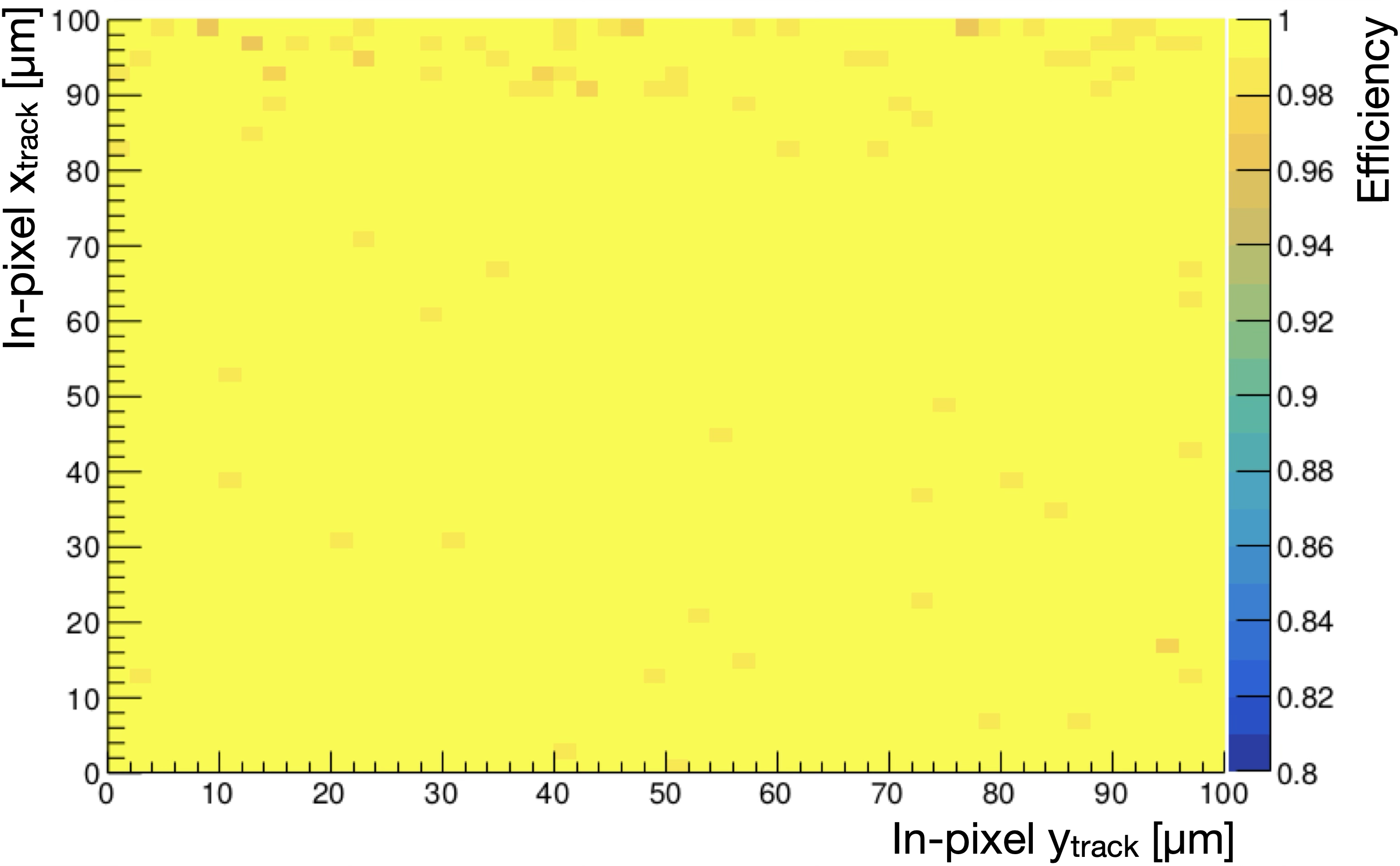}
\caption{\label{3D25effmang} Hit detection efficiency maps for a \mbox{$4\times1$} pixel grid of a non-irradiated FBK sensor rotated by $\SI{6}{\degree}$ and biased at $\SI{30}{V}$.}
\end{figure}

\clearpage
The hit detection efficiency can be further increased by rotating the DUT relative to the beam, as described in Section~\ref{sec:observables}, thereby shortening the path of the tracks through the electrodes. Figure~\ref{3D25effmang} shows the hit detection efficiency map for a \mbox{$4\times 1$} pixel grid with the sensor fully depleted and the DUT rotated by 6\textdegree. In this case, the inefficiencies corresponding to the \mbox{$p^+$ columns} disappear. Detector simulations have also demonstrated that for a track angular distribution similar to that found in minimum bias events from proton-proton collisions, the column inefficiency is negligible due to the tilted trajectories of the tracks. 

\begin{figure}[htb!]
\centering
\includegraphics[width=0.47\textwidth]{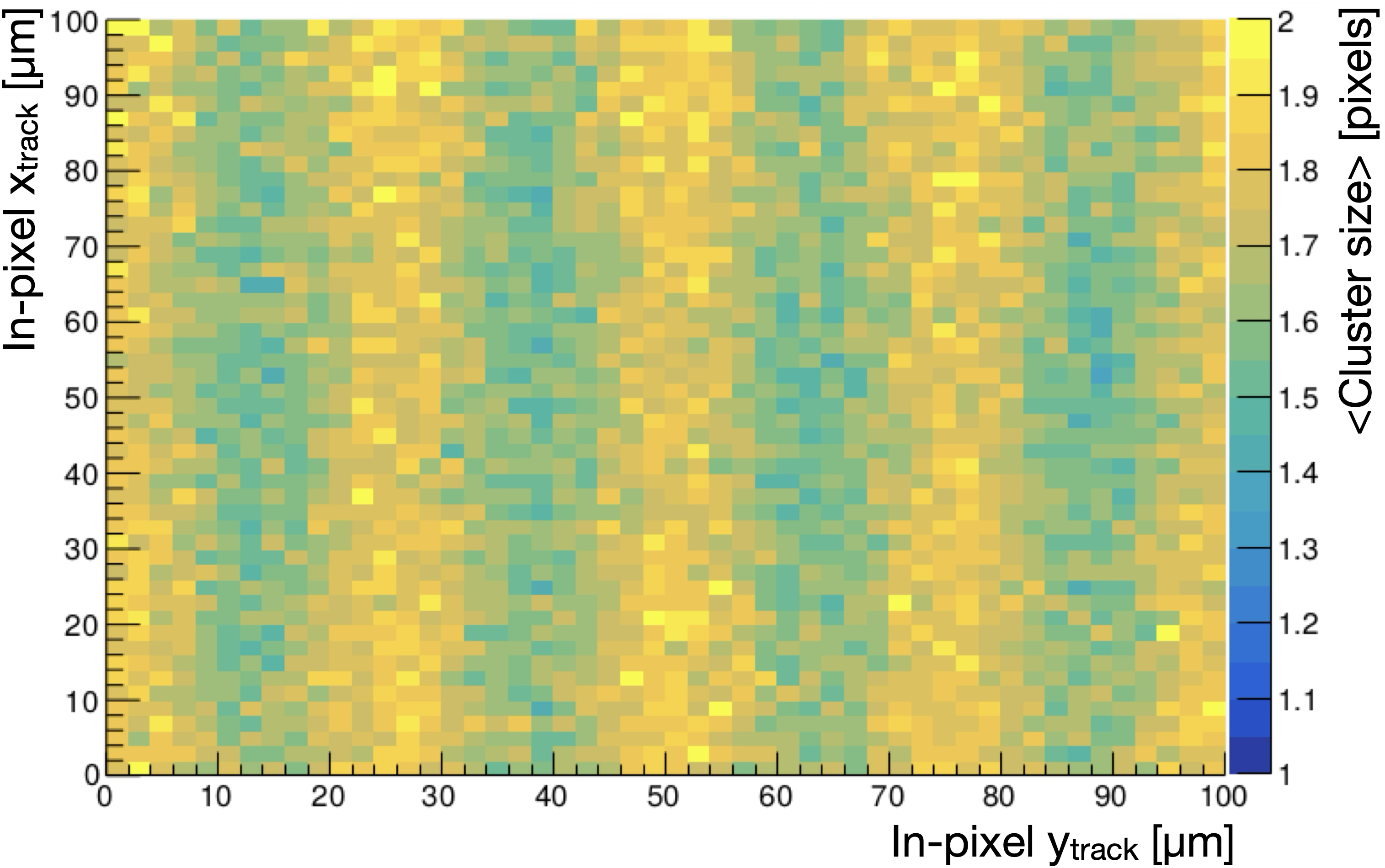}
\caption{\label{3D25clsmang} Average cluster size maps for a \mbox{$4\times1$} pixel grid of a non-irradiated FBK sensor rotated by $\SI{6}{\degree}$ and biased at $\SI{30}{V}$.}
\end{figure}

\begin{figure}[htb!]
\centering
\includegraphics[width=7.8cm]{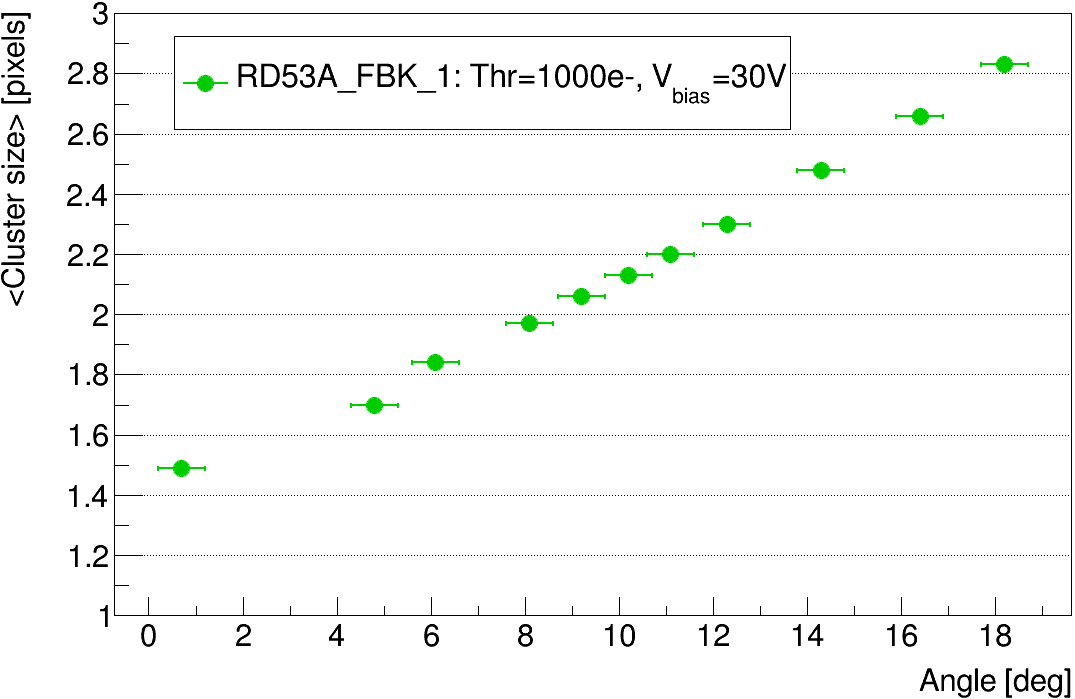}
\caption{\label{3D25clsang} Average cluster size of a non-irradiated FBK sensor biased at $\SI{30}{V}$ as a function of the rotation angle.}
\end{figure}

Figure~\ref{3D25clsmang} shows the average cluster size map for the same DUT configuration. The change in average cluster size as a function of rotation angle at a bias voltage of $\SI{30}{V}$ is shown in Fig.~\ref{3D25clsang}. The cluster size increases with angle, since particles are more likely to traverse two or more pixels.

\subsection{CROCv1 modules}
\label{sec:freshCROC}
Beam tests of CROCv1 devices have also been carried out, and their performance has been compared to that of RD53A samples as well as to the requirements described in Table~\ref{tab:sensorRequirements}. Four CROCv1 modules, built using 3D pixel sensors with a $25\times100$~$\SI{}{\micro m^{2}}$ pixel size from either FBK or CNM productions, were characterized. In addition, results from a CNM device with square pixels of \mbox{$\SI{50}{\micro m}$ pitch} are provided for comparison.

The devices were tuned to a pixel threshold ranging from 1000 to 1200 electrons, depending on the sample, at a temperature of $\SI{-10}{\degree C}$. In all cases, the total number of masked pixels was lower than 1\%.

\subsubsection{Hit detection efficiency and cluster charge}
\label{sec:hit_eff_fresh-CROC}
\vspace{0.1cm}

\begin{figure}[htb!]
    \centerline{\includegraphics[width=7.8cm]{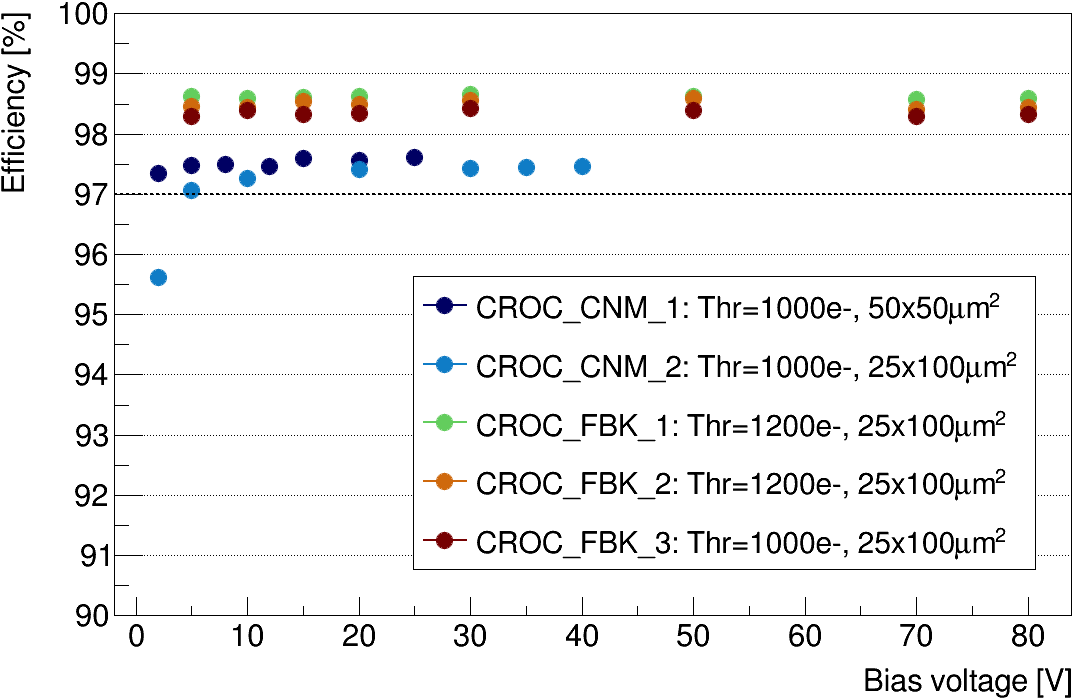}}
    \caption{Hit detection efficiency at normal incidence as a function of the bias voltage applied to several non-irradiated sensors. There are three samples from FBK and one from CNM with a $25\times100$~$\SI{}{\micro m^{2}}$ pixel size, as well as one CNM sensor with a $\SI{50}{\micro m}$ pitch. The dashed line indicates 97\% efficiency.}
    \label{fig:eff_bias_scan_fresh}
\end{figure}

The hit detection efficiency was measured at normal incidence as a function of the bias voltage, up to a maximum of $\SI{80}{V}$, with full depletion occurring below $\SI{10}{V}$. Figure~\ref{fig:eff_bias_scan_fresh} shows that the efficiency is well above 97\% once full depletion is reached.

The higher efficiency of FBK sensors compared to CNM sensors is due to the smaller radius of their columnar electrodes. Furthermore, the comparison to a CNM sample with a $50\times50$~$\SI{}{\micro m^{2}}$ pixel size shows that both cell designs asymptotically achieve the same performance.

\begin{figure}[htb!]
\centering
\subfloat[]{\includegraphics[width=7.81cm]{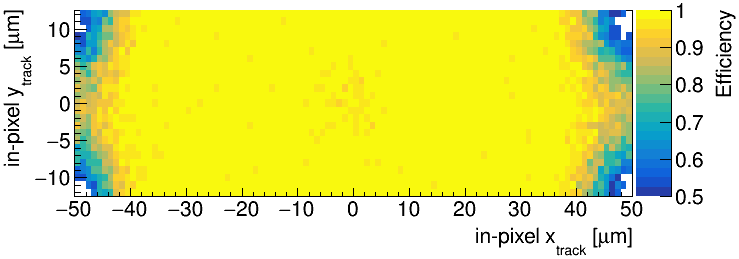}\label{fig:eff_CNM_2V}}
\hspace{0.5cm}
\subfloat[]{\includegraphics[width=7.81cm]{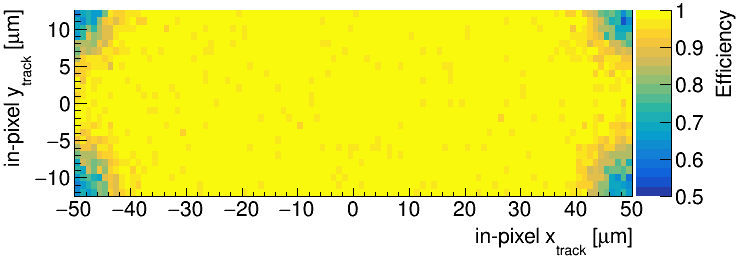}\label{fig:eff_CNM_30V}}
\hspace{0.5cm}
\subfloat[]{\includegraphics[width=7.8cm]{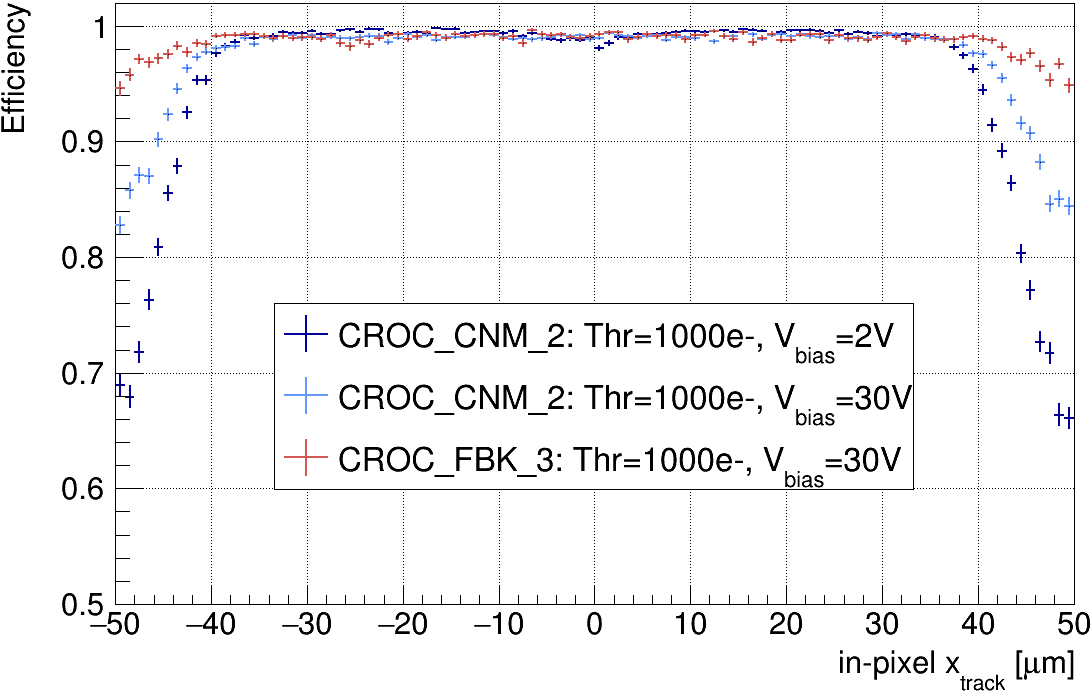}\label{fig:prof_eff_comparison_fresh}}
\vspace{-0.2cm}
\caption{Hit detection efficiency cell map of a non-irradiated CNM sensor biased at (a) $\SI{2}{V}$ and (b) $\SI{30}{V}$. White regions indicate an efficiency below \SI{0.5}{}. The efficiency profile along the long pitch of these maps and the corresponding one for a FBK sensor biased at $\SI{30}{V}$ are shown in (c).}
\label{fig:pixel_cell_eff_fresh}
\end{figure}

The efficiency cell maps corresponding to the CNM sample with a $25\times100$~$\SI{}{\micro m^{2}}$ pixel size are shown in Figs.~\figsubref{fig:pixel_cell_eff_fresh}{fig:eff_CNM_2V} and~\figsubref{fig:pixel_cell_eff_fresh}{fig:eff_CNM_30V} for a bias voltage below ($\SI{2}{V}$) and above ($\SI{30}{V}$) full depletion. Different efficiency regions exist because depletion occurs progressively from the central $n^+$ column towards the $p^+$ columns positioned at the corners of the pixel cell. The profile of these maps along the long edge of the pixel has been included in Fig.~\figsubref{fig:pixel_cell_eff_fresh}{fig:prof_eff_comparison_fresh} to facilitate the comparison between the two bias voltages. The efficiency profile of an FBK sensor biased at $\SI{30}{V}$ has been added to show how the smaller radius of its electrodes with respect to those of CNM increases the hit detection efficiency near the cell corners.

\begin{figure}[htb!]
    \centerline{\includegraphics[width=7.8cm]{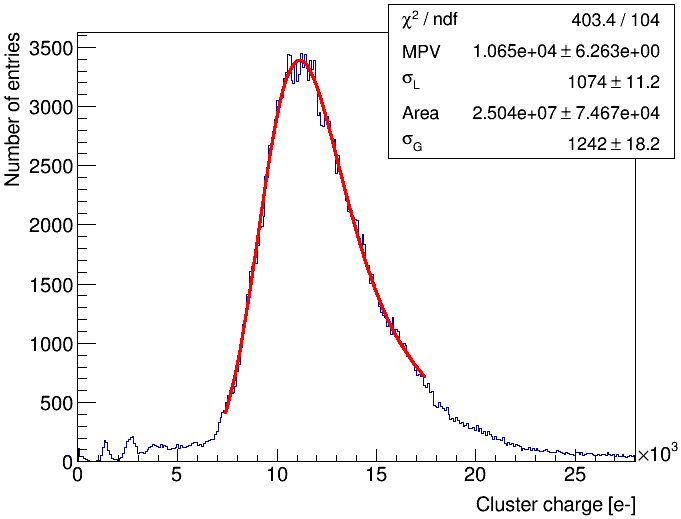}}
    \caption{Cluster charge distribution measured for a non-irradiated module biased at $\SI{30}{V}$, under normal incidence of a pion beam at $\SI{120}{GeV/c}$. It is fitted to a Landau distribution with most probable value \rm{MPV} and width $\sigma_{\rm{L}}$, convoluted with a Gaussian distribution with width $\sigma_{\rm{G}}$.}
    \label{fig:CROC_fresh_charge}
\end{figure}

A typical cluster charge distribution at normal beam incidence and full depletion of the sensor is shown in Fig.~\ref{fig:CROC_fresh_charge}. This distribution corresponds to module CROC\_FBK\_2 and is very similar to those of CROC\_FBK\_1 and CROC\_FBK\_3, as expected. The MPV extracted from a fit to the central region of the distribution is around $\SI{10600}{}$ electrons, consistent with simulations for a non-irradiated sensor of $\SI{150}{\micro m}$ thickness.


\subsubsection{Cluster size and spatial resolution}
\label{sec:resolution_fresh-CROC}
\vspace{0.1cm}

\begin{figure}[htb!]
\centering
\subfloat[]{\includegraphics[width=7.8cm]{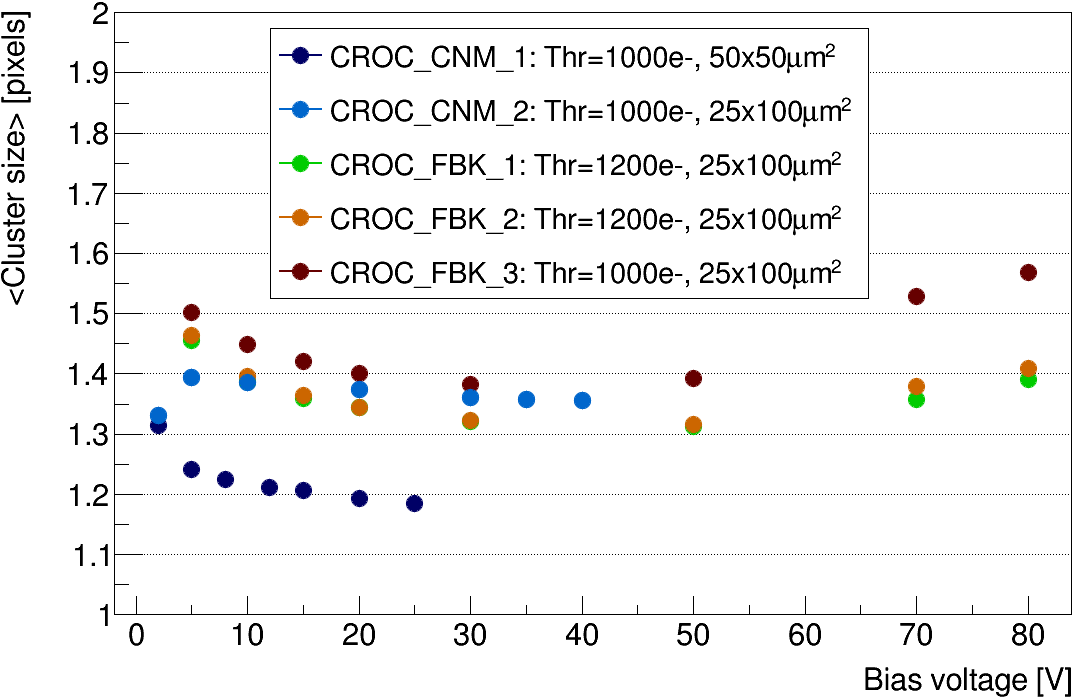}\label{fig:clsize_fresh}}
\hspace{0.5cm}
\vspace{-0.29cm}
\subfloat[]{\includegraphics[width=7.81cm]{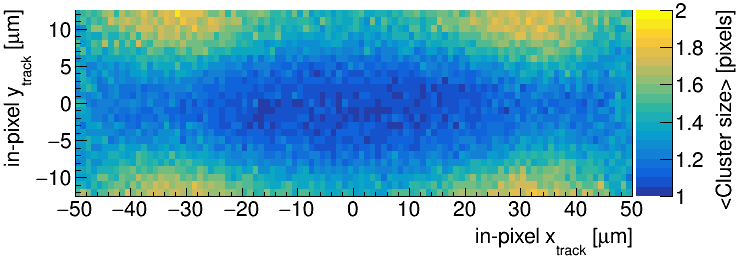}\label{fig:clsize_CNM_2V}}
\hspace{0.5cm}
\vspace{-0.29cm}
\subfloat[]{\includegraphics[width=7.80cm]{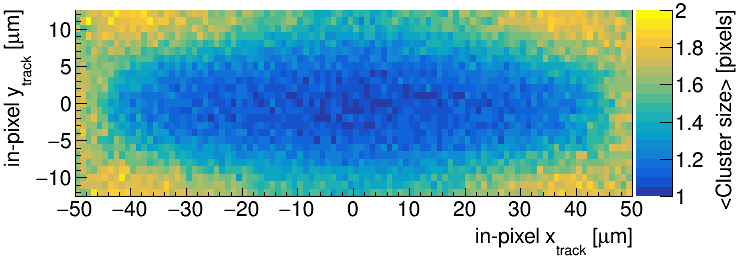}
\label{fig:clsize_CNM_5V}}
\hspace{0.5cm}
\vspace{-0.29cm}
\subfloat[]{\includegraphics[width=7.81cm]{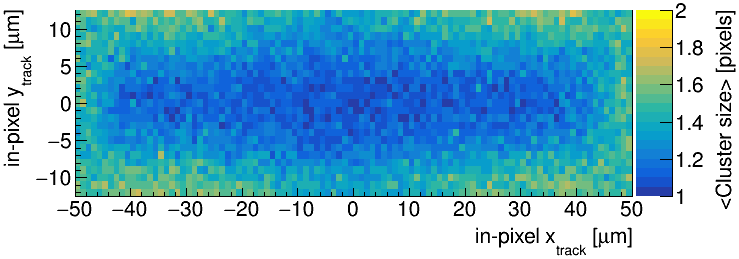}\label{fig:clsize_FBK_30V}}
\hspace{0.5cm}
\subfloat[]{\includegraphics[width=7.81cm]{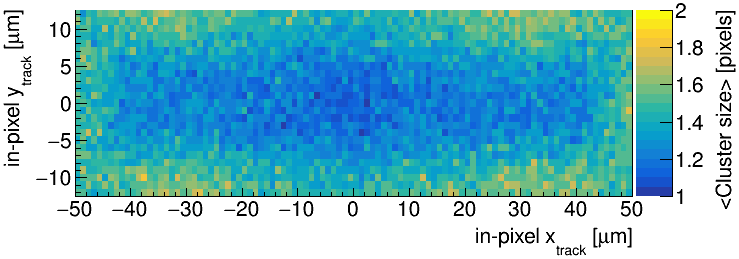}\label{fig:clsize_FBK_80V}}
\caption{(a) Average cluster size at normal incidence as a function of the bias voltage applied to several non-irradiated devices. The cluster size cell maps correspond to the CNM sensor with $25\times100$~$\SI{}{\micro m^{2}}$ pixel size biased at (b) $\SI{2}{V}$ and (c) $\SI{5}{V}$, as well as to one of the FBK sensors biased at (d) $\SI{30}{V}$ and (e) $\SI{80}{V}$.}
\label{fig:clsize_bias_scan_fresh}
\end{figure}

The average cluster size as a function of the bias voltage for non-irradiated CROC modules is shown in Fig.~\figsubref{fig:clsize_bias_scan_fresh}{fig:clsize_fresh}. As described earlier in this section, the cluster size depends on the electric field strength. It decreases with increasing bias voltage and gradually increases again once the voltage exceeds full depletion. The minimum cluster size occurs when the electric field is strong enough to considerably reduce the charge sharing caused by diffusion.
All sensors with a $25\times100$~$\SI{}{\micro m^{2}}$  pixel size show comparable cluster sizes, regardless of the manufacturer. Figures~\figsubref{fig:clsize_bias_scan_fresh}{fig:clsize_CNM_2V}~\dhyphen~\figsubref{fig:clsize_bias_scan_fresh}{fig:clsize_FBK_80V} display pixel cell maps of the cluster size at different bias voltages. The maps corresponding to $\SI{2}{V}$ and $\SI{5}{V}$ belong to module CROC\_CNM\_2 tuned to a threshold of 1000 electrons, while those corresponding to $\SI{30}{V}$ and $\SI{80}{V}$ refer to module CROC\_FBK\_1 tuned to a threshold of 1200 electrons. The increase of cluster size at higher bias voltages, as seen in Fig.~\figsubref{fig:clsize_bias_scan_fresh}{fig:clsize_fresh}, is homogeneously spread throughout the pixel cell.  One possible explanation for this observation is that for bias voltages significantly above full depletion, the electric field is strong enough to produce microavalanches in the material; whenever this happens in or near the struck pixel, the cluster size is artificially increased. 

\begin{figure}[htb!]
\centering
\subfloat[]{\includegraphics[width=7.8cm]{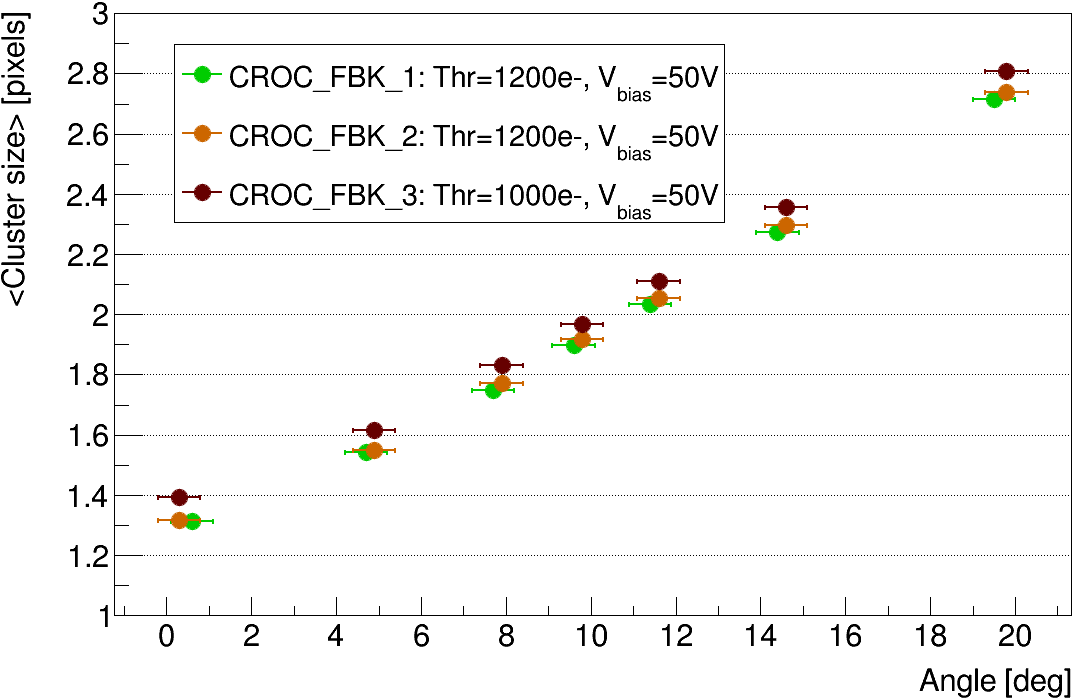}\label{fig:clsize_angle_fresh}}
\hspace{0.5cm}
\subfloat[]{\includegraphics[width=7.81cm]{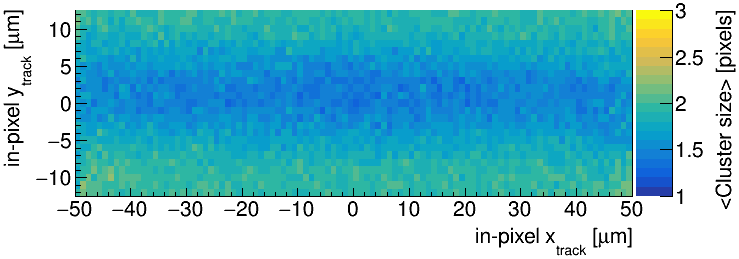}\label{fig:clsize_8deg_fresh}}
\hspace{0.5cm}
\subfloat[]{\includegraphics[width=7.80cm]{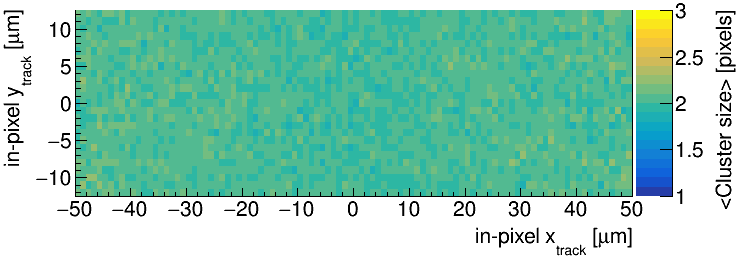}
\label{fig:clsize_12deg_fresh}}
\hspace{0.5cm}
\subfloat[]{\includegraphics[width=7.81cm]{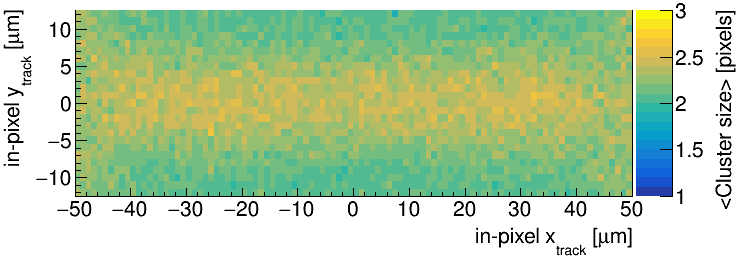}\label{fig:clsize_14deg_fresh}}
\hspace{0.5cm}
\caption{(a) Average cluster size of several non-irradiated FBK sensors biased at $\SI{50}{V}$ as a function of the rotation angle. Cluster size cell maps for one of the samples rotated by (b) $\SI{8}{\degree}$, (c) $\SI{12}{\degree}$, and (d) $\SI{15}{\degree}$.}
\label{fig:clsize_angle_scan_fresh}
\end{figure}

The evolution of the cluster size as a function of the rotation angle, as defined in Section~\ref{sec:observables}, has been studied for FBK sensors biased at $\SI{50}{V}$ and the results are shown in Fig.~\ref{fig:clsize_angle_scan_fresh}. As can be seen from the cluster size maps corresponding to angles of around $\SI{8}{\degree}$, $\SI{12}{\degree}$, and $\SI{15}{\degree}$, tracks generate a larger cluster size in different regions of the pixel cell depending on the angle. As expected, values are higher in the central region for larger angles, because particles crossing close to the center of a pixel cell can deposit charge that is spread among three pixels.

\begin{figure}[htb!]
    \centering
    \subfloat[]{\includegraphics[width=7.8cm]{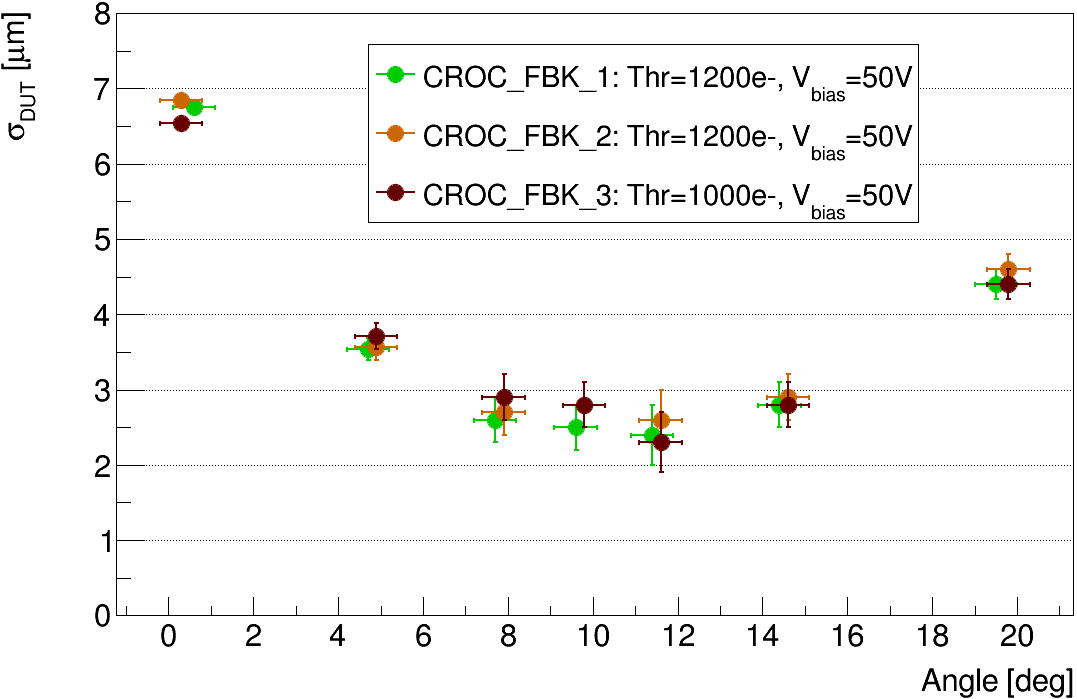}\label{fig:fresh_DUT_resolution}}
    \hspace{0.5cm}
    \centering
    \subfloat[]{\includegraphics[width=7.8cm]{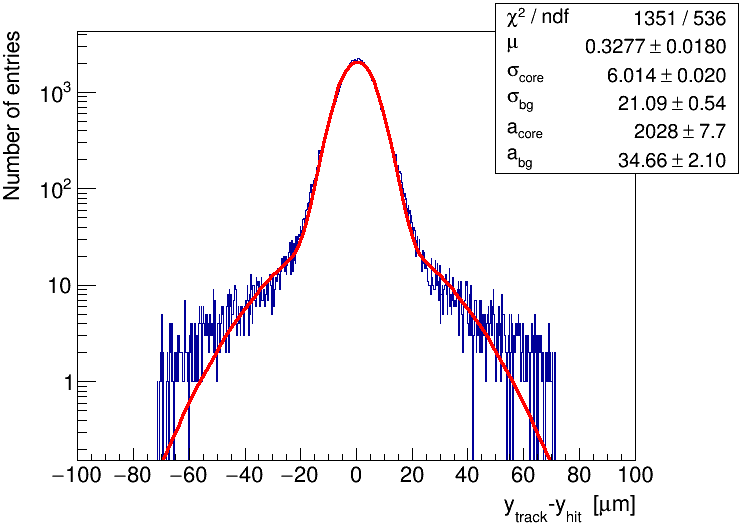}\label{fig:fresh_DUT_residualsX}}
    \hspace{0.5cm}
    \centering
    \subfloat[]{\includegraphics[width=7.8cm]{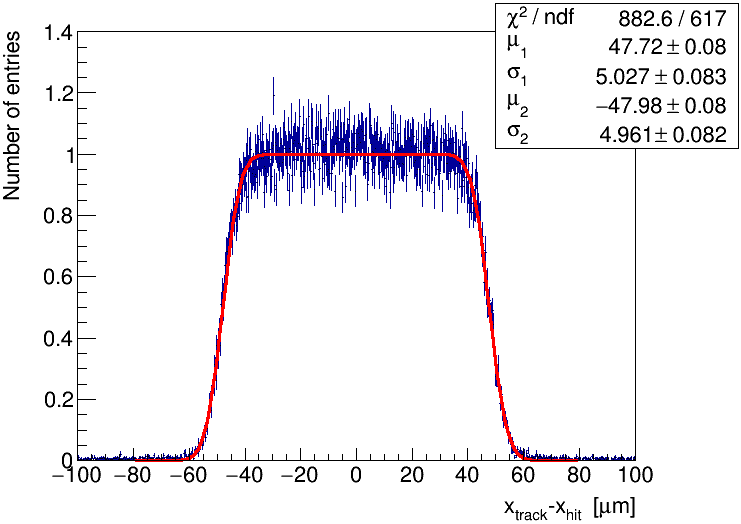}\label{fig:fresh_DUT_residualsY}}
    \caption{(a) Resolution $\sigma_{\rm{DUT}}$ of several non-irradiated FBK sensors, biased at $\SI{50}{V}$, as a function of the rotation angle. The residual distributions along (b) the $\SI{25}{\micro m}$ pixel pitch and (c) the $\SI{100}{\micro m}$ pixel pitch for one of the samples rotated by $\SI{10}{\degree}$ are shown. The latter is normalized to the mean number of entries in the central region. The residual distribution along the short pitch is fitted to the sum of two Gaussian distributions, while the edges of the residual distribution along the long pitch are fitted to the CDF of a Gaussian distribution.}
    \label{fig:CROC_fresh_resolution}
    \vspace{0.5cm}
\end{figure}

From the same set of measurements, the spatial resolution of FBK sensors as a function of the angle has been estimated and is shown in Fig.~\ref{fig:CROC_fresh_resolution}. The best resolution of around $\SI{2.5}{\micro m}$ is obtained for rotation angles where the average cluster size reaches a value of 2 (Fig.~\ref{fig:clsize_angle_scan_fresh}). Residual distributions from one of the modules, used to estimate the spatial resolution, are shown in Figs.~\ref{fig:fresh_DUT_residualsX} and~\ref{fig:fresh_DUT_residualsY}, corresponding to the short and long pixel pitches, respectively. The DUT residuals along the \SI{25}{\micro m} pitch are fitted to the sum of two Gaussians, accounting for a background contribution mainly due to track-hit misassociations. The standard deviation of the Gaussian that fits the core of the distribution represents the combined resolution of the DUT and the telescope, as described in Section~\ref{sec:observables}. For the DUT residuals along the \SI{100}{\micro m} pitch, the edges are fitted with the CDF of a Gaussian, which provides an approximate telescope resolution of~\SI{5}{\micro m}. Since the telescope resolution degrades for rotated modules, a correction proportional to the cosine of the angle is applied to determine the DUT resolution along the short pixel pitch.

%%%%%%%%%%%%%%%%%%%%%%%%%%%%%%%%%
%%%%%%%%%%%%%%%%%%%%%%%%%%%%%%%%%

\clearpage
\section{Results for irradiated single-chip modules}
\label{sec:irradiated_results}
For the results in this section, single-chip modules were irradiated at KIT, FNAL, and CERN, and their post-irradiation performance was evaluated in test beams. The results are presented separately for each ROC flavor. A summary of the devices can be found in~\ref{sec:DUTsummary}.

Although the CMS efficiency requirement for normal incidence after irradiation (96\%) is looser than for inclined incidence (97\%), the stricter 97\% criterion is adopted consistently throughout this analysis to ensure uniform comparisons across all measurements, including those at normal incidence and for non-irradiated samples.

\subsection{RD53A modules}

Three sensors with 3D pixels interconnected with RD53A chips were irradiated at KIT with fluences ranging from \mbox{\SI{1.4e16}{n_{eq}/\cm^{2}}} to \mbox{\SI{1.8e16}{n_{eq}/\cm^{2}}}, and their performance was tested in beam at DESY. Module RD53A\_FBK\_2 is from the Stepper-1 production, while RD53A\_FBK\_3 and RD53A\_FBK\_4 are from the Stepper-2 production. The modules were tuned to average pixel thresholds between 1400 electrons and 1700 electrons, at a temperature of $\SI{-27}{\degree C}$. 

Additional sensors bonded to the RD53A readout chip were sent to the Fermilab ITA facility for irradiation and subsequently tested at the FTBF. The results for a single device, referred to as RD53A\_CNM\_1 in the following, are presented here. The fluence received by the module is estimated to be \mbox{\SI{1.2e16}{n_{eq}/\cm^{2}}}, comparable to that of the modules irradiated at KIT and tested at DESY.  Two test beam campaigns were conducted with the device tuned to two different thresholds of 1200 and 1600 electrons at temperatures between $\SI{-30}{\degree C}$ and $\SI{-25}{\degree C}$. 

\begin{figure}[htb!]
    \centering
    \subfloat[]{\includegraphics[width=7.8cm]{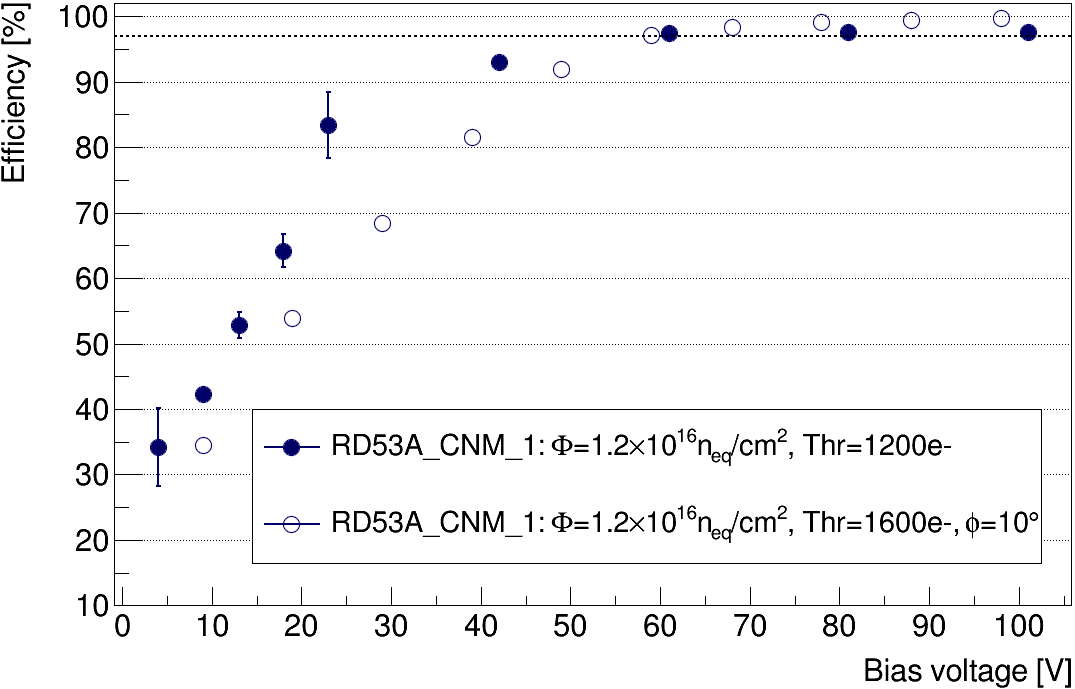}\label{fig:RD53A_FTBF_irrad_eff}}
    \hspace{0.5cm}
    \subfloat[]{\includegraphics[width=7.8cm]{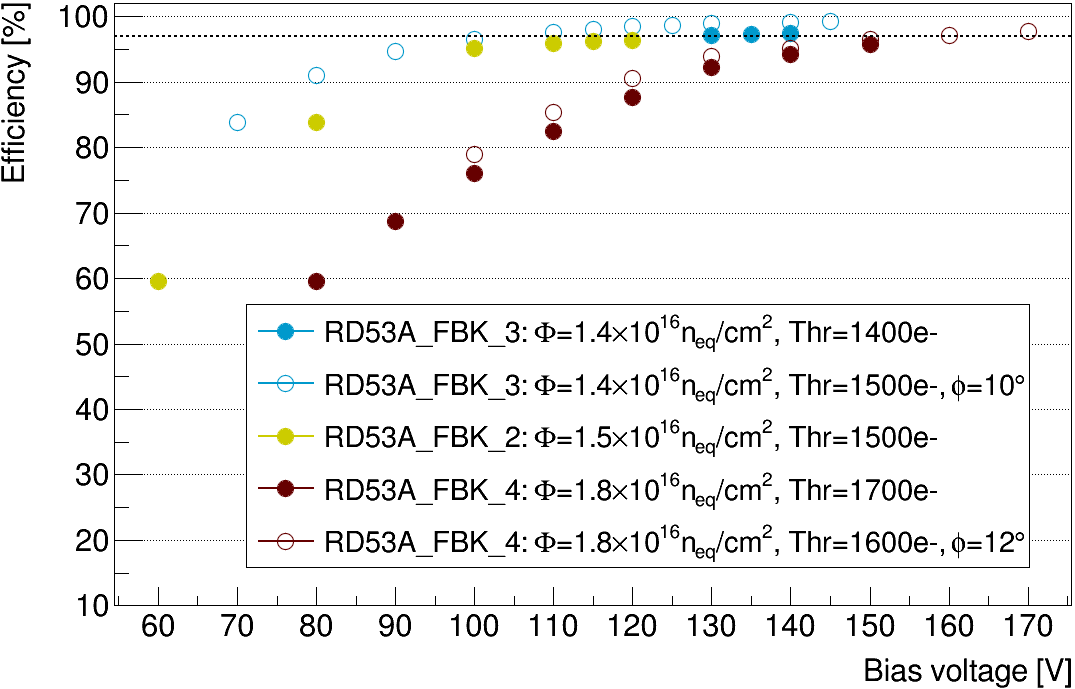}\label{fig:RD53A_DESY_irrad_eff}}
    \caption{Hit detection efficiency as a function of the bias voltage applied to irradiated 3D pixel modules tested at (a) FTBF and (b) DESY. The full circles refer to normal incidence, while the open circles indicate a rotation angle. The dashed line corresponds to the 97\% efficiency requirement from Table~\ref{tab:sensorRequirements}.}
    \label{fig:RD53A_irrad_eff}
\end{figure}

Figure~\ref{fig:RD53A_irrad_eff} shows the hit detection efficiency of all irradiated modules as a function of the bias voltage. For a fluence of \mbox{\SI{1.2e16}{n_{eq}/\cm^{2}}} (RD53A\_CNM\_1), a hit detection efficiency greater than 97\% is obtained above $\SI{70}{V}$ (Fig.~\figsubref{fig:RD53A_irrad_eff}{fig:RD53A_FTBF_irrad_eff}). Higher bias voltages are required to reach the same efficiency as the fluence increases, with $\SI{110}{V}$ needed at a fluence of \mbox{\SI{1.5e16}{n_{eq}/\cm^{2}}} and \mbox{$\SI{150}{V}$ for \SI{1.8e16}{n_{eq}/\cm^{2}}} (Fig.~\figsubref{fig:RD53A_irrad_eff}{fig:RD53A_DESY_irrad_eff}). Results are also shown for small rotation angles. In all cases, the rotated samples show higher efficiency compared to normal beam incidence, as the rotation reduces the probability of a particle going straight through the $p^+$ column.

\newpage
In module RD53A\_FBK\_2 (Stepper-1), a sudden increase in the number of masked channels was observed at bias voltages above $\SI{130}{V}$. A similar increase occurred in modules RD53A\_FBK\_3 and RD53A\_FBK\_4 (Stepper-2), albeit at higher bias voltages: $\SI{150}{V}$ for the less irradiated module and $\SI{170}{V}$ for the more irradiated one. The shorter \mbox{$n^+$ column} in the Stepper-2 modules may explain the different behavior of the two productions. If the tip of the \mbox{$n^+$ column} (kept at ground potential by the readout chip) is positioned too close to the backside of the sensor (at bias voltage potential), electric discharges can occur, potentially compromising the functionality of the sensor~\cite{DallaBetta}. For all data points shown in Fig.~\figsubref{fig:RD53A_irrad_eff}{fig:RD53A_DESY_irrad_eff}, the number of masked channels is below 1\%. More studies on the number of masked channels were carried out on CROC assemblies, as reported in Section~\ref{sec:irradCROC}.

\begin{figure}[htb!]
    \centering
    \includegraphics[width=7.8cm]{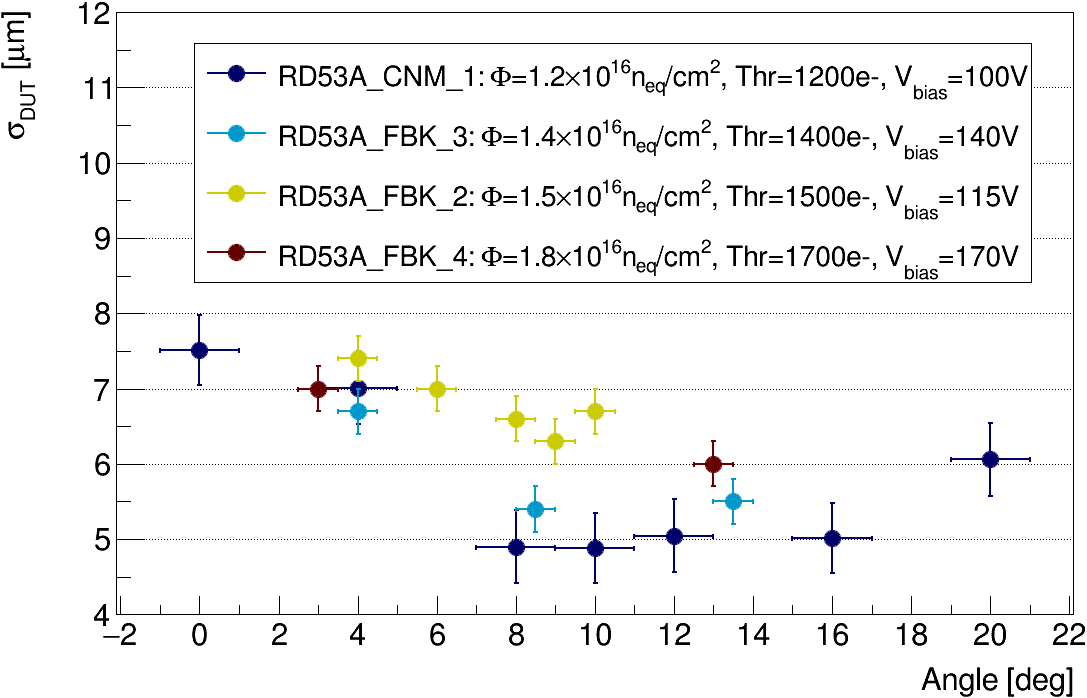}
    \caption{Resolution $\sigma_{\rm{DUT}}$ of the irradiated 3D pixel modules tested at DESY and FTBF as a function of the rotation angle away from normal incidence.}
    \label{fig:RD53A_irrad_res}
\end{figure}

\newpage
Fig.~\ref{fig:RD53A_irrad_res} shows the spatial resolution of the RD53A devices as a function of the rotation angle, as defined in Section~\ref{sec:observables}. The observed resolution degrades with increasing fluence as expected, with best values ranging between 5 and $\SI{6}{\micro m}$.

\subsubsection{Results for fluences beyond specifications}
\vspace{0.1cm}

As explained in Section~\ref{sec:irradiation} the irradiations performed at the CERN PS are non-uniform along one direction. The estimation of the fluence relies on the spectroscopy measurements performed on the aluminum foils glued to the samples, which are split in eight segments to obtain more granular information. A two-dimensional Gaussian profile is used to fit the eight measurements. The standard deviations are determined using the beam profile given by the IRRAD personnel, and the mean value along the coordinate with non-uniform irradiation is fixed to the position where the hit detection efficiency is lowest. From this fit, the fluences received by the central regions of two samples were estimated to be \mbox{\SI{2.1e16}{n_{eq}/\cm^{2}}} (RD53A\_FBK\_5) and \mbox{\SI{2.6e16}{n_{eq}/\cm^{2}}} (RD53A\_FBK\_6). The results presented below are based on the analysis of pixels from these highly irradiated areas.

\begin{figure}[htb!]
    \centering
    \includegraphics[width=7.8cm]{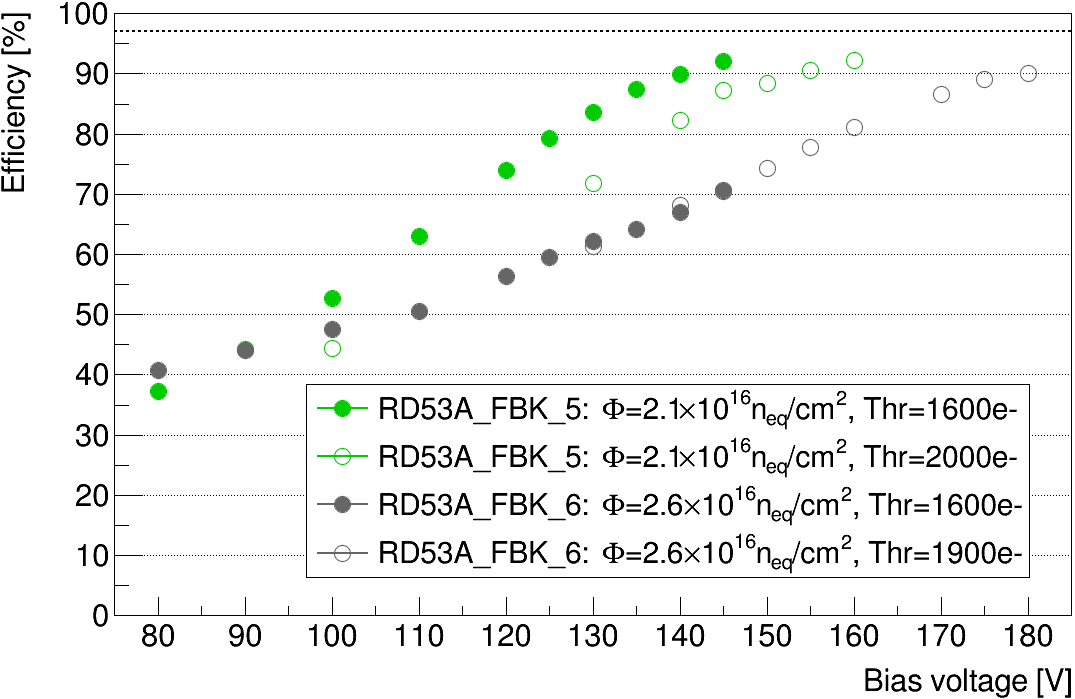}
    \caption{Hit detection efficiency as a function of the bias voltage applied to two highly irradiated 3D pixel modules at normal beam incidence. The full and open circles represent low and high average pixel thresholds, respectively.}
    \label{fig:RD53A_veryHighFluence}
\end{figure}

Due to the significant irradiation dose and highly non-uniform irradiation profile, it was not possible to perform a proper gain calibration of the ADC. As a result, charge and resolution studies could not be performed. The samples were studied in two separate test beam campaigns using different global threshold settings. Figure~\ref{fig:RD53A_veryHighFluence} shows the hit detection efficiency of the two samples as a function of the applied bias voltage with a global threshold of about 1600 and 2000 electrons. Even the most irradiated sensor reaches an efficiency of 90\% at $\SI{180}{V}$ with 1\% masked pixels. These irradiation fluences are well beyond the operational conditions of the IT at the HL-LHC, but the measurements presented here can be relevant for possible future applications of these sensors.

\vspace{-0.15cm}
\subsection{CROCv1 modules}
\label{sec:irradCROC}
Three modules (CROC\_FBK\_2, CROC\_FBK\_3, CROC\_CNM\_2) that were characterized before irradiation (Section~\ref{sec:freshCROC})  were irradiated to a fluence of \mbox{\SI{1e16}{n_{eq}/\cm^{2}}} at the CERN PS and retested in beam at the SPS. An additional  module (CROC\_FBK\_4) was irradiated to a fluence of \mbox{\SI{1.6e16}{n_{eq}/\cm^{2}}} at KIT and tested in beam at DESY. 

\begin{figure}[htb!]
    \centerline{\includegraphics[width=7.8cm]{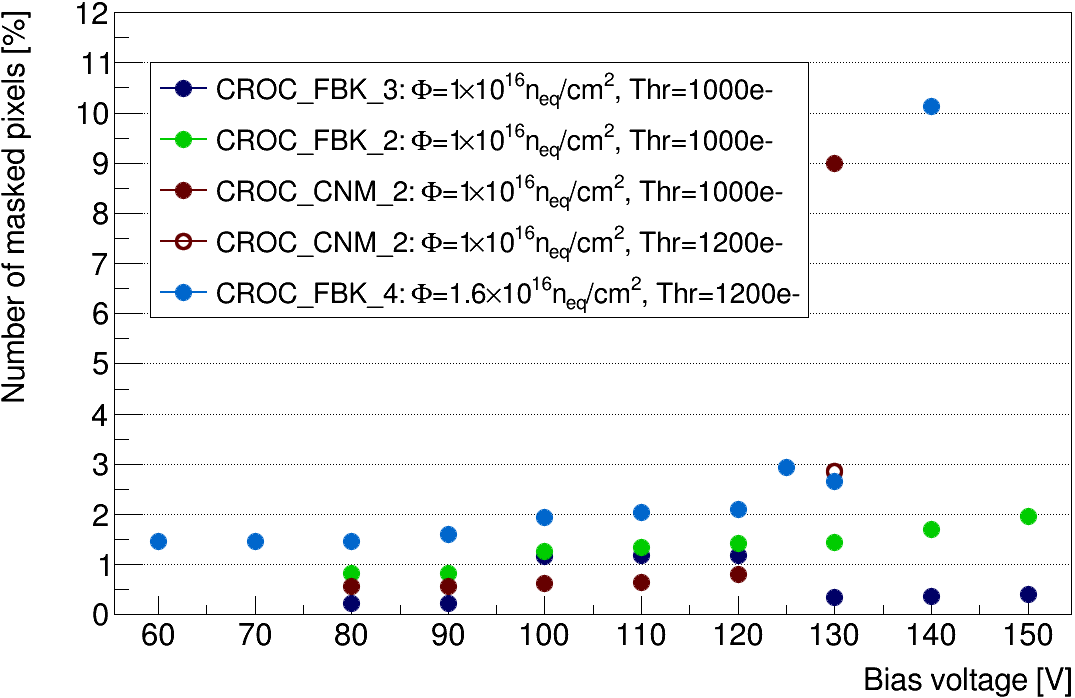}}
    \caption{Percentage of masked pixels as a function of the bias voltage applied to sensors irradiated to \mbox{\SI{1e16}{n_{eq}/\cm^{2}}} and \mbox{\SI{1.6e16}{n_{eq}/\cm^{2}}}.}
    \label{fig:noise_scan_irrad}
\end{figure}

The modules irradiated to the lower fluence have been tuned to an average pixel threshold of 1000~electrons at a temperature of $\SI{-30}{\degree C}$, unless otherwise specified. For the module exposed to the largest fluence, the tuning has been done to an average pixel threshold of 1200~electrons at a similar temperature.  The percentage of masked pixels as a function of the bias voltage for all the devices is shown in Fig.~\ref{fig:noise_scan_irrad}. The occupancy cut applied for the determination of noisy channels is the same as the one used in the tuning of the modules before irradiation. The less irradiated FBK samples are stable up to a bias voltage of $\SI{150}{V}$, with the percentage of masked pixels staying below 2\%.  The CNM module has a steep increase at $\SI{130}{V}$.  Raising the threshold to 1200 electrons in the CNM module reduces the percentage of masked channels at that bias voltage from 9\% to 3\%. Therefore, the upcoming plots for this module at $\SI{130}{V}$ will show only the data corresponding to a threshold of 1200 electrons. For the more irradiated FBK device, the fraction of masked pixels remains below 3\% up to $\SI{130}{V}$, but rises sharply to 10\% at $\SI{140}{V}$.
\clearpage

\subsubsection{Hit detection efficiency and cluster charge}
\label{sec:hit_eff_irrad-CROC}
\vspace{0.1cm}

\begin{figure}[htb!]
\centering
\subfloat[]{\includegraphics[width=7.8cm]{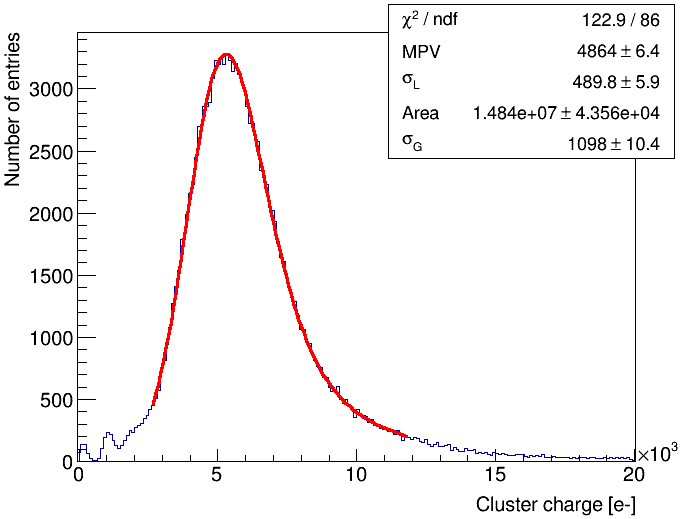}\label{fig:CROC_irrad_charge}}
\hspace{0.5cm}
\vspace{-0.01cm}
\subfloat[]{\includegraphics[width=7.8cm]{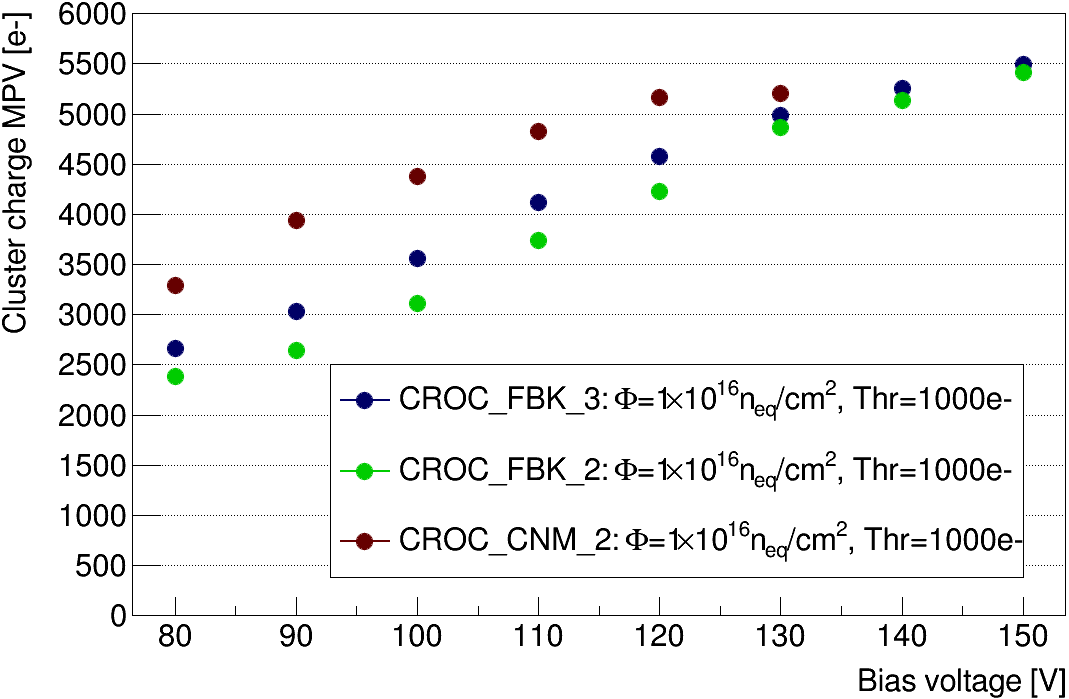}\label{fig:charge_bias_scan_1e16}}
\hspace{0.5cm}
\vspace{-0.01cm}
\subfloat[]{\includegraphics[width=7.8cm]{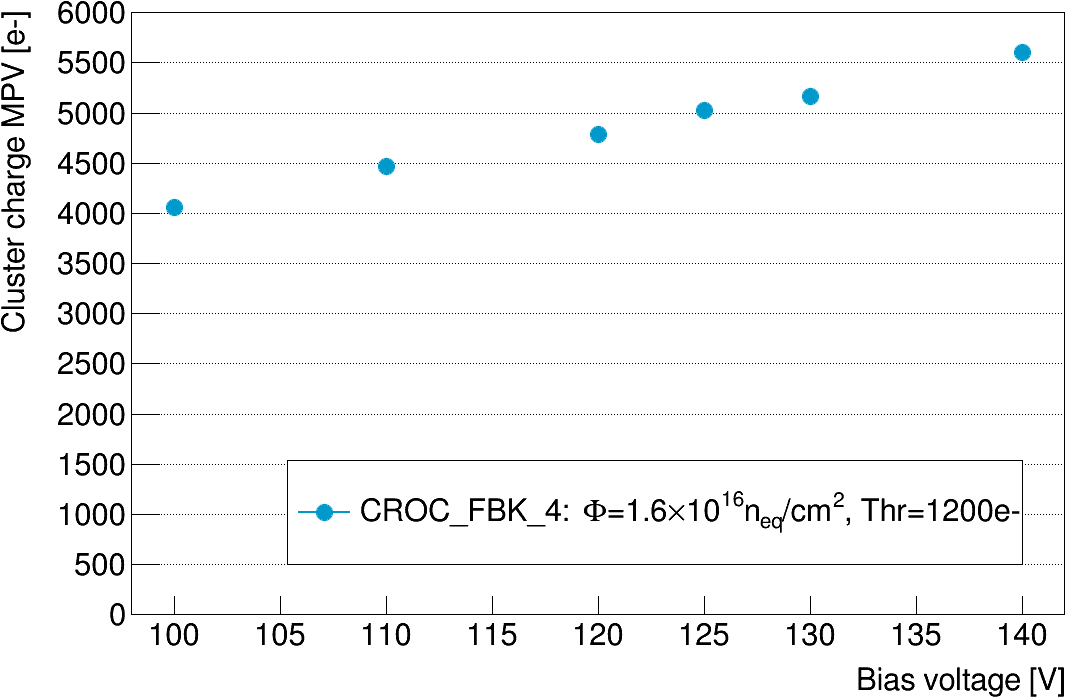}\label{fig:charge_bias_scan_1d6e16}}
\caption{(a) Cluster charge distribution measured for a module irradiated to \mbox{\SI{1e16}{n_{eq}/\cm^{2}}} and biased at $\SI{130}{V}$, under normal incidence of a pion beam at $\SI{120}{GeV/c}$. It is fitted to a Landau distribution with most probable value \rm{MPV} and width $\sigma_{\mathrm{L}}$, convoluted with a Gaussian distribution with width $\sigma_{\mathrm{G}}$. The cluster charge is shown as a function of the bias voltage applied to modules irradiated to (b) \mbox{\SI{1e16}{n_{eq}/\cm^{2}}} and (c) \mbox{\SI{1.6e16}{n_{eq}/\cm^{2}}}, under normal incidence of a pion beam at $\SI{120}{GeV/c}$ and an electron beam at $\SI{5.2}{GeV/c}$, respectively.}
\label{fig:charge_irrad}
\end{figure}

Figure~\figsubref{fig:charge_irrad}{fig:CROC_irrad_charge} shows the fitted cluster charge distribution for module CROC\_FBK\_2 at normal incidence and a bias voltage of $\SI{130}{V}$. The MPV, which is around 5500 electrons at the highest bias voltage, indicates a reduction to approximately half the charge observed before irradiation~(Fig.~\ref{fig:CROC_fresh_charge}). The MPVs of the cluster charge distributions as a function of the bias voltage are shown in Figs.~\figsubref{fig:charge_irrad}{fig:charge_bias_scan_1e16} and~\figsubref{fig:charge_irrad}{fig:charge_bias_scan_1d6e16} for modules irradiated to \mbox{\SI{1e16}{n_{eq}/\cm^{2}}} and \mbox{\SI{1.6e16}{n_{eq}/\cm^{2}}}, respectively. The MPV as a function of bias voltage exhibits an increasing trend for both fluences, with a feeble saturation observed in the devices irradiated at the lower fluence. The MPV values themselves are also comparable between the two fluences, which seems compatible with TCAD simulations based on the ``CERN" radiation damage model reported in~\cite{Boughedda_2021}, where minimal changes in charge collection efficiency are predicted between \mbox{\SI{1e16}{n_{eq}/\cm^{2}}} and \mbox{\SI{1.5e16}{n_{eq}/\cm^{2}}}. An important caveat to this comparison is that the device irradiated to the higher fluence was exposed to \SI{23}{MeV} protons at KIT, while the lower-fluence irradiations were performed at CERN using \SI{23}{GeV} protons. Hence, the radiation damage conditions may differ between the devices. This can affect the observed charge collection and should be considered when interpreting the results.

\begin{figure}[htb!]
\centering
\subfloat[]{\includegraphics[width=7.8cm]{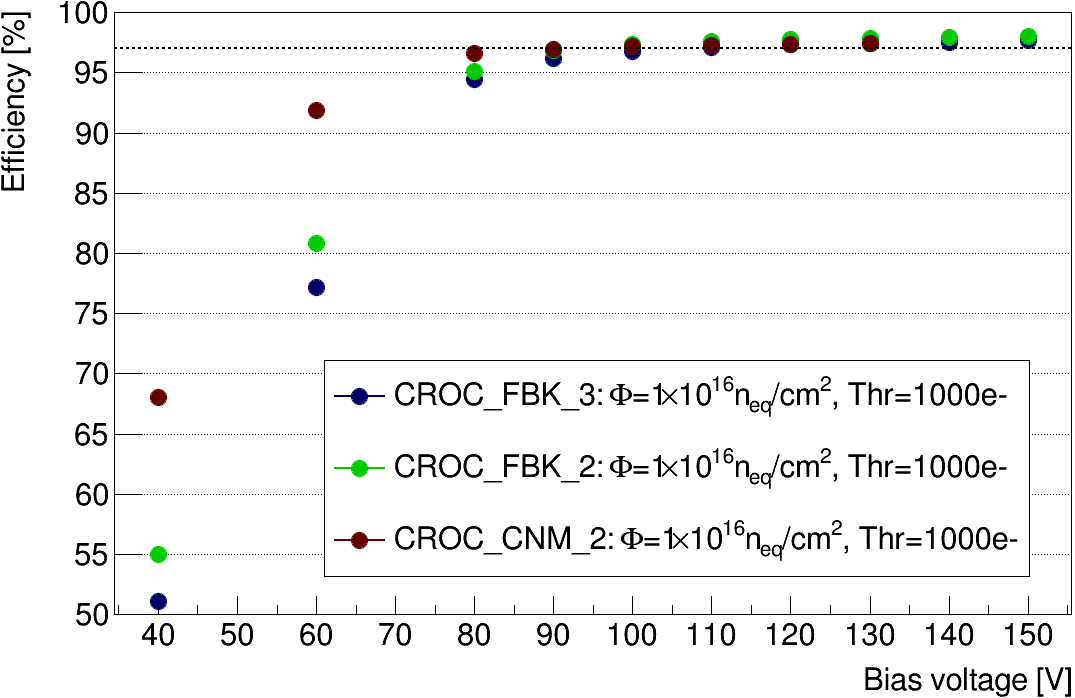}\label{fig:CROC_irrad_eff}}
\hspace{0.5cm}
\subfloat[]{\includegraphics[width=7.8cm]{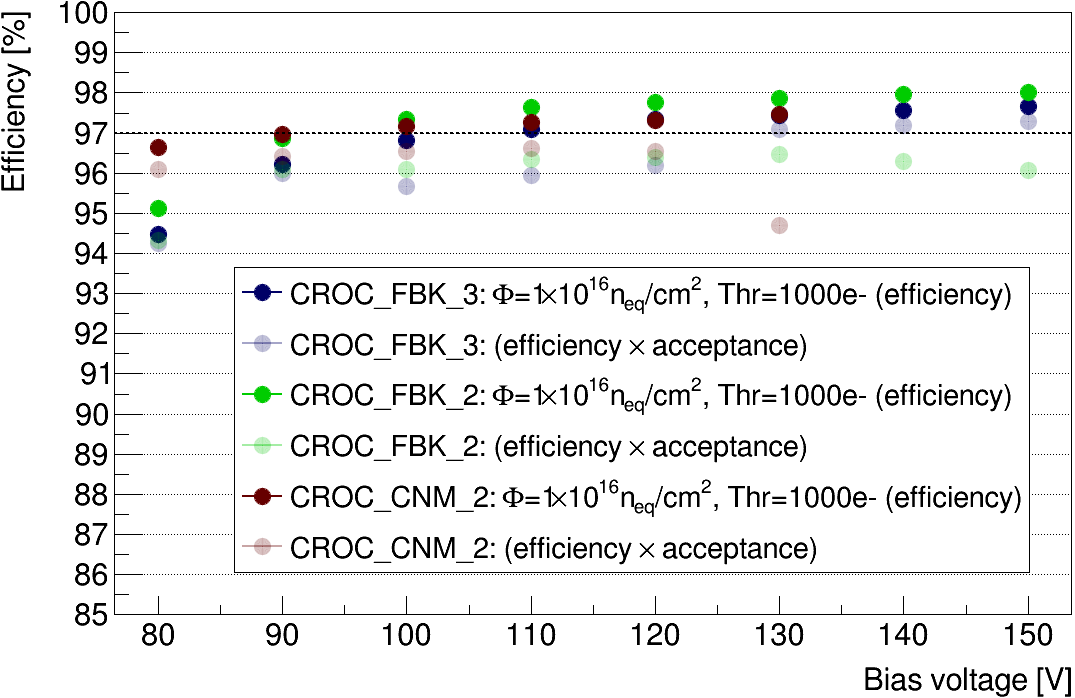}\label{fig:CROC_irrad_effxacc}}
\hspace{0.5cm}
\subfloat[]{\includegraphics[width=7.8cm]{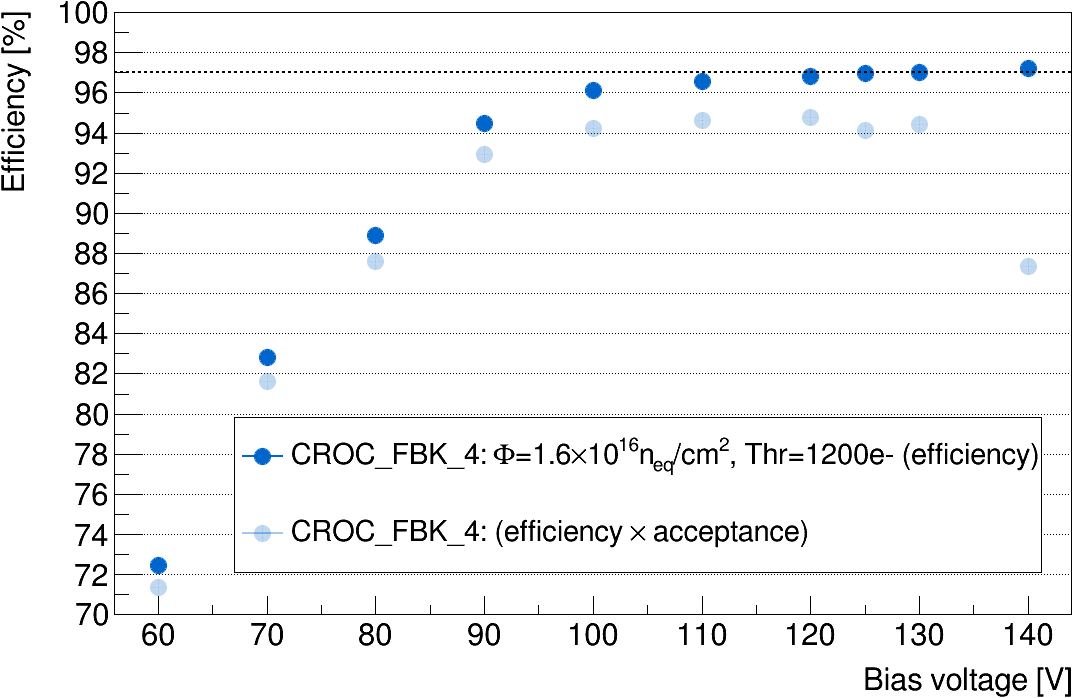}\label{fig:CROC_irrad_eff_1d6e16}}
\caption{(a) Hit detection efficiency as a function of the bias voltage in modules irradiated to \mbox{\SI{1e16}{n_{eq}/\cm^{2}}}. (b) Hit detection efficiency for the same devices on the approximate efficiency plateau, with and without the correction for the acceptance. (c) Hit detection efficiency with and without the acceptance correction as a function of the bias voltage in a module irradiated to \mbox{\SI{1.6e16}{n_{eq}/\cm^{2}}}. The solid colors represent the efficiency while the lighter colors indicate the efficiency times the acceptance. The dashed line corresponds to the 97\% efficiency requirement stated in Table~\ref{tab:sensorRequirements}.}
\label{fig:eff_irrad}
\end{figure}

\clearpage
The hit detection efficiency after irradiation at normal incidence is more than 97\% for all modules, as illustrated in Fig.~\ref{fig:eff_irrad}. At a fluence of \mbox{\SI{1e16}{n_{eq}/\cm^{2}}}, the efficiency plateau is reached from around $80\dhyphen\SI{90}{V}$, while at \mbox{\SI{1.6e16}{n_{eq}/\cm^{2}}}, about $\SI{100}{V}$ are needed. This corresponds to an operational range of $30\dhyphen\SI{50}{V}$ in which the modules can be operated with excellent performance and low noise, meeting the CMS requirements shown in Table~\ref{tab:sensorRequirements}. To assess the impact of masked pixels on the efficiency, the acceptance with the definition given by Eq.~\ref{eq:3} is used. The efficiency corrected for the acceptance is depicted in Figs.~\figsubref{fig:eff_irrad}{fig:CROC_irrad_effxacc} and~\figsubref{fig:eff_irrad}{fig:CROC_irrad_eff_1d6e16} for the low and high fluences, respectively.

\begin{figure}[htb!]
\centering
\subfloat[]{\includegraphics[width=7.8cm]{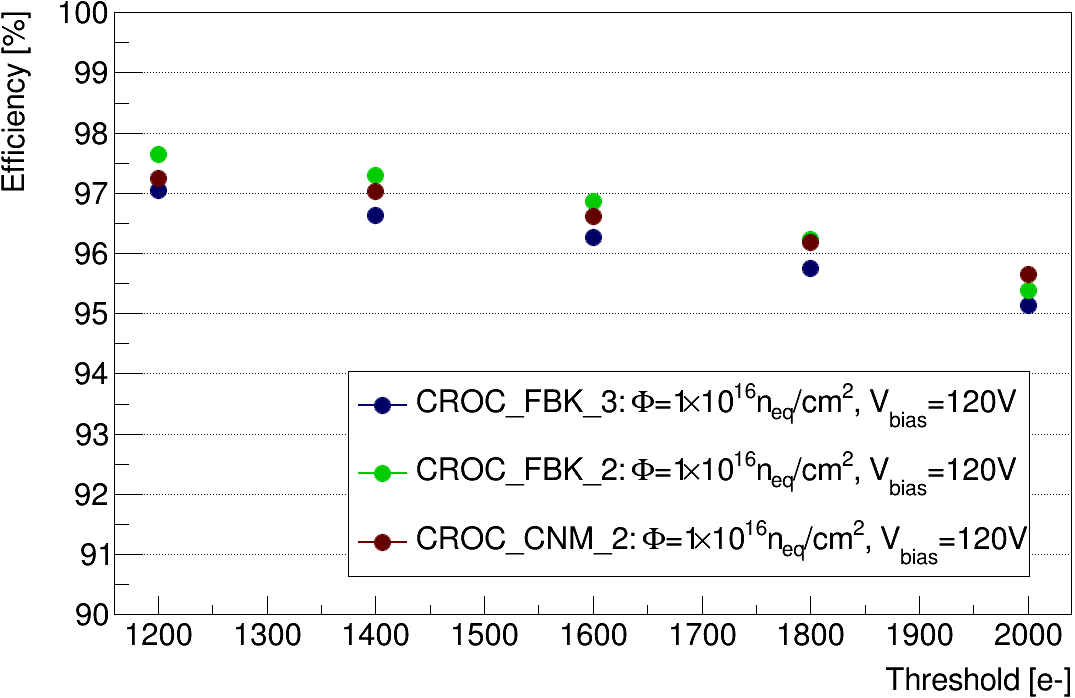}\label{fig:thresholdScan_irrad_1e16}}
\hspace{0.5cm}
\subfloat[]{\includegraphics[width=7.8cm]{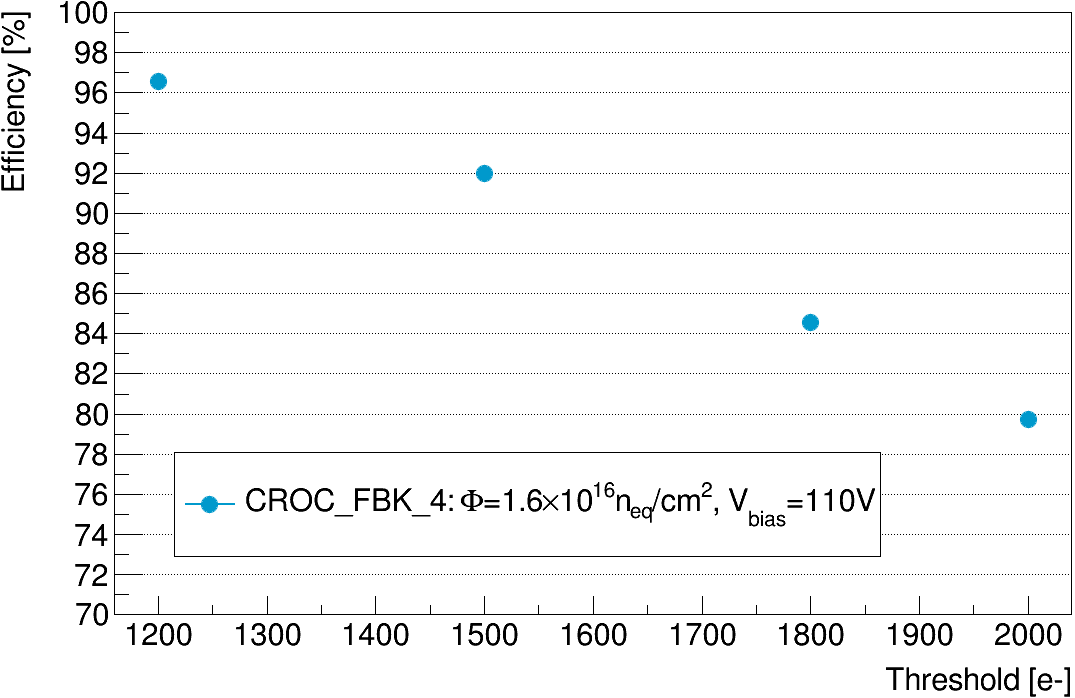}\label{fig:thresholdScan_irrad_1d6e16}}
\caption{Hit detection efficiency at normal incidence as a function of the average pixel threshold to which the modules irradiated at (a) \mbox{\SI{1e16}{n_{eq}/\cm^{2}}} and (b) \mbox{\SI{1.6e16}{n_{eq}/\cm^{2}}} were tuned. These measurements were performed without a noise mask and at bias voltages of 120 and $\SI{110}{V}$, respectively.}
\label{fig:thresholdScan_irrad}
\end{figure}

The efficiency also depends on the threshold to which the modules are tuned. This was studied for modules irradiated to \mbox{\SI{1e16}{n_{eq}/\cm^{2}}} and \mbox{\SI{1.6e16}{n_{eq}/\cm^{2}}}, and the outcome is summarized in Fig.~\ref{fig:thresholdScan_irrad}.  For these measurements, no noise mask was applied and the beam was normally incident. The samples irradiated to \mbox{\SI{1e16}{n_{eq}/\cm^{2}}} and \mbox{\SI{1.6e16}{n_{eq}/\cm^{2}}} were biased at $\SI{120}{V}$ and $\SI{110}{V}$, respectively. An efficiency loss of around 2\% is seen when doubling the threshold in the less irradiated samples. This reduction becomes more significant as the fluence increases, reaching a drop of around 16\% in the sample irradiated at the higher fluence.

\begin{figure}[htb!]
    \centerline{\includegraphics[width=7.8cm]{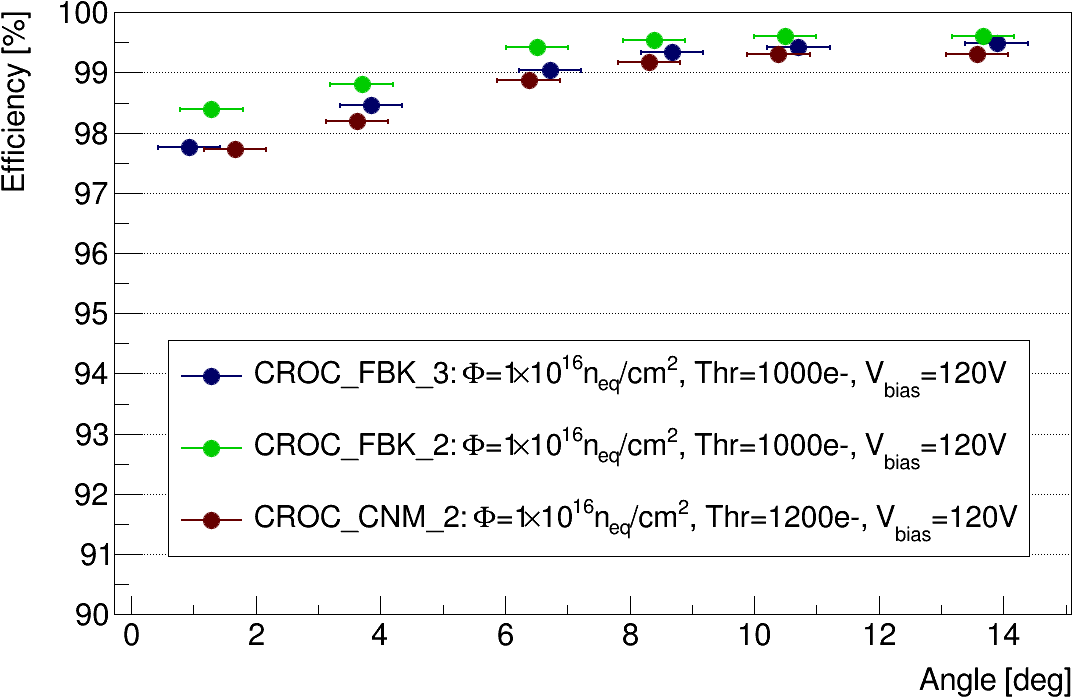}}
    \caption{Hit detection efficiency for modules irradiated to \mbox{\SI{1e16}{n_{eq}/\cm^{2}}} as a function of their angle with respect to the beam. These measurements were performed at a bias voltage of $\SI{120}{V}$ and without a noise mask.}
    \label{fig:eff_angular_scan}
\end{figure}

Measurements of the efficiency as a function of the rotation angle, as defined in Section~\ref{sec:observables}, were made.  The samples irradiated to \mbox{\SI{1e16}{n_{eq}/\cm^{2}}} were biased at $\SI{120}{V}$ and no noise mask was applied. The FBK and CNM devices were tuned to thresholds of 1000 and 1200 electrons, respectively. Figure~\ref{fig:eff_angular_scan} illustrates the recovery of the inefficiency with angle, which is attributed to the effect of the columnar electrodes discussed earlier. The efficiency exceeds 99\% from around $\SI{8}{\degree}$.

\begin{figure}[htb!]
\centering
\subfloat[]{\includegraphics[width=7.81cm]{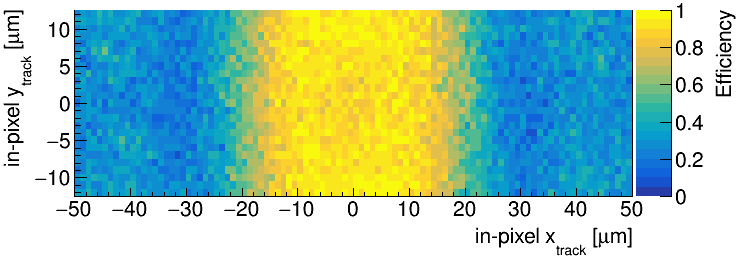}\label{fig:eff_FBK_40V}}
\hspace{0.5cm}
\subfloat[]{\includegraphics[width=7.81cm]{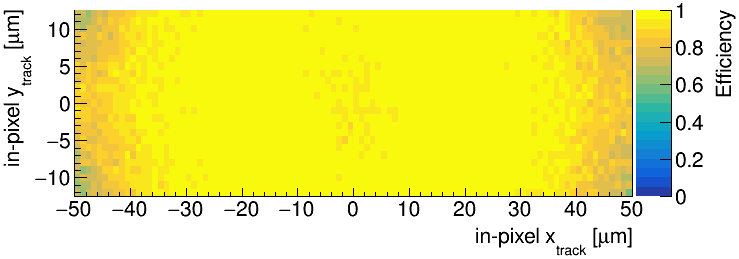}\label{fig:eff_FBK_80V}}
\hspace{0.5cm}
\subfloat[]{\includegraphics[width=7.81cm]{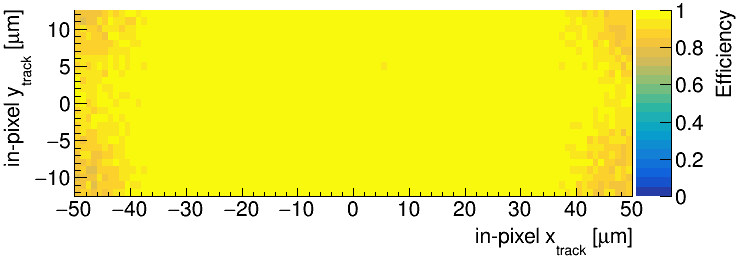}\label{fig:eff_FBK_120V}}
\hspace{0.4cm}
\subfloat[]{\includegraphics[width=7.82cm]{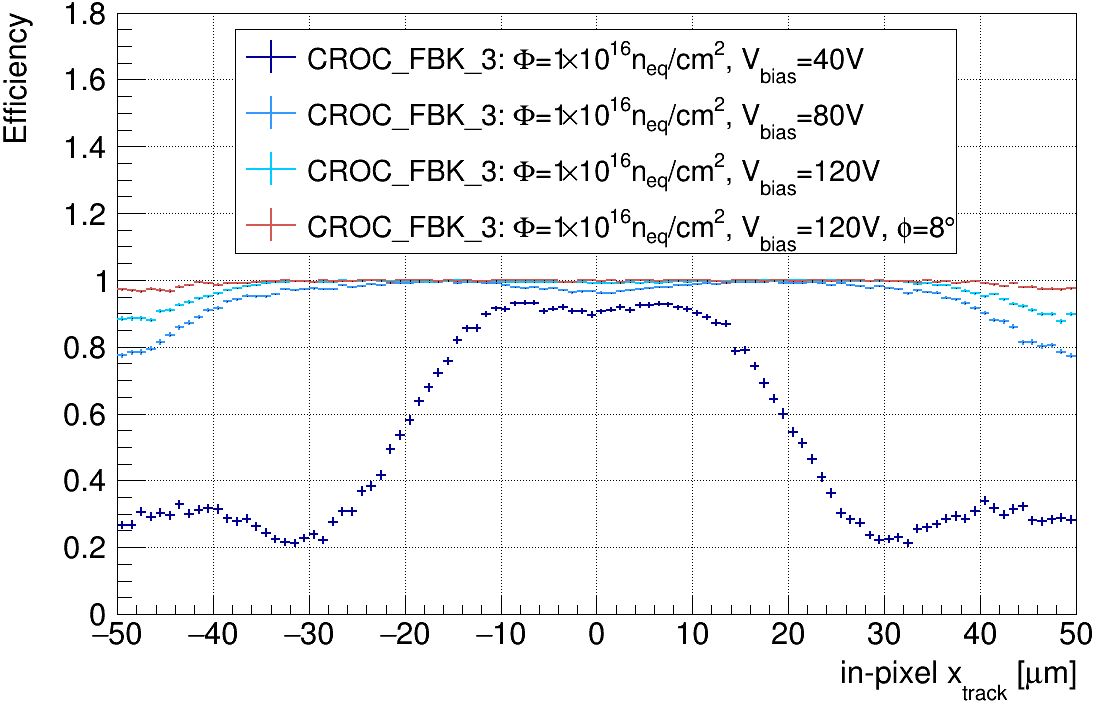}\label{fig:prof_eff_comparison_irrad}}
\caption{Hit detection efficiency cell map of an FBK sensor irradiated to \mbox{$\SI{1e16}{n_{eq}/\cm^{2}}$} and biased at (a) $\SI{40}{V}$, (b) $\SI{80}{V}$ and (c) $\SI{120}{V}$ with normal beam incidence. Their corresponding efficiency profiles along the long pitch direction together with that of the module biased at $\SI{120}{V}$ and rotated by roughly $\SI{8}{\degree}$ are shown in (d).}
\label{fig:pixel_cell_eff_irrad}
\end{figure}

The efficiency cell maps for an FBK sensor irradiated to \mbox{\SI{1e16}{n_{eq}/\cm^{2}}} at three bias voltages ($\SI{40}{V}$, $\SI{80}{V}$, $\SI{120}{V}$) are shown in Fig.~\ref{fig:pixel_cell_eff_irrad} to illustrate the progress in the depletion of the sensor after irradiation. The change of the efficiency in the pixel cell with the bias voltage can also be observed through the profiles of these maps along the long pitch shown in Fig.~\figsubref{fig:pixel_cell_eff_irrad}{fig:prof_eff_comparison_irrad}. In addition, the profile coming from the same sensor biased at $\SI{120}{V}$ and rotated by an angle of about $\SI{8}{\degree}$ has been included for comparison to those at normal beam incidence and as an example of the columnar inefficiency recovery.

\clearpage
\subsubsection{Cluster size}
\label{sec:resolution_irrad-CROC}
\vspace{0.1cm}

\begin{figure}[htb!]
\centering
\subfloat[]{\includegraphics[width=7.8cm]{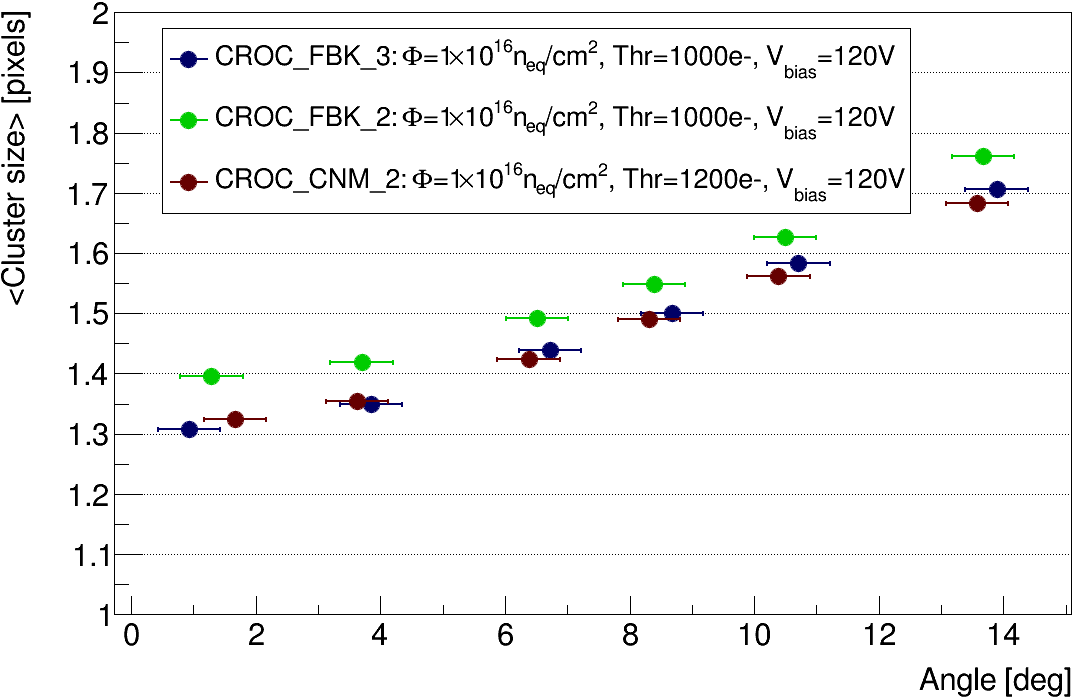}
\label{fig:clsize_angle_irrad}}
\hspace{0.5cm}
\vspace{-0.35cm}
\subfloat[]{\includegraphics[width=7.8cm]{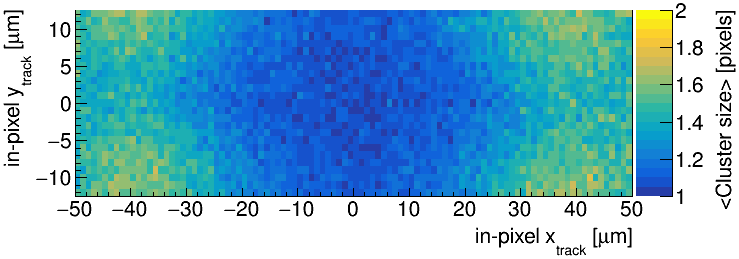}
\label{fig:clsize_120V_irrad}}
\hspace{0.5cm}
\vspace{-0.35cm}
\subfloat[]{\includegraphics[width=7.8cm]{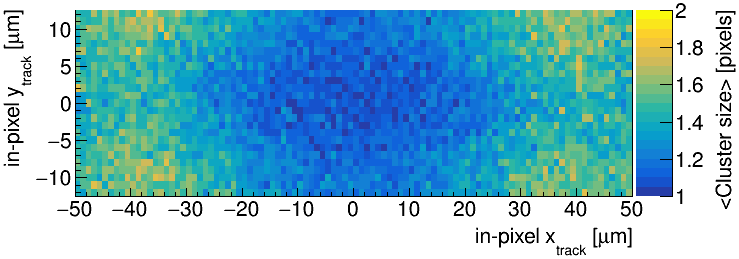}
\label{fig:clsize_4deg_irrad}}
\hspace{0.5cm}
\vspace{-0.35cm}
\subfloat[]{\includegraphics[width=7.8cm]{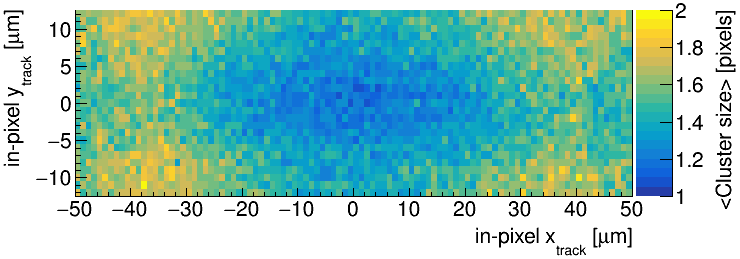}
\label{fig:clsize_8deg_irrad}}
\hspace{0.5cm}
\vspace{-0.1cm}
\subfloat[]{\includegraphics[width=7.8cm]{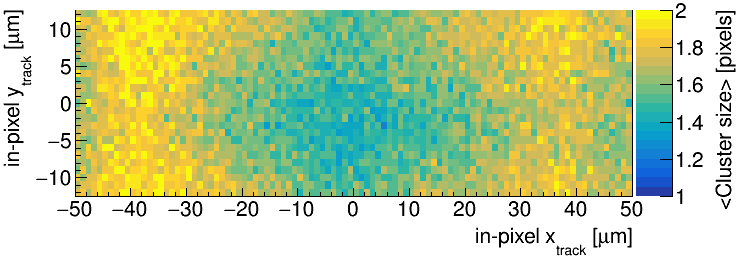}
\label{fig:clsize_14deg_irrad}}
\caption{(a) Average cluster size as a function of the rotation angle for several modules irradiated to \mbox{$\SI{1e16}{n_{eq}/\cm^{2}}$} and biased at $\SI{120}{V}$. Cluster size cell maps at an angle of around (b) $\SI{1}{\degree}$, (c) $\SI{4}{\degree}$, (d) $\SI{8}{\degree}$ and (e) $\SI{14}{\degree}$ from one of the FBK samples.}
\label{fig:clsize_angle_scan_irrad}
\vspace{0.5cm}
\end{figure}

The dependence of the mean cluster size on the rotation angle and threshold was also studied for the irradiated CROC modules, given the importance of the cluster size for the spatial resolution. Figure~\figsubref{fig:clsize_angle_scan_irrad}{fig:clsize_angle_irrad} presents the evolution of the average cluster size with increasing rotation angle.  This scan was performed with the modules irradiated to \mbox{\SI{1e16}{n_{eq}/\cm^{2}}} and biased at $\SI{120}{V}$. In comparison to non-irradiated modules, the increase in cluster size with the angle is substantially lower. Nevertheless, the mean cluster size increases by roughly 25\% when the module is rotated by about $\SI{14}{\degree}$. Maps of the mean cluster size for a pixel cell at four different angles are shown in Figs.~\figsubref{fig:clsize_angle_scan_irrad}{fig:clsize_120V_irrad}~\dhyphen~\figsubref{fig:clsize_angle_scan_irrad}{fig:clsize_14deg_irrad}, for module CROC\_FBK\_3.

\begin{figure}[htb!]
\centering
\subfloat[]{\includegraphics[width=7.8cm]{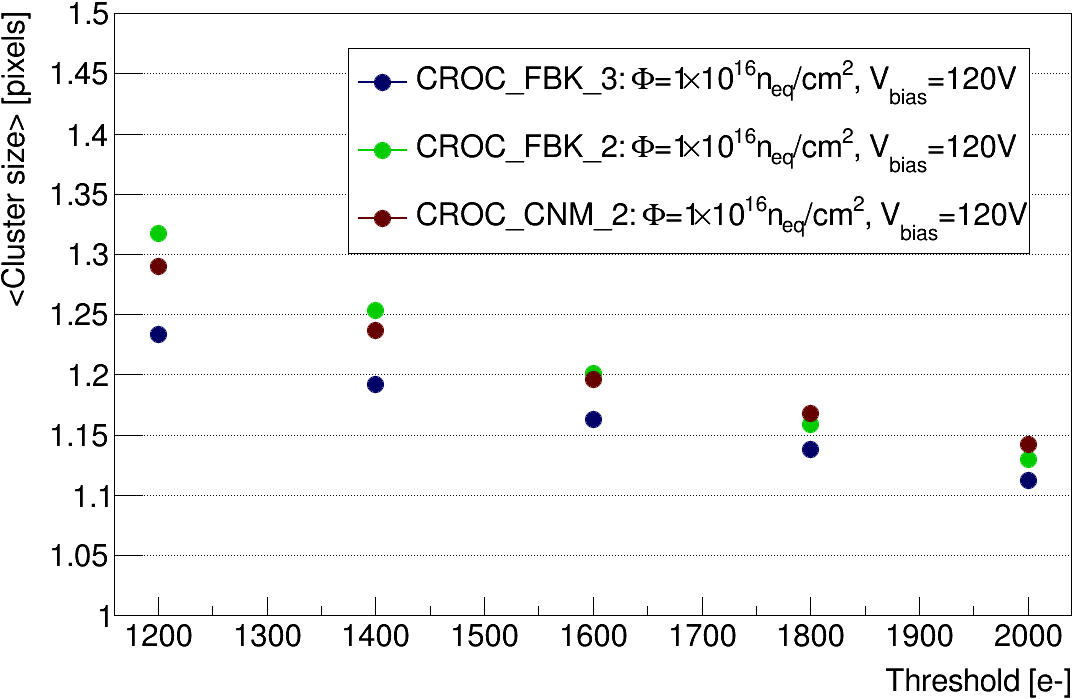}\label{fig:clsize_thr_1e16}}
\hspace{0.5cm}
\subfloat[]{\includegraphics[width=7.8cm]{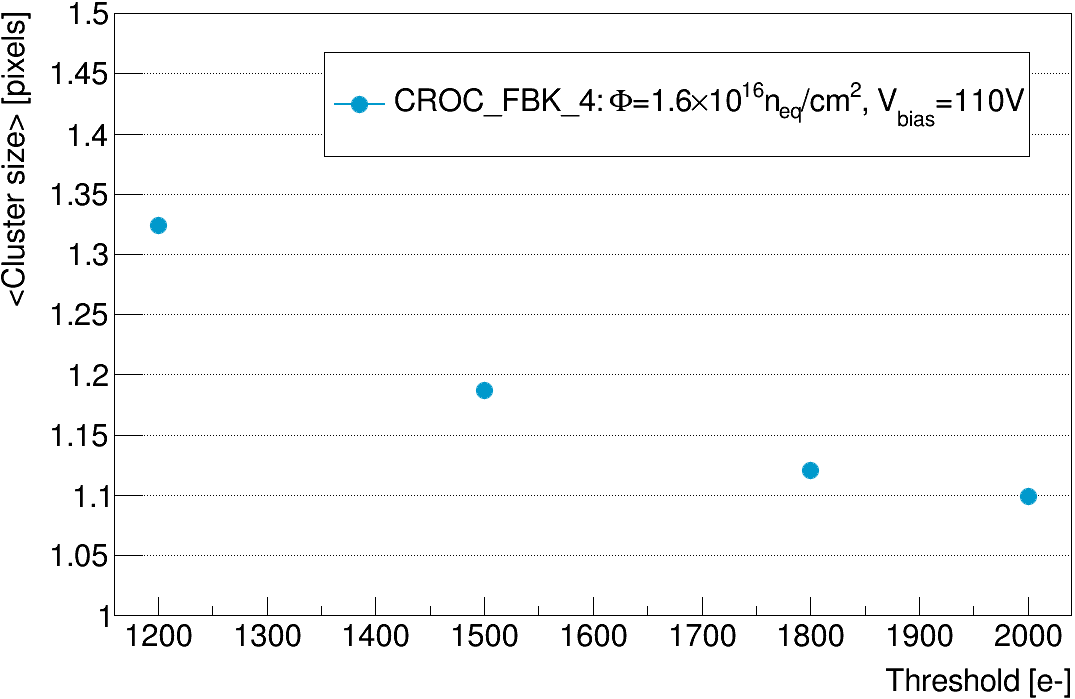}\label{fig:clsize_thr_1d6e16}}
\caption{Average cluster size as a function of the average pixel threshold to which the modules irradiated at (a) \mbox{\SI{1e16}{n_{eq}/\cm^{2}}} and (b) \mbox{\SI{1.6e16}{n_{eq}/\cm^{2}}} have been tuned.}
\label{fig:clsize_thr_scan_irrad}
\end{figure}

The importance of tuning the sensors to a threshold as low as possible is evident not only from the hit detection efficiency but also from the cluster size. Figure~\ref{fig:clsize_thr_scan_irrad} shows the average cluster size as a function of the average pixel threshold of modules irradiated to \mbox{\SI{1e16}{n_{eq}/\cm^{2}}} and \mbox{\SI{1.6e16}{n_{eq}/\cm^{2}}}. These measurements were performed at normal beam incidence, without noise mask and at bias voltages of 120 and $\SI{110}{V}$, respectively. The cluster size decreases by around 20\% at both fluences when doubling the threshold, since shared charge between neighboring pixels increasingly falls below the detection threshold.

%%%%%%%%%%%%%%%%%%%%%%%%%%%%%%%%%
%%%%%%%%%%%%%%%%%%%%%%%%%%%%%%%%%

\section{Conclusions}
\label{sec:conclusions}
The tracker group of the CMS Collaboration has carried out an extensive R\&D program aimed at qualifying 3D pixel sensors for deployment in the innermost layer of the Inner Tracker, which will be operated during the High-Luminosity LHC phase. Two manufacturers were involved in the program: Fondazione Bruno Kessler (FBK, Italy) and Centro Nacional de Microelectrónica (CNM, Spain). 

Current technology enables the production of thin active pixel layers with a thickness of approximately $\SI{150}{\micro m}$, supported on a low material budget substrate, resulting in a total sensor thickness of $\SI{250}{\micro m}$. The final cell design features columnar electrodes with diameters approximately one-fifth of the small pixel pitch.

Sensors provided by the two foundries were bump-bonded to the RD53A (small-area prototype) and CROCv1 (full-size prototype) readout chips. The assemblies were subjected to beam tests both before and after irradiation with protons of different energies, up to fluences of \mbox{\SI{2.6e16}{n_{eq}/\cm^{2}}}.

Non-irradiated modules show excellent performance, reaching full depletion at bias voltages below $\SI{10}{V}$ and spatial resolutions as low as $\SI{2.5}{\micro m}$. The loss in hit detection efficiency observed for normally incident particles is completely recovered by a small rotation of the device. Therefore, this effect is not a concern for the final detector, where particles will strike the sensors under a variety of incident angles. CNM sensors exhibit a slightly lower detection efficiency than FBK sensors at normal incidence, due to their larger column diameter.

Measurements of irradiated assemblies show a hit detection efficiency greater than 96\% at normal incidence after exposure to a fluence of \mbox{\SI{1e16}{n_{eq}/\cm^{2}}}, with fewer than 2\% masked channels for applied bias voltages between $\SI{90}{V}$ and $\SI{150}{V}$, well within specifications. CNM and FBK sensors show comparable performance at this irradiation level. The operation margin after irradiation to \mbox{\SI{1.6e16}{n_{eq}/\cm^{2}}} remains reasonably broad, spanning from $\SI{100}{V}$ to $\SI{130}{V}$. The spatial resolution of irradiated devices is measured to be between 5 and $\SI{6}{\micro m}$, an acceptable decrease compared to non-irradiated modules. 

The manufacturing of the production wafers is currently ongoing at FBK, with the integration and checkout of the CMS Inner Tracker planned to start at the end of 2026.

The results presented in this paper are consistent with those reported by the ATLAS Collaboration in~\cite{3datlas} and prove the soundness of the choice of 3D pixel sensors for the innermost layer of the CMS Inner Tracker. Additional laboratory and test beam measurements are planned on $1\times2$ modules to confirm the excellent performance of single-chip assemblies and to assess their stability over long periods of operation.

%\end{linenumbers}

\section*{Acknowledgments}
This work has been developed in the framework of the CERN RD50 Collaboration and has received funding from the Spanish Ministry of Science and Innovation (MCNI/AEI/10.13039/501100011033), with grant reference PID2020-113705RB-C31. It has been supported by the European Union funding programs Horizon 2020 Research and Innovation, under grant agreement No. 101004761 (AIDAInnova) and Horizon Europe, under grant agreement No. 101057511(EURO-LABS). Moreover, it has been developed in the framework of ``Ayudas María Zambrano para la atracción de talento internacional" co-funded by the Ministry of University of Spain and the European Union NextGenerationEU, with reference code: C12.I4.P1. This work is also supported by the Complementary Plan in Astrophysics and High-Energy Physics (CA25944), project C17.I02.P02.S01.S03 CSIC CERN, funded by the Next Generation EU funds, RRF and PRTR funds, and the Autonomous Community of Cantabria.
Additionally, partial support is provided by ICSC – Centro Nazionale di Ricerca in High Performance Computing, Big Data and Quantum Computing, funded by European Union – NextGenerationEU.

This document was prepared by the tracker group of the CMS Collaboration using the resources of the Fermi National Accelerator Laboratory (Fermilab), a U.S. Department of Energy, Office of Science, Office of High Energy Physics HEP User Facility. Fermilab is managed by Fermi Forward Discovery Group, LLC, acting under Contract No. 89243024CSC000002.

The tracker groups gratefully acknowledge financial support from the following funding agencies: BMWFW and FWF (Austria); FNRS and FWO (Belgium); CERN; MSE and CSF (Croatia); Academy of Finland, MEC, and HIP (Finland); CEA and CNRS/IN2P3 (France); BMFTR, DFG, and HGF (Germany); GSRT (Greece); NKFIH K143477 and VLAB at HUN-REN Wigner RCP (Hungary); DAE and DST (India); INFN (Italy); PAEC (Pakistan); SEIDI, CPAN, PCTI and FEDER (Spain); Swiss Funding Agencies (Switzerland); MST (Taipei); STFC (United Kingdom); DOE and NSF (U.S.A.). This project has received funding from the European Union’s Horizon 2020 research and innovation programme under the Marie Sk\l odowska-Curie grant agreement No 884104 (PSI-FELLOW-III-3i) and the European Union NextGenerationEU/PRTR subproject C17.I02.P02.S01.S03 CSIC CERN. Individuals have received support from HFRI (Greece). 

\bibliographystyle{elsarticle-num} 
\bibliography{references}

%%%%%%%%%%%%%%%%%%%%%%%%%%%%%%%%%
%%%%%%%%%%%%%%%%%%%%%%%%%%%%%%%%%

\clearpage
\onecolumn
\appendix
\setcounter{table}{0}
\renewcommand{\thetable}{A\arabic{table}}
\section{DUT summary}
\label{sec:DUTsummary}
\begin{table*}[!h]
    \centering
    \caption{Summary of the devices that have been tested on beam and whose results have been included in the paper. FBK sensors coupled with the RD53A readout chip come from two different productions: modules RD53A\_FBK\_1 and RD53A\_FBK\_2 from the ``Stepper-1'' production and modules RD53A\_FBK\_3, RD53A\_FBK\_4, RD53A\_FBK\_5 and RD53A\_FBK\_6 from the ``Stepper-2'' production. The pixel design of the two productions was the same except the length of the $n^+$ columns, which was shortened by $\SI{15}{\micro m}$ in the Stepper-2 production.}
    \begin{tabular}{lccccccc}
    \hline\hline
     \hspace{0.5cm}Module & Readout & Sensor & Pixel size & Fluence & Irradiation & \hspace{0.7cm}Test beam area &\\
     \hspace{0.65cm}name & chip & manufacturer & [$\SI{}{\micro m^{2}}$] & [$\SI{}{n_{eq}/\,cm^{2}}$] & facility & \hspace{-1cm}non-irradiated & \hspace{-1cm}irradiated\\ 
    \hline\hline
     RD53A\_CNM\_1 & RD53A & CNM & $25\times100$ & $\SI{1.2e16}{}$ & ITA & \hspace{-1cm} - & \hspace{-1cm} FTBF\\
     RD53A\_FBK\_1 & RD53A & FBK & $25\times100$ & - & - & \hspace{-1cm}DESY & \hspace{-1cm} - \\
     RD53A\_FBK\_2 & RD53A & FBK & $25\times100$ & $\SI{1.5e16}{}$ & KIT & \hspace{-1cm} - & \hspace{-1cm}DESY\\
     RD53A\_FBK\_3 & RD53A & FBK & $25\times100$ & $\SI{1.4e16}{}$ & KIT & \hspace{-1cm} - & \hspace{-1cm}DESY\\
     RD53A\_FBK\_4 & RD53A & FBK & $25\times100$ & $\SI{1.8e16}{}$ & KIT & \hspace{-1cm} - & \hspace{-1cm}DESY\\
     RD53A\_FBK\_5 & RD53A & FBK & $25\times100$ & $\SI{2.1e16}{}$ & CERN PS & \hspace{-1cm} - & \hspace{-1cm}SPS\\
     RD53A\_FBK\_6 & RD53A & FBK & $25\times100$ & $\SI{2.6e16}{}$ & CERN PS & \hspace{-1cm} - & \hspace{-1cm}SPS\\
     CROC\_CNM\_1  & CROC  & CNM & $50\times50$  & - & - & \hspace{-1cm}SPS & \hspace{-1cm}-\\
     CROC\_CNM\_2  & CROC  & CNM & $25\times100$ & $\SI{1e16}{}$ & CERN PS & \hspace{-1cm}SPS & \hspace{-1cm}SPS\\
     CROC\_FBK\_1  & CROC  & FBK & $25\times100$ & - & - & \hspace{-1cm}SPS & \hspace{-1cm}-\\
     CROC\_FBK\_2  & CROC  & FBK & $25\times100$ & $\SI{1e16}{}$ & CERN PS & \hspace{-1cm}SPS & \hspace{-1cm}SPS\\
     CROC\_FBK\_3  & CROC  & FBK & $25\times100$ & $\SI{1e16}{}$ & CERN PS & \hspace{-1cm}SPS & \hspace{-1cm}SPS\\
     CROC\_FBK\_4  & CROC  & FBK & $25\times100$ & $\SI{1.6e16}{}$ & KIT & \hspace{-1cm}- & \hspace{-1cm}DESY\\
     \hline\hline
    \end{tabular}
    \label{tab:DUTsummary}
\end{table*}

\newpage

\setlength{\parindent}{0pt}
\setlength{\parskip}{6pt plus 2pt minus 1pt}

\input{TrackerAuthorList_2025_3DPixelSensors}

\end{document}

%% file: TrackerAuthorList_2025_3DPixelSensors.tex
\begin{center}
{\Large \textbf{The Tracker Group of the CMS Collaboration}\par}
\end{center}

\vspace{1cm}

\textcolor{black}{\textbf{Institut~f\"{u}r~Hochenergiephysik, Wien, Austria}\\*[0pt]
W.~Adam, T.~Bergauer, M.~Dragicevic, R.~Fr\"{u}hwirth\cmsAuthorMark{1}, H.~Steininger}

\textcolor{black}{\textbf{Universiteit~Antwerpen, Antwerpen, Belgium}\\*[0pt]
W.~Beaumont, T.~Janssen, H.~Kwon, D.~Ocampo Henao, T.~Van Laer, P.~Van~Mechelen
% ...
}

\textcolor{black}{\textbf{Texas~A\&M~University, College~Station, USA}\\*[0pt]
R.~Eusebi}

\textcolor{black}{\textbf{Vanderbilt~University, Nashville, USA}\\*[0pt]
P.~D'Angelo, W.~Johns}

% ...
1: Also at Vienna University of Technology, Vienna, Austria \\
% ...
